\documentclass[a4paper,11pt]{article}
\pdfoutput=1

\usepackage{jinstpub}
\usepackage[utf8]{inputenc}
\usepackage{amsmath}
\usepackage{gensymb}
\usepackage{graphicx}
\usepackage{verbatim}
\usepackage{lineno}
\usepackage{float}
\usepackage{siunitx}
\usepackage{textcomp}

\title{Construction of precision wire readout planes for the Short-Baseline Near Detector (SBND)}

\collaboration{SBND Collaboration}

\author[15]{R. Acciarri}
\author[1]{C. Adams}
\author[21]{C. Andreopoulos}
\author[33]{J. Asaadi}
\author[6]{M. Babicz}
\author[34]{C. Backhouse}
\author[15]{W. Badgett}
\author[15]{L. F. Bagby}
\author[29]{D. Barker}
\author[24]{C. Barnes}
% ...

\affiliation[1]{Argonne National Laboratory, Lemont, IL 60439, USA}
% ...
\affiliation[6]{CERN, European Organization for Nuclear Research 1211 Geneve 23, Switzerland, CERN}
% ...
\affiliation[15]{Fermi National Accelerator Laboratory, Batavia, IL 60510, USA}
% ...
\affiliation[21]{University of Liverpool, Liverpool L69 7ZE, United Kingdom}
% ...
\affiliation[24]{University of Michigan, Ann Arbor, MI 48109, USA}
% ...
\affiliation[29]{University of Sheffield, Department of Physics and Astronomy, Sheffield S3 7RH, United Kingdom}
% ...
\affiliation[33]{University of Texas at Arlington, TX 76019, USA}
\affiliation[34]{University College London, London WC1E 6BT, United Kingdom}
% ...

\emailAdd{sbnd\_info@fnal.gov}

\abstract{The Short-Baseline Near Detector time projection chamber is unique in the design of its charge readout planes. These anode plane assemblies (APAs) have been fabricated and assembled to meet strict accuracy and precision requirements: wire spacing of 3~mm $\pm$ 0.5 mm and wire tension of 7~N $\pm$ 1~N across 3,964 wires per APA, and flatness within 0.5~mm over the 4~m $\times$ 2.5~m extent of each APA. This paper describes the design, manufacture and assembly of these key detector components, with a focus on the quality assurance at each stage.}

\arxivnumber{} 

\begin{document}

\maketitle
\flushbottom


\section{Introduction}


The Fermilab Short-Baseline Neutrino Program~\cite{Machado:2019oxb}\cite{Antonello:2015lea} will make a broad range of neutrino physics measurements in the Booster Neutrino Beam (BNB)~\cite{Stancu:2001cpa} using a suite of three liquid argon time projection chambers (LAr TPCs)~\cite{Rubbia:1977zz}. These measurements will include searches for short-baseline oscillations indicative of the existence of sterile neutrinos, measurements of neutrino-argon interactions, and searches for physics beyond the standard model such as heavy neutral leptons and millicharged particles. The closest detector to the beam dump in SBN will be SBND, the Short-Baseline Near Detector, which is currently under construction and will join the existing downstream detectors MicroBooNE~\cite{Acciarri:2016smi} and ICARUS T-600~\cite{Amerio:2004ze}.

\begin{figure}[h!]
\centering
\includegraphics[width=0.45\textwidth]{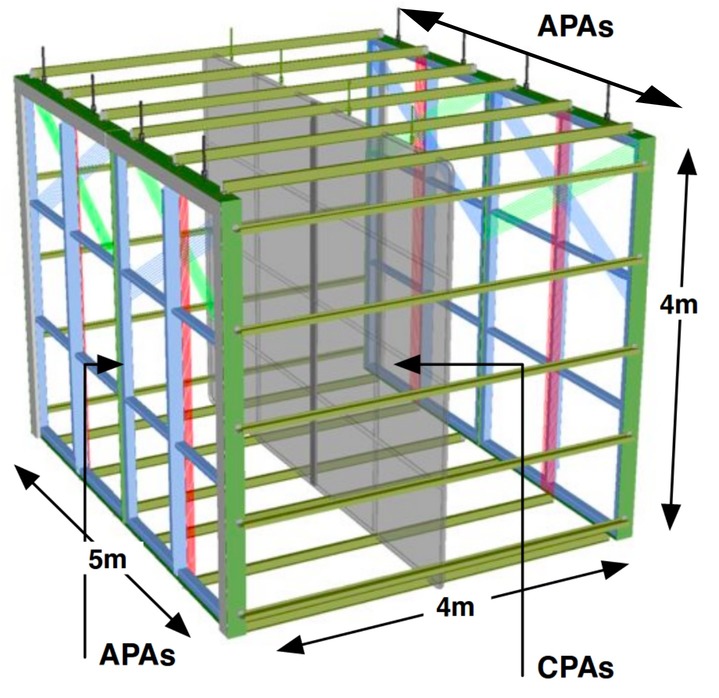}
\caption{A model of the SBND TPC, showing the positions of the anode plane and cathode plane assemblies (APAs and CPAs).}
\label{fig:SBNDDetector}
\end{figure}

The SBND detector is a single phase LAr TPC with an active mass of 112 t, shown in figure~\ref{fig:SBNDDetector}. A detailed description of the full detector and its design justifications can be found in \cite{Antonello:2015lea}. The TPC, which is fully immersed in liquid argon, consists of a central 5~m $\times$ 4~m cathode wall formed of two 2.5~m $\times$ 4~m cathode plane assemblies (CPAs) in between two anode walls, each 2~m away from the cathode. The anode walls are also 5~m $\times$ 4~m in dimension, and each consists of two 2.5~m $\times$ 4~m anode plane assemblies (APAs). The CPAs are held at $-100$~kV with the APAs at ground. The resulting 500~V/cm electric field causes ionisation electrons produced by the passage of the charged particles through the liquid argon to drift to the APAs, where they are read out. 

\begin{figure}[h!]
\centering
\includegraphics[height=0.225\textheight]{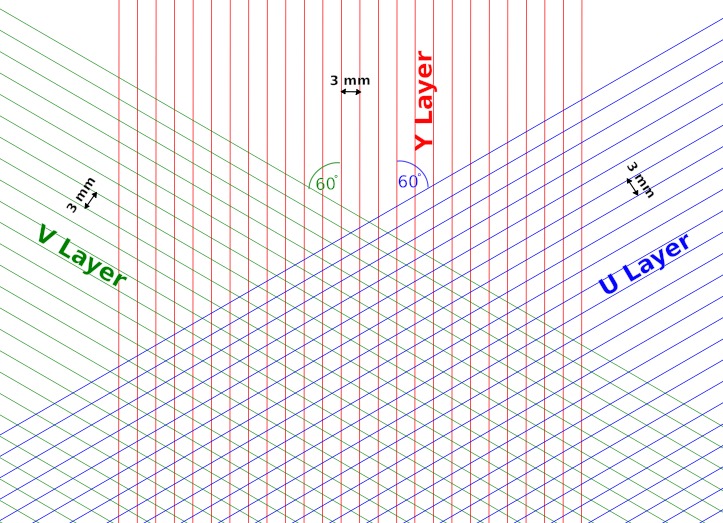}
\includegraphics[height=0.225\textheight]{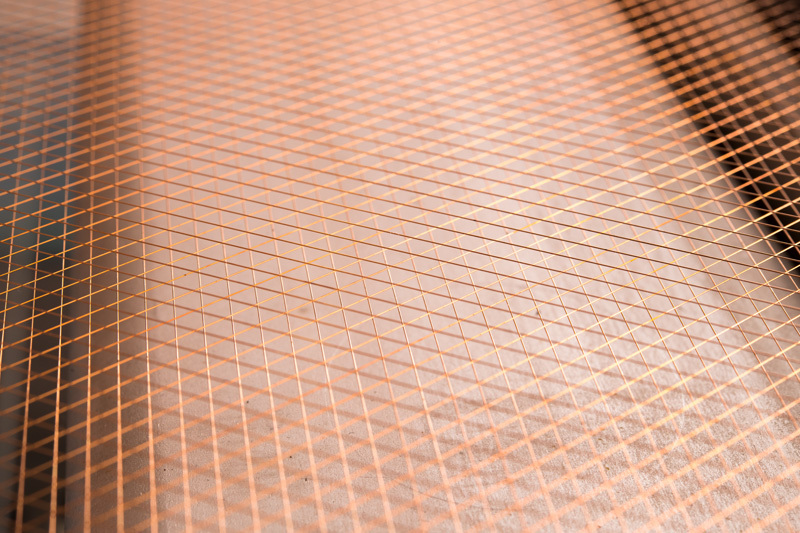}
\caption{Left: a diagram showing the layering of the wires (with the beam going from left to right). Right: a photograph of the wire layers on a completed APA.}
\label{fig:DoesntExist}
\end{figure}

To achieve the readout, the side of the APA that is facing the CPA is covered with three layers of wires, forming a grid as shown in figure \ref{fig:DoesntExist}. 

\begin{figure}[h!]
    \centering
    \includegraphics[width=0.7\textwidth]{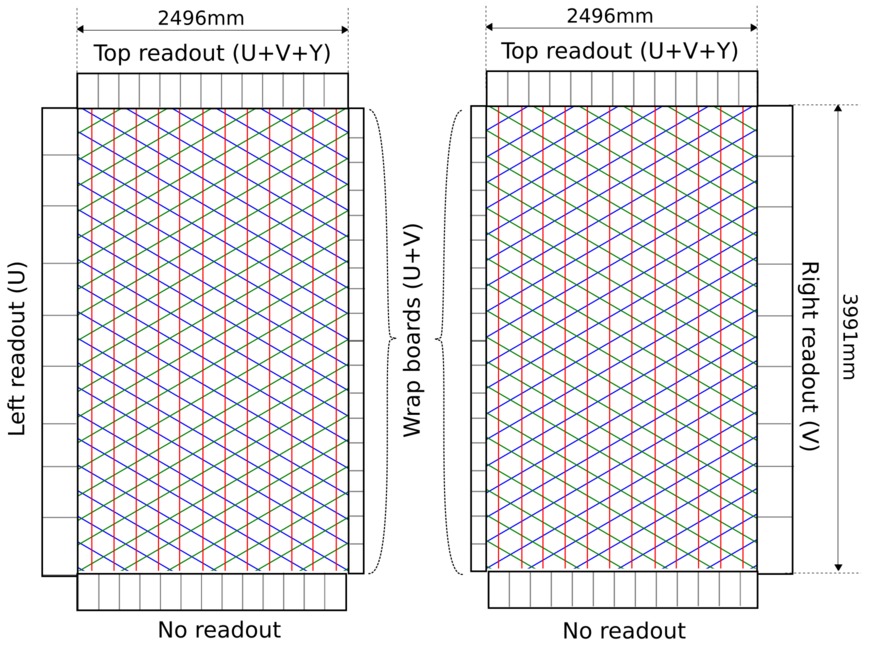}
    \caption{A diagram showing the positioning of the boards around an APA pair.}
    \label{fig:GeomBoardsOverhead}
\end{figure}

The two layers closest to the cathode, called the U layer and V layer, have wires spaced by 3 mm and at ±60 degrees to the vertical. These are induction layers, and the wires produce a differential current pulse as the ionisation electrons drift past. The layer furthest from the cathode, called the Y layer, has its wires running vertically, again spaced by 3 mm. This is the collection layer, and the wires produce a unipolar current pulse as the ionisation electrons are collected onto them~\cite{Adams:2018gbi}. As shown in figure \ref{fig:GeomBoardsOverhead}, the wires are read out by electronics boards on the top and outside vertical edges of the APAs. Where the two APAs in each wall meet each other, `wrap' boards form a one-to-one connection between induction wires on each APA, so that the entire anode wall acts as a single grid.

The pattern of observed current pulses allows the position of the ionisation on the anode wall to be reconstructed with millimeter precision. To reconstruct the distance from the anode wall at which the ionisation occurred, the arrival time of the ionisation electrons at the anode (which takes approximately 1~ms from the events closest to the cathode) is compared with the time of a prompt flash of scintillation light that accompanies the ionisation (the fast component of which takes less than 10 ns to arrive). This scintillation light is detected by an array of photon detectors placed behind the APAs. Through these means, full three-dimensional reconstruction of the passage of charged particles through the TPC is achieved, which enables properties of the interacting neutrinos, such as their flavours and energies, to be measured.

\begin{figure}[t]
    \centering
    \includegraphics[width=0.8\textwidth]{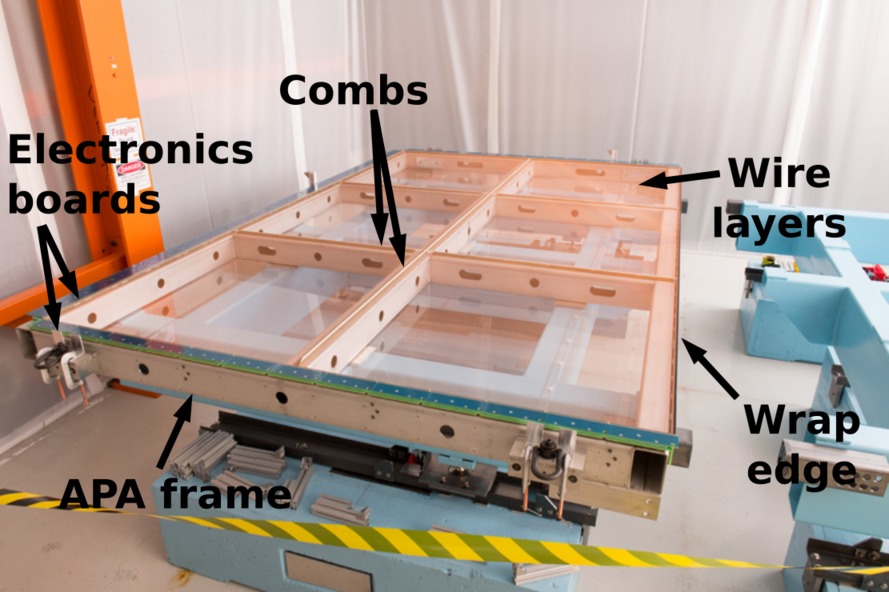}
    \caption{A photograph of one of the SBND APAs.}
    \label{fig:APAphoto}
\end{figure}

\begin{figure}[b]
    \centering
    \includegraphics[height=0.22\textheight]{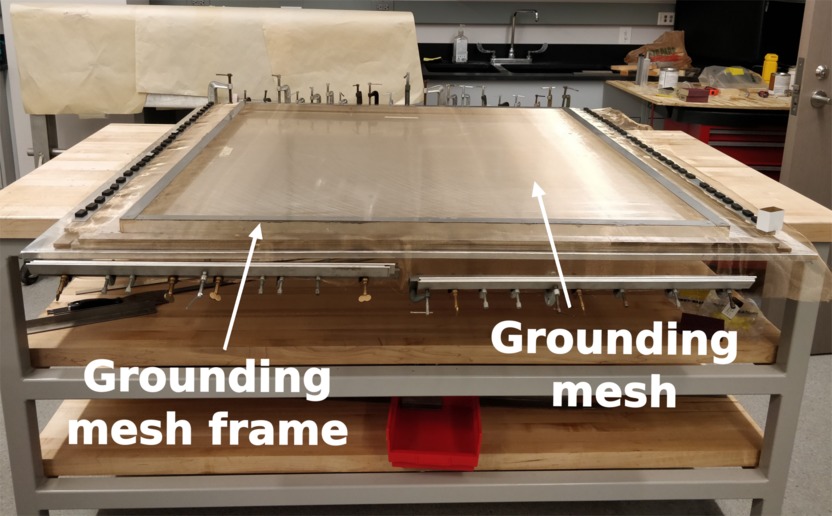}
    \includegraphics[height=0.22\textheight]{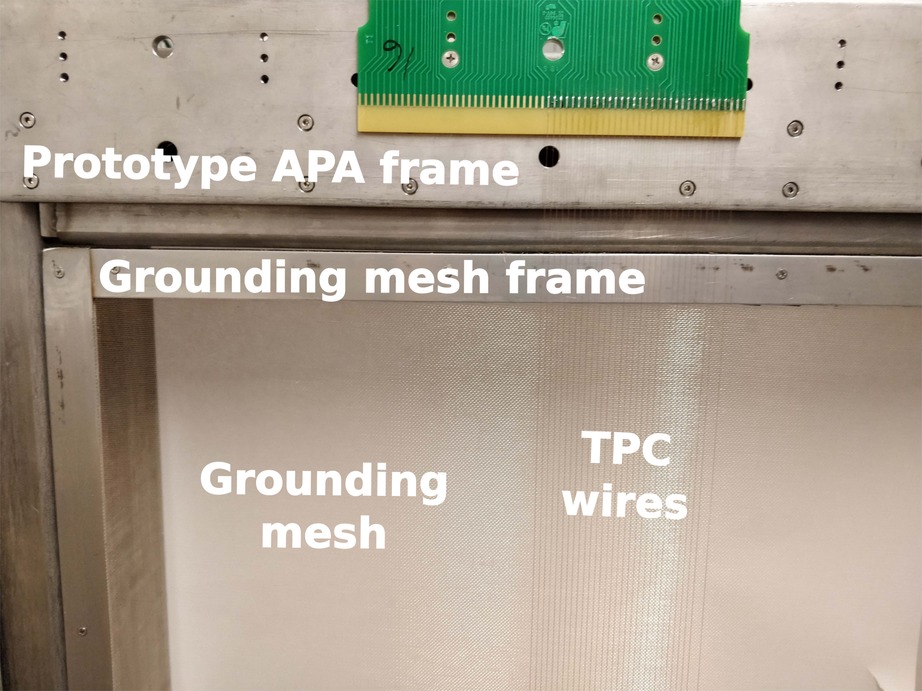}
    \caption{Photographs of an APA grounding mesh unit under construction (left) and being test-fitted into a prototype APA frame opening (right).}
    \label{fig:meshframe}
\end{figure}

Figure \ref{fig:APAphoto} shows one of the APAs that bears the readout planes, which is composed of several key parts. First is the steel APA frame that provides a rigid structure on which to mount the readout, and controls the levelling of the wire layers. Second are the wire layers themselves, and the electronics boards that anchor them. Third are the wire support combs that cross the wire layers at regular intervals, providing mechanical support and maintaining the wire spacing across the full area of the readout plane. These components are explored in detail in section \ref{sec:APAdetails}. Also, in order to produce a uniform ground plane behind the TPC wires, a bronze wire woven mesh screen is installed flush with the frame surface closest to the wires. 
The mesh, which is 85\% transparent to allow photons to reach the photon detectors mounted behind the APAs, is stretched and secured in 24 stainless steel sub-frames that are installed into the six "windows" of each APA frame from the backside during TPC construction.  Figure \ref{fig:meshframe} shows one of the grounding mesh frames being stretched and secured into the steel sub-frame and later installed into a prototype APA frame just behind the TPC wires.     

The uniformity of each APA is critical for operation of the detector, since the electric fields in this region are determined by the position of each wire. This motivates the construction of the APA to exacting specifications. This paper will detail the specifications to which the SBND APAs were constructed, and the methods by which those specifications were met.

\section{SBND Wire Plane Specifications}\label{sec:SBNDSpecs}

The specifications for the SBND wire planes were set to ensure that the readout wires and thus the detector response) are uniform in space, and to optimise the signal-to-noise ratio and spatial resolution: decreasing wire spacing gives greater resolution at the price of a reduced signal-to-noise ratio (since the signal received by each wire scales with the number of ionisation electrons produced by a particle traversing the wire spacing). The specifications are summarised in table \ref{tab:WireSpecs}.  During each stage of construction the adherence to the design criteria was assessed, using quality control techniques described in section \ref{sec:Testing}.

\begin{table}[htbp]
    \centering
    \caption{\label{tab:WireSpecs} The key specifications for SBND APA wiring.}
    \smallskip
    \begin{tabular}{|p{4cm}|p{10cm}|}
        \hline
        Test & Requirement \\
        \hline
        \hline
        Wire tension at time of wire laying & $7 \pm 1$~N \\
        \hline
        Equilibrium wire tension (after slackening) & $5 \pm 1$~N \\
        \hline
        Wire spacing & 3 $\pm$ 0.5~mm \\
        \hline
        Layer spacing & 3 $\pm$ 0.5~mm \\
        \hline
        Electrical continuity & $<$ 500~$\Omega$ \\
        \hline
        Electrical isolation & $>$ 10~M$\Omega$
        between adjacent wires\\
        \hline
        Cold test & All properties hold after being cooled to approximately 150~K and then returned to room temperature\\
        \hline
    \end{tabular}
\end{table}

The uniformity of wire plane spacing and the tension on the wires are two key parameters whose stringent specification has driven many of the choices made in the design of the APAs. The motivations for these specifications are discussed below.

\subsection{APA Flatness Specification} 




Each wire layer in an APA is held at an electrical potential (as shown in table \ref{tab:BoardComponents}) such that the electrons pass by the induction planes, inducing a charge signal. They are then collected by the wires of the collection plane. The wire layer bias voltages were chosen to fulfil the criteria for electron transparency, such that the drifting electrons are not collected on the two induction layers.  
The ratio of electric fields needed to achieve electron transparency is given by the formula

\begin{equation}
    \frac{E_B}{E_A} \geq\frac{1+\rho}{1-\rho}~ ,
\end{equation} ~\cite{Bevilacqua}

\noindent where $E_A$ and $E_B$ are the electric fields before and after each wire plane, $\rho = 2\pi\frac{r}{d}$, $d$ is the wire spacing, and $r$ is the wire radius. 

Distortions of the wire layers (either from layer to layer, or from wire to wire within a layer) lead to changes in $d$, reducing the electron transparency and therefore the collection efficiency of the readout plane. Simulation studies showed that wire layer non-uniformity greater than 0.5~mm would allow electron transparency to fall below 90\% in certain regions of the detector. As a result of these studies a tolerance of $\pm$ 0.5~mm was set on the flatness of the wire planes. This overall flatness tolerance encompasses all flatness deviations due to twist, bow and folds in the APA frame material. It has been verified at every stage of APA construction, including after cooling to cryogenic temperatures (as discussed in section \ref{sec:ColdTest}).



\subsection{Wire Tension Specification}\label{subsec:TensionOverTime}
The wire used on the SBND APAs is tempered copper-beryllium alloy ($98 \%$ Cu, $2 \%$ Be), with a diameter of $150~\mu$m. The nominal tension specification of 7~N for this wire arose from two main considerations. Firstly, the wire tension must be high enough to keep the wires rigid and at the correct spacing at all times, temperatures and voltages, since argon circulation in the cryostat and the application of the bias voltage have the potential to induce oscillation in slack wires. Oscillations could induce noise on the wires or in the most extreme case allow neighbouring wires to touch. Secondly, the wire tension must be low enough to pose no risk of breakage at any point during installation or operations (with the breaking strength of the wire used on SBND being approximately 25~N). This is most relevant as the APA cools to liquid argon temperatures, as the wires have a lower thermal coefficient than the stainless steel of the frame and a greater ratio of surface area to volume, so they contract more rapidly in the cold, leading to a temporary increase in tension on the order of 1~N.

The tension required for the wires to remain taut and evenly-spaced is dependent on the wire length. For the longest wires (approximately 4~m long in the Y layers), this tension was determined to be 5~N, using computational fluid dynamics and finite element analysis simulations, assuming a maximum argon flow of 50~mm/s. For ease of assessment at the wiring sites, no attempt was made to set varying tension specifications as a function of wire length. Instead, we require all wires to meet the tension requirements of the longest wires.

\begin{figure}[h!]
\centering
\includegraphics[width=0.6\textwidth]{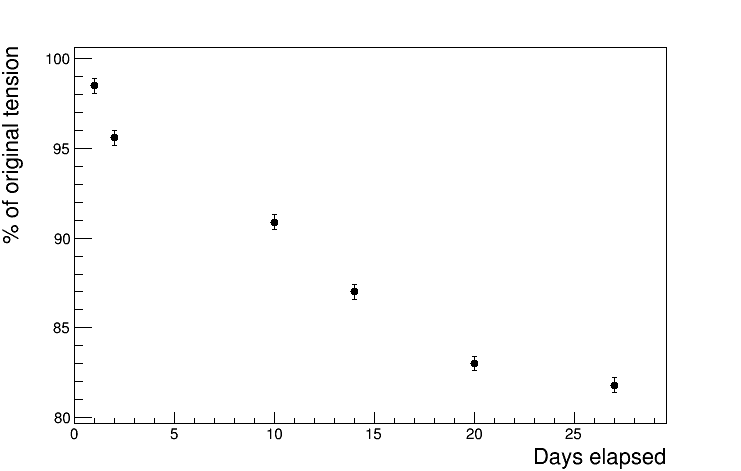}
\caption{Tension loss over time as seen on the prototype frame, as sampled over 64 wires with a mean initial tension of 6.21 N. The tensions were measured and the uncertainties determined using the laser photodiode system discussed in section \ref{subsec:TensionTests}.}
\label{fig:TensionLossOverTime}
\end{figure}

During preliminary wiring tests, the wire tension was observed to decrease as a function of time, as the wires physically stretch under the load. The decrease over the course of approximately one month is shown in figure \ref{fig:TensionLossOverTime}.  

This data was taken from a prototype frame with wires of up to 1.5~m in length, which was wired at the Daresbury laboratory in the UK using the wiring process as described in section \ref{subsec:UK-Winding-Procedure}. The wires were observed to lose approximately $20\%$ of their original tension before settling into equilibrium. The 7~N specification was chosen to ensure that all wires would remain above 5~N once their tension had reached equilibrium, under the assumption that the wires on the final frames would experience the same proportional tension loss.



\section{The Anode Plane Assemblies}
\label{sec:APAdetails}

The APAs are composed of four main components: the frame, the wires, the electronics boards, and the combs. This section discusses the construction of the frames, boards, and combs; the construction of the wire layers is discussed in section \ref{sec:WiringMethods}.

\subsection{APA Frame}\label{subsec:APAFrame}

\begin{figure}[h!]
    \centering
    \includegraphics[width=0.99\textwidth]{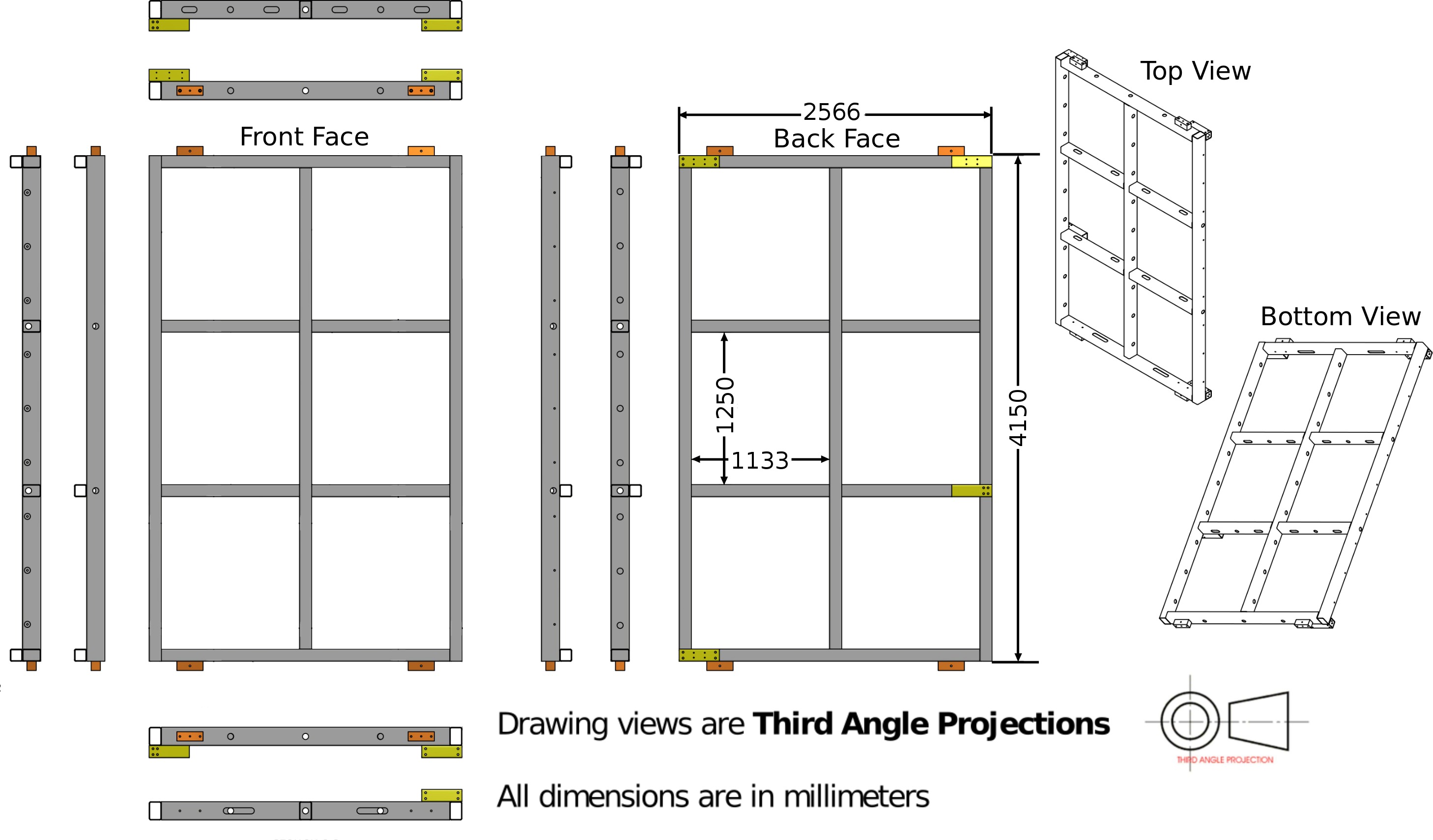}
    \caption{An engineering drawing of the stainless steel structure of a single APA.}
    \label{fig:APAOverview}
\end{figure}

\begin{figure}[h!]
    \centering
    \includegraphics[width=0.6\linewidth]{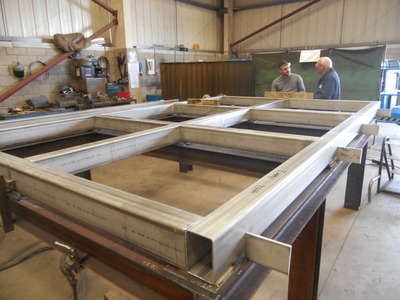}
    \caption{The tack welded APA frame in manufacturing jig.}
    \label{fig:FramePreWeld}
\end{figure}

The SBND APA frame is a stainless steel structure (shown in figures \ref{fig:APAOverview} and \ref{fig:FramePreWeld}) with outer dimensions of 2166~mm $\times$ 4160~mm, composed of welded stainless steel 304 rectangular hollow section (RHS) tubes, with dimensions 150~mm $\times$ 100~mm, and wall thickness 5~mm.  

Since the three layers of readout wires are supported by this frame, its curvature and flatness have direct consequences for the flatness and spacing of the wires. To achieve the 0.5~mm flatness constraint discussed in section \ref{sec:SBNDSpecs}, adjustable levelling bars are attached across the front and wrap-side surfaces of the frame (where the electronics boards that bear the wires attach). The boards that bear the wire layers then mount to the bars.

The production process to achieve this precision flatness was a multi-step procedure. At the material stage, the RHS is selected for flatness, choosing the flattest pieces available for the long vertical tubes. The welding procedure was developed to maintain flatness, with initial tack welding at the corners followed by full fillet welds. 

\begin{figure}[h!]
    \centering
    \includegraphics[angle=-90,origin=c,width=0.25\linewidth]{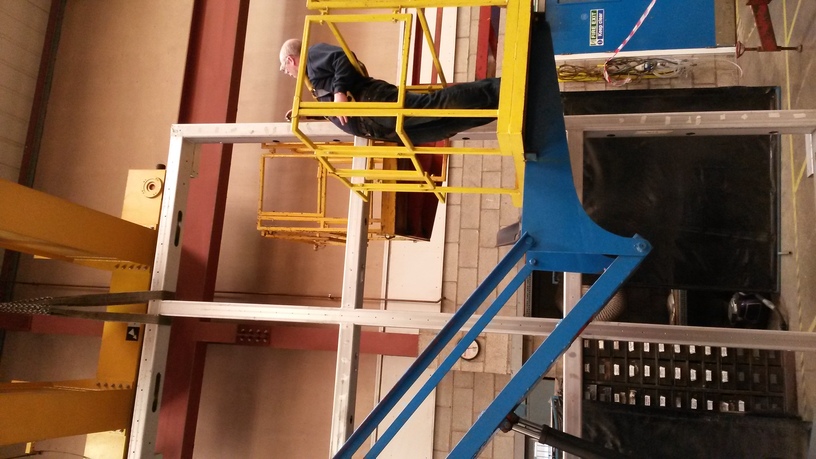}
    \includegraphics[angle=-90,origin=c,width=0.25\linewidth]{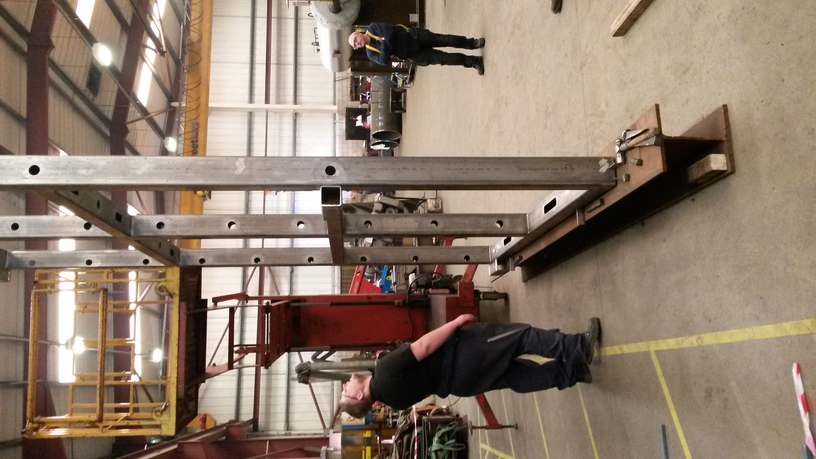}
    \caption{An APA frame suspended vertically for the laser tracker ball survey of the front face and wrap edge flatness.}
    \label{fig:Survey}
\end{figure}
\begin{figure}[h!]
    \centering
    \includegraphics[width=0.6\textwidth]{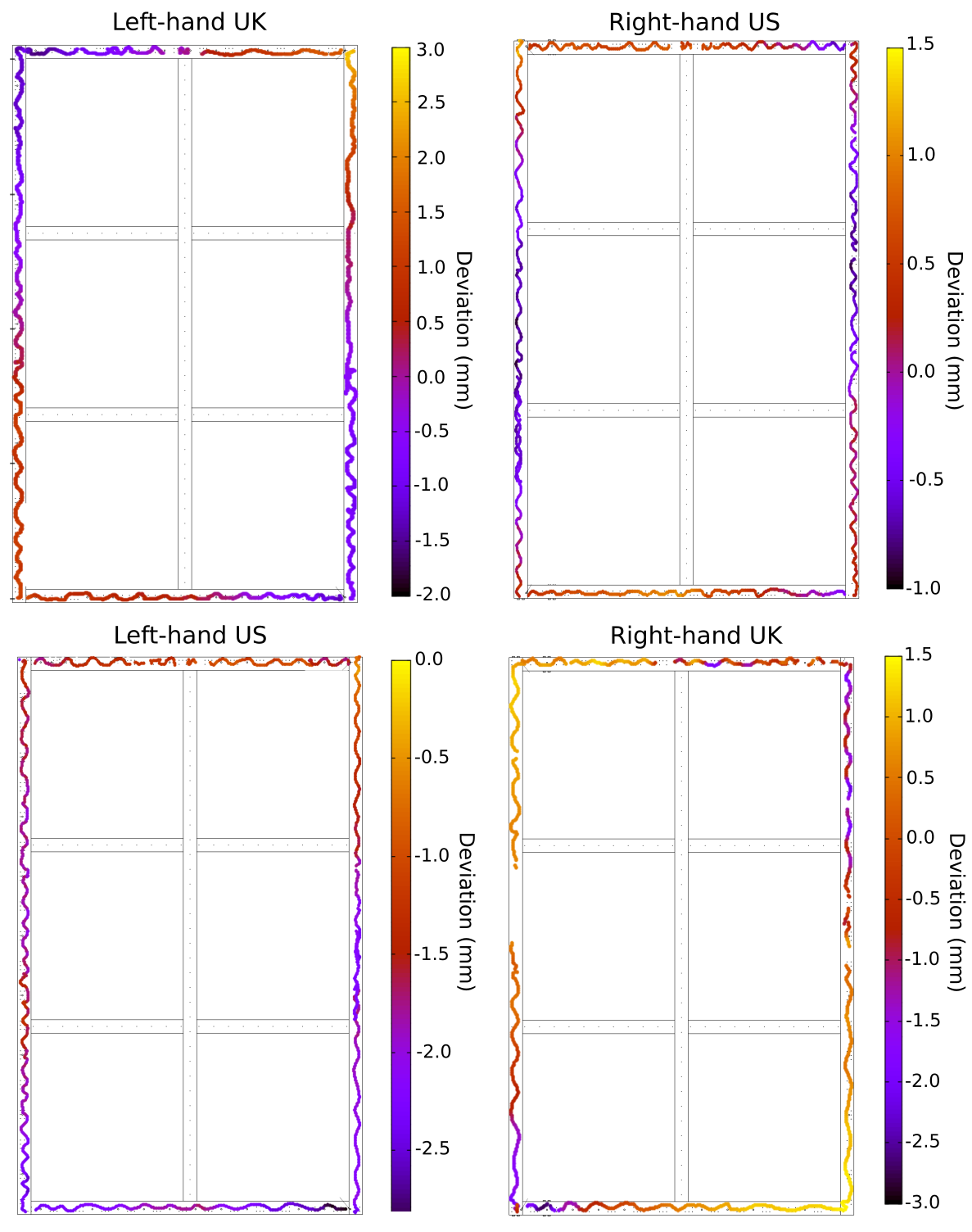}
    \caption{The deviation of the bare APA frames from their respective best fit planes. The data points correspond to a continuous measurement taken across the surface of the APA frame using the laser tracker.}
    \label{fig:FrameFlatnessPreLevelling}
\end{figure}

A high precision measurement of the bare frame flatness was made using a FARO\textsuperscript{\textregistered} laser tracker ball, with the APA suspended in the vertical orientation (as shown in figure \ref{fig:Survey}) to mimic the position of the APA in the TPC. The overall flatness of the initial welded structure was $\pm$2~mm for all frames.  These results are presented in figure \ref{fig:FrameFlatnessPreLevelling}. The welded frame labels given here are used to distinguish between the frame chirality and the wiring site location. 

The laser survey data was used to create a map of the shims needed to produce a level surface across the front face and wrap edge of each frame.  This shim map relates the flatness of the bare frame to the levelled frame. 

\begin{figure}
     \centering
     \includegraphics[height=0.25\textheight]{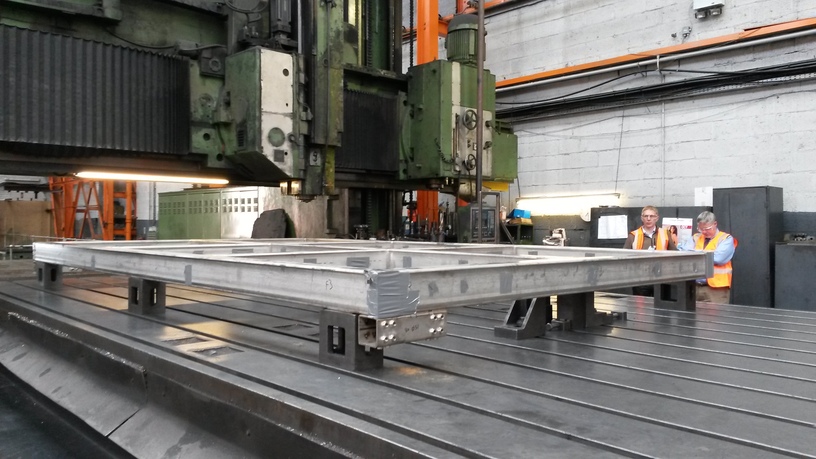}
     \includegraphics[height=0.25\textheight]{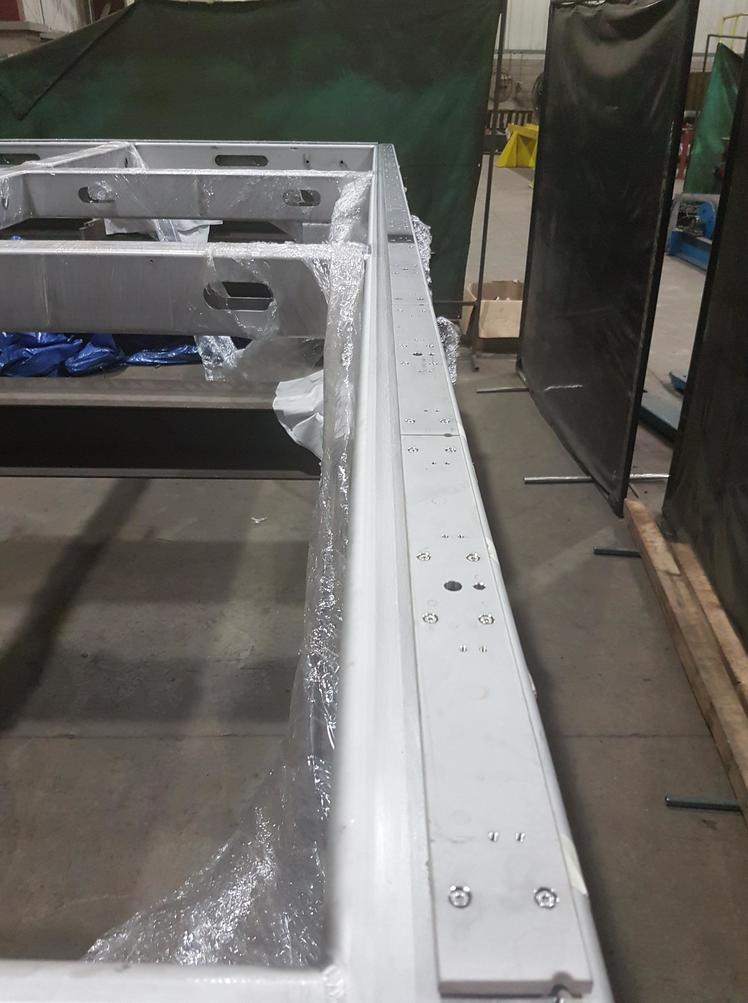}
     \caption{Left: APA frame machining. Right: APA frame with levelling bars attached.}
     \label{fig:LevelledFrame}
 \end{figure}

  Post-welding, there are 911 holes which are machined into the frame (shown in figure \ref{fig:LevelledFrame}), which form the basis for the detector assembly using the APA frame.  They allow for mounting not only the levelling bars (and hence the wires), but also the connections to the rest of the TPC, photon detection system, and the cryostat. The hole positions to be machined in the bare frame were determined using the shim map. The holes must be positioned on the APA frame relative to the final, levelled frame surface, therefore the hole pattern machined on the bare frame is a transformation of these locations, based on the laser survey information.  This transformation enables the holes to be precisely located relative to the final levelled frame surface rather than the bare frame itself. Figure \ref{fig:LevelledFrame} shows this machining process, and levelling bars attached in position on the front face of the frame.
 
 
The frame was then chemically cleaned, before attaching the levelling bars with the calculated thicknesses of shims.  This method takes into consideration all deviations from flatness in the bare frame due to twist, bow or folds in the RHS, and produced a levelled surface.  Due to certain localised features on the frame, such as weld seams, it was necessary to cut these shim packs in certain places in order to achieve the desired flatness: between levelling bars and at corners, the flatness was measured with a straight edge and feeler gauges and the shims were iteratively adjusted to create a flatness within tolerance. 

\begin{figure}[h!]
    \centering
    \includegraphics[width=0.7\textwidth]{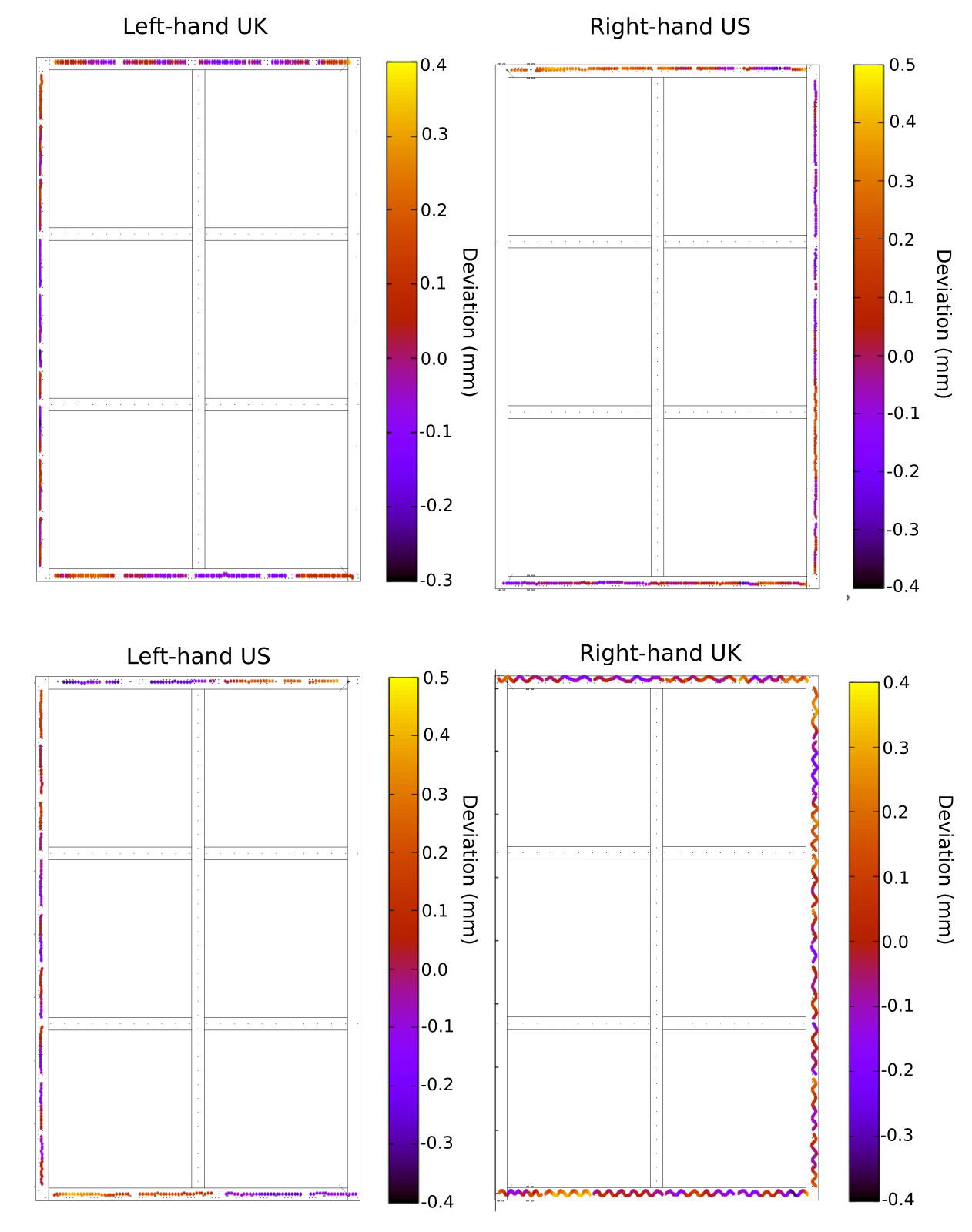}
    \caption{The deviation of the levelled APA frames from their respective best fit planes.}
    \label{fig:levelledAPAsurveys}
\end{figure}

The levelled frame was then surveyed as before, and where necessary the shim packs were adjusted to optimise the flatness.  The final results are shown in figure \ref{fig:levelledAPAsurveys}. Table \ref{tab:flatnessresults} shows the maximal deviation from flatness across the levelled front face and wrap edges of the frame respectively. The wiring surfaces on all four levelled frames are shown to have flatness well within the specified tolerance of $\pm 0.5$ mm.

\begin{table}[hbtp]
   \centering
   \caption{\label{tab:flatnessresults} The levelled flatness with maximum deviation observed, with the APA frames supported in a horizontal position as shown in figure \ref{fig:Survey}.}
   \begin{tabular}{|c|c|c|}
   \hline
     Frame & Front face flatness (mm) & Wrap edge flatness (mm) \\
       \hline 
       \hline
     UK left-hand & $\pm 0.30$ & $\pm 0.30$ \\
     \hline
     US right-hand & $\pm 0.36$ & $\pm 0.27$ \\
       \hline
     US left-hand & $\pm 0.37$ & $\pm 0.25$ \\
       \hline
     UK right-hand & $\pm 0.37$ & $\pm 0.28$ \\
     \hline
\end{tabular}
\end{table}



\subsection{Electronics Boards}\label{subsec:GeomBoards}

\begin{figure}[h!]
    \centering
    \includegraphics[width=0.95\textwidth]{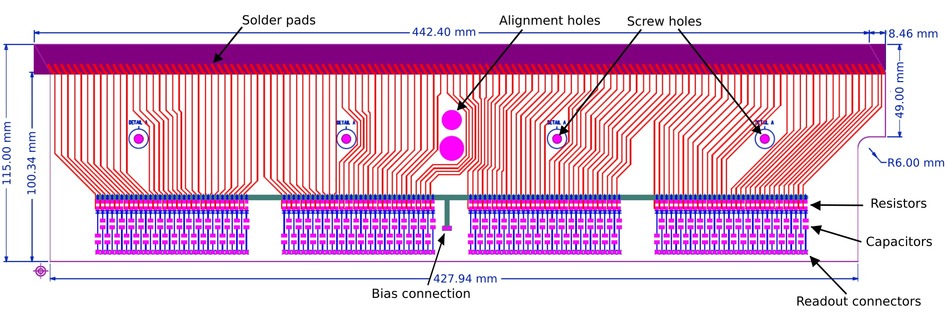}
    \caption{A board layout design of a U layer side readout board.}
    \label{fig:BoardDiagram}
\end{figure}

\begin{figure}[h!]
    \centering
    \includegraphics[width=0.325\textwidth]{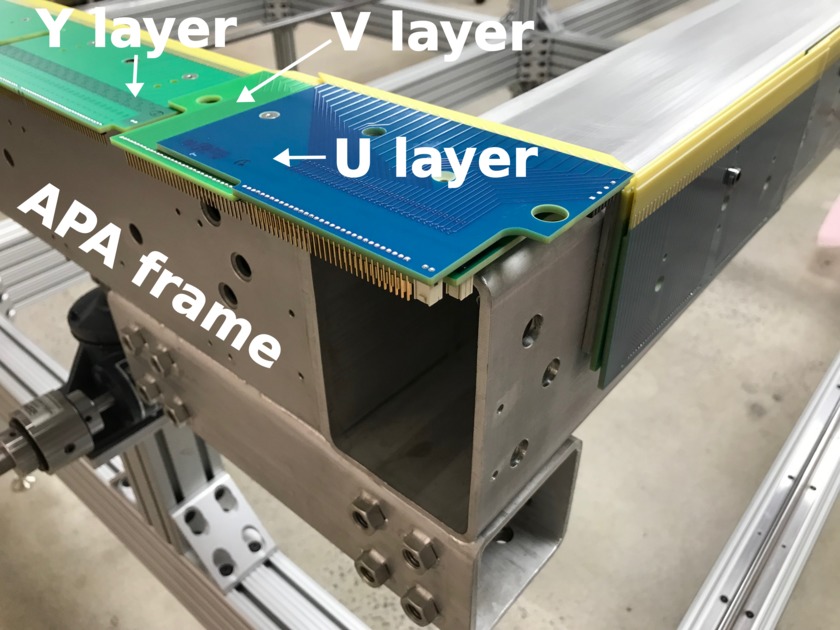}
    \includegraphics[width=0.325\textwidth]{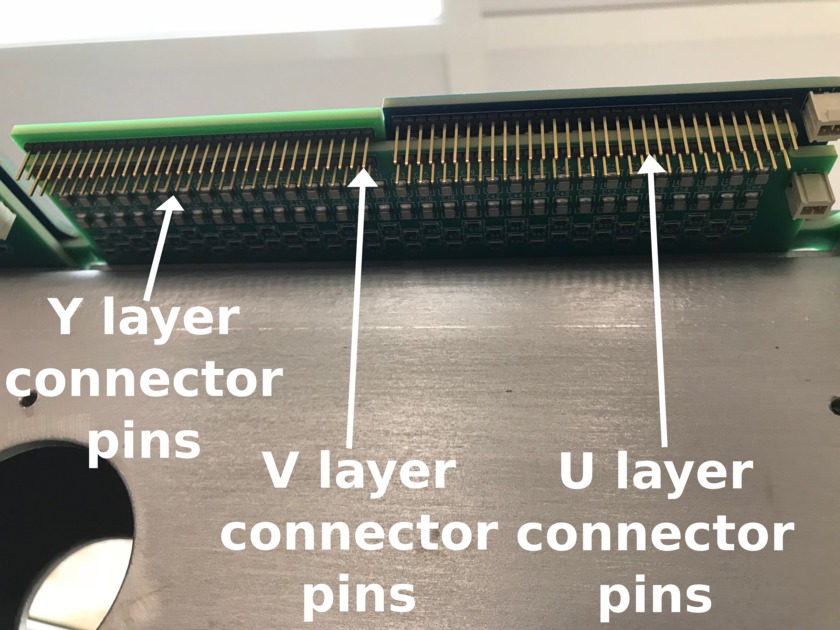}
    \includegraphics[width=0.325\textwidth]{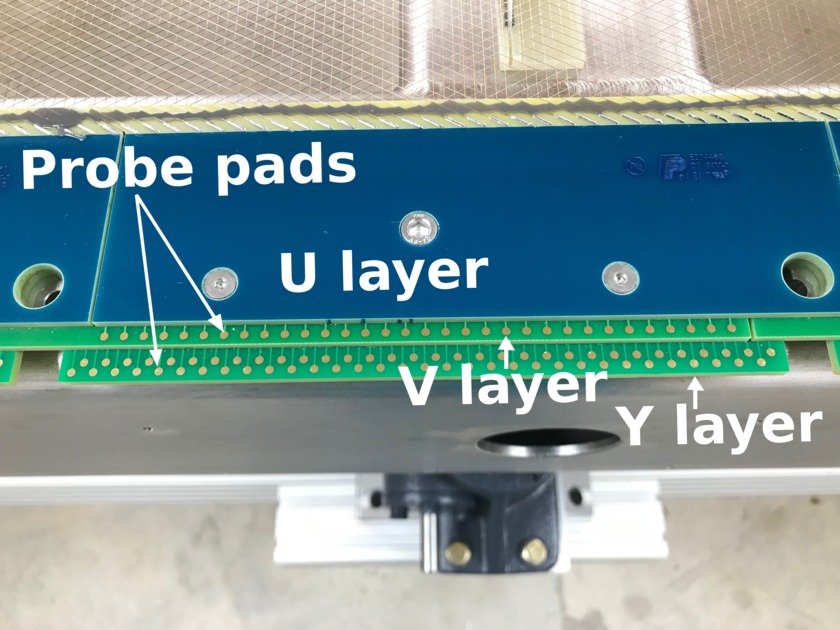}
    \caption{Photos showing all 3 layers of readout boards (left, centre) and non-readout boards (right) on the top side of an APA.}
    \label{fig:GeomBoardPhotos}
\end{figure}

The wires are mounted to the levelled APA frame structure using G10 printed circuit boards, which fix the wire position and provide an interface to the readout electronics. They can be divided into two classes: readout boards (which connect to the readout electronics to extract signals from the wires) and non-readout boards (which mechanically fix the wires in place). These classes subdivide further into eighteen types of board, ranging from approximately 180~mm to 440~mm in length. A board layout design for one of the boards is shown in figure \ref{fig:BoardDiagram}, and photographs of the boards in use at the US wiring site are shown in figure~\ref{fig:GeomBoardPhotos}.

\begin{figure}[h!]
    \centering
    \includegraphics[width=0.95\textwidth]{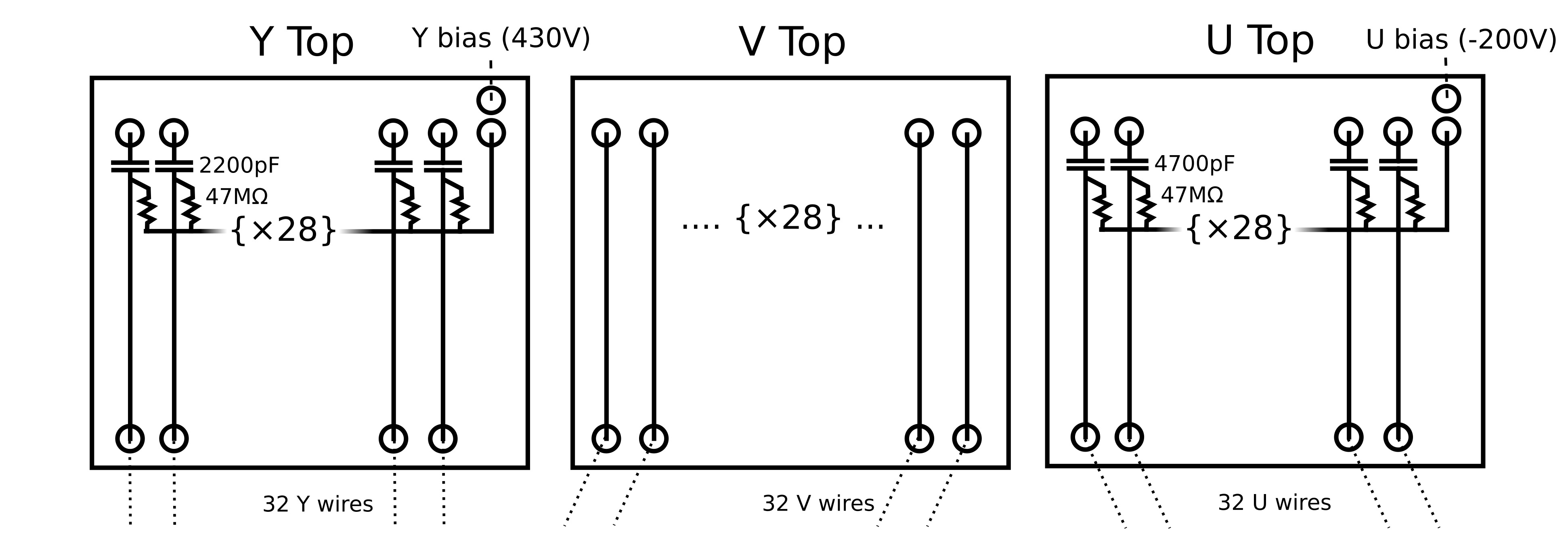}
    \caption{Schematics showing the layout of readout boards for each of the 3 planes.}
    \label{fig:GeomBoardLayouts}
\end{figure}

All boards (whether readout or non-readout) bear a number of solder pads for affixing wires, manufactured with the appropriate angle and spacing for their respective wire layers, with a trace leading from the solder pad to the back of the board. For readout boards, the traces terminate in connector pins that mate with the readout electronics (shown in figure \ref{fig:GeomBoardPhotos}). The readout on the U and Y planes is capacitively coupled to the bias, as is shown in figure \ref{fig:GeomBoardLayouts}; since the V plane is biased at 0~V, it is directly coupled to the readout. The specifications of the components and the bias voltages of the different layers are summarised in table \ref{tab:BoardComponents}.

\begin{table}[hbtp]
   \centering
   \caption{\label{tab:BoardComponents} Specifications for the different electronics boards used in the SBND APAs. All resistors and capacitors are manufactured to 5\% tolerance.}
   \begin{tabular}{|c|c|c|c|}
        \hline
        Plane & Resistance (M$\Omega$) & Capacitance (pF) & Bias Voltage (V) \\
        \hline
        \hline
        Y & 47 & 2200 & 430 \\ 
        \hline
        V & n/a & n/a & 0 \\
        \hline
        U & 47 & 4700 & -200 \\
        \hline
    \end{tabular}
\end{table}

For non-readout boards, the traces terminate in conductive probe pads that give an accessible connection for testing. When the APA planes are fully coupled, each wire in a plane has one end that terminates on a readout board and one end that terminates on a non-readout board.

Alignment holes in the centre of each board fix the board's position with respect to the levelling plate on which it is mounted, verified by use of an alignment pin that passes through both board and plate. Once a board is aligned, it is secured tightly to the plate with screws.

\begin{figure}[h!]
    \centering
    \includegraphics[width=0.55\textwidth]{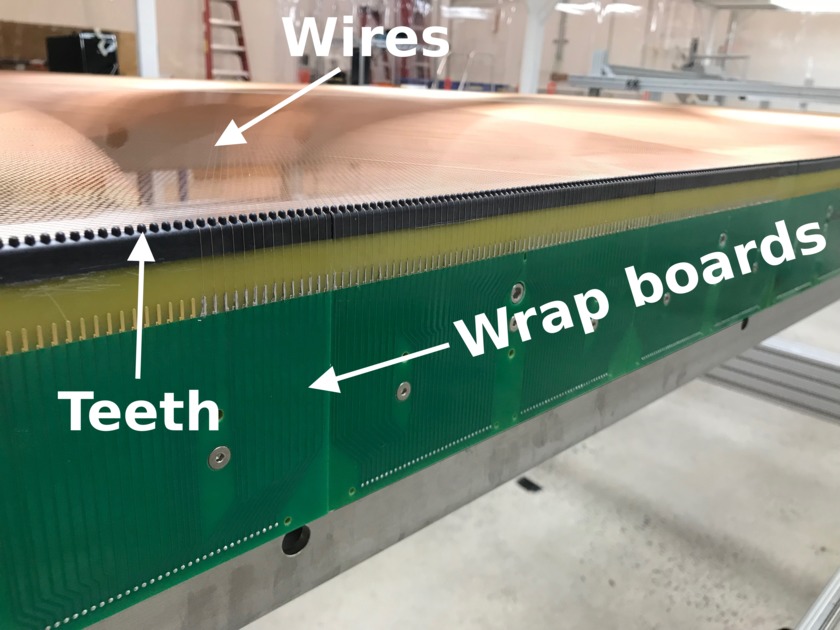}
    \caption{A view of wired wrap boards, showing the injection-moulded teeth.}
    \label{fig:Teeth}
\end{figure}

The boards that lie in the seam between APA frames are referred to as wrap boards. These non-readout boards have no resistor or capacitor components; however, rather than terminating in a conductive test pad the traces for wrap boards terminate in sockets for jumper connections that allow the two neighbouring APAs to be jumpered together. These jumper connectors are a unique feature of SBND: they give an electrical connection between induction plane wires in two adjacent APA frames, allowing readout as shown in figure \ref{fig:GeomBoardLayouts}, thereby reducing the total number of readout channels. Wrap boards are also fitted at the wiring sites with injection-moulded plastic `teeth' that maintain the wire spacing once the wires are folded down flush to the surface of the board (perpendicular to the wire plane itself), as shown in figure \ref{fig:Teeth}.

\subsection{Wire Support Combs}

\begin{figure}[h!]
    \centering
    \includegraphics[height=0.21\textheight]{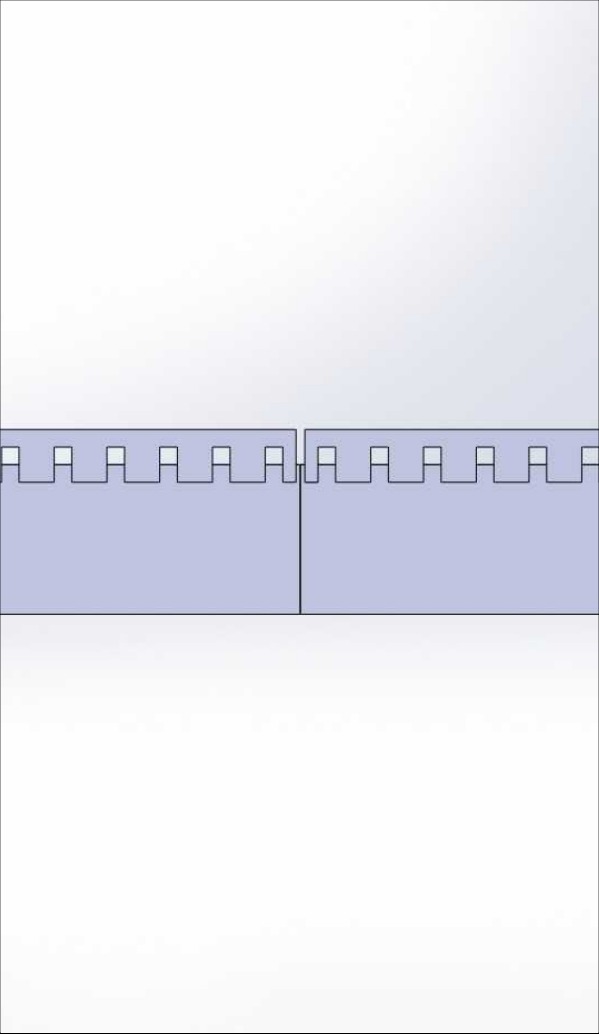}
    \includegraphics[height=0.21\textheight]{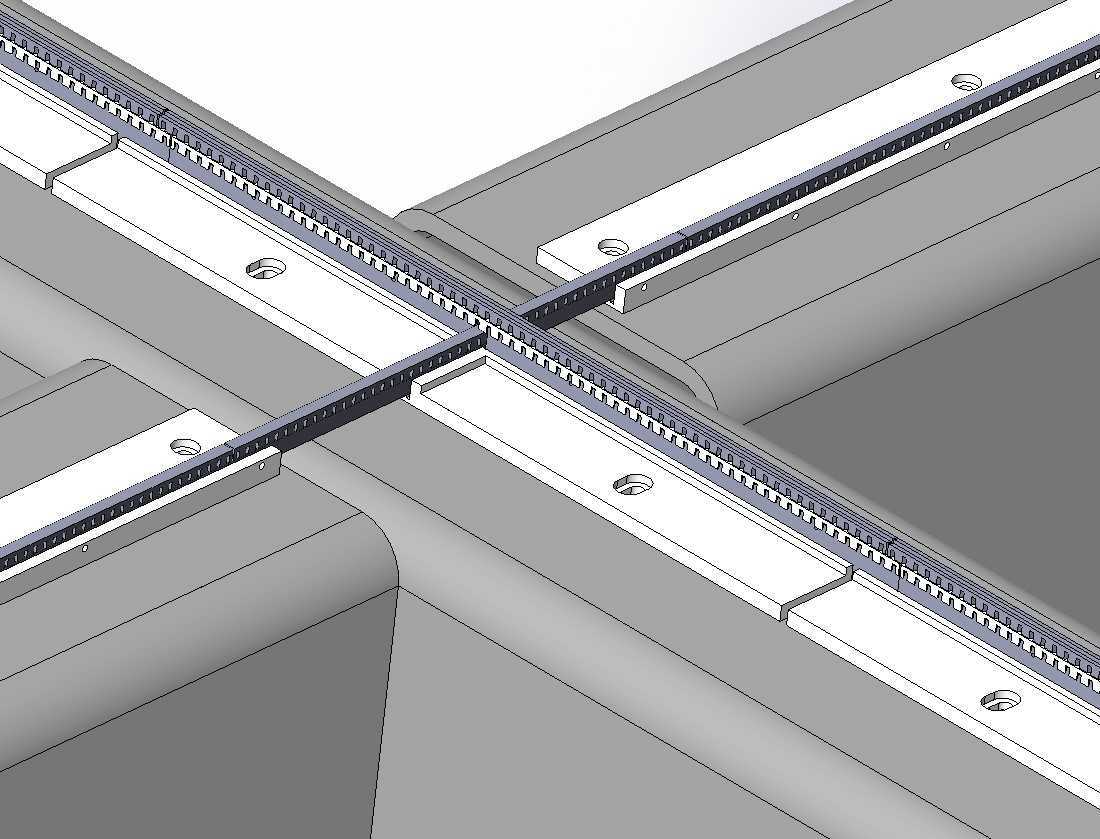}
    \includegraphics[height=0.21\textheight]{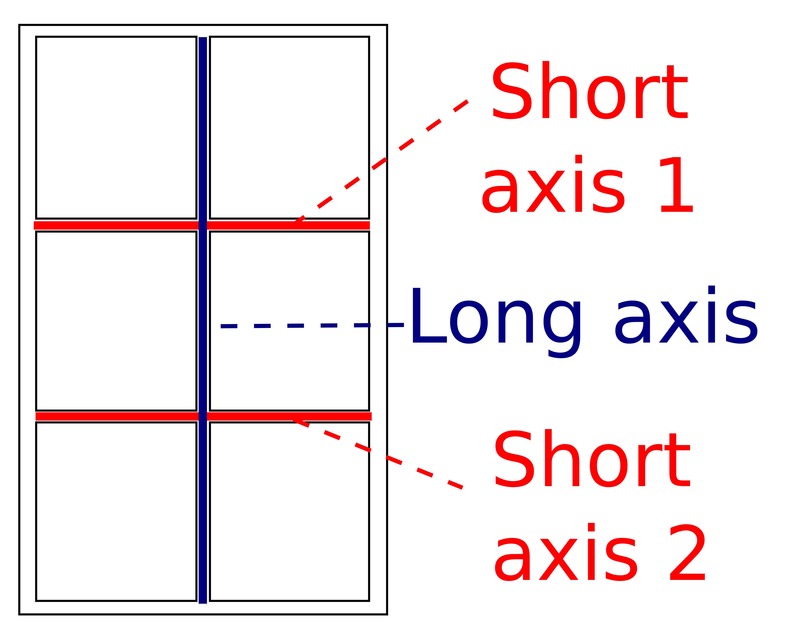}
    \caption{Left: 3D models of the combs showing the wire-capturing apertures. Centre: the intersection of the Y combs with the stacked U and V combs. Right: the axes of the APA frame along which the combs lie --- U and V along the long axis, Y along the short axes.}
    \label{fig:CombModels}
\end{figure}

\begin{figure}[h!]
    \centering
    \includegraphics[height=0.235\textheight]{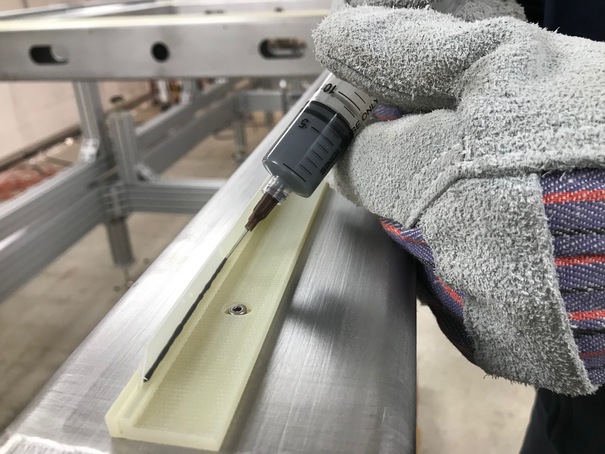}
    \includegraphics[height=0.235\textheight]{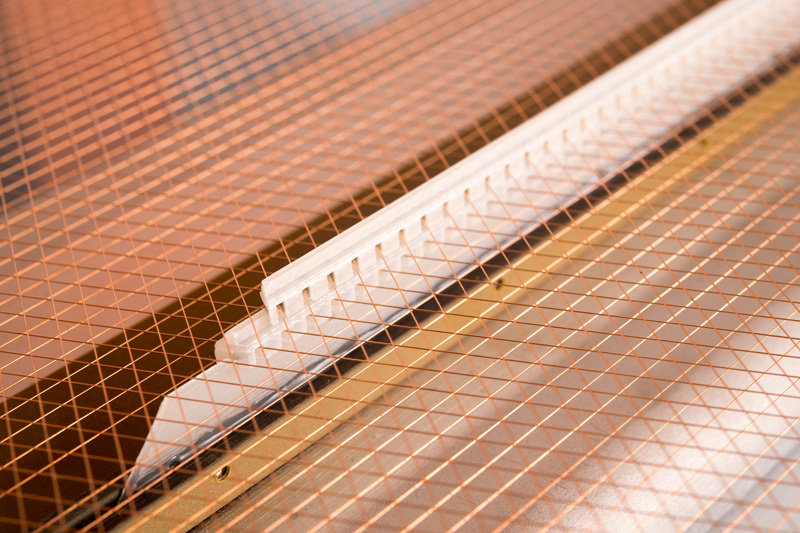}
    \caption{Left: the adjustable base slats of the combs being epoxied in place after their level has been set. Right: a finished stack of combs encapsulating the wires.}
    \label{fig:Slats}
\end{figure}

SBND uses G10 `combs' (as shown in figures \ref{fig:CombModels} and \ref{fig:Slats}) to hold the wires at intermediate points between where they are soldered to the boards. The wires are held with epoxy where they pass through the combs, ensuring the uniformity of the wire spacing across the wire plane and insuring against the possibility of a wire breaking -- should a wire break, its broken end will be constrained between the nearest combs, rather than leaving the entire length of the broken wire free to drift in the liquid argon.

The combs are rectangular lengths of G10 with apertures for the wires to pass through. A model showing the design of the combs and their positioning on an APA frame is shown in figure \ref{fig:CombModels}, with photographs in figure \ref{fig:Slats}. Two rows of three layers of combs capture the Y plane wires along the short axis struts of the APA frame, while a single row of two layers of combs captures the V and U wires along the long axis -- since these combs are parallel to the Y wires, the third layer is not necessary. A single Y wire is omitted from the Y layer to allow space for this row of combs. The end result of this scheme is to ensure that no wire in SBND has an unconstrained length longer than 1.6~m.

The combs themselves are designed in stacking layers, beginning with an adjustable slat that allows the entire stack to be levelled to the plane of the wires. In order to level each slat to support but not deform the wire layers, a series of test wires are laid, and the slat position is adjusted to sit below the test wire, and fixed in place with a grub screw while epoxy is applied.  The subsequent comb layers need no further levelling, since they are designed and manufactured to provide support at precisely the wire spacing of 3~mm.  After each wire layer is laid, the layers of combs are affixed, capturing each wire inside a 1.0~mm $\times$ 1.0 mm~aperture that is filled with epoxy, gripping the wire without deflecting it.

\section{Wiring Methods}\label{sec:WiringMethods}
The four SBND APAs were wired using two different techniques, at two separate sites - Daresbury Laboratory near Warrington in the UK, and Wright Laboratory at Yale University in the US. Each production site delivered two completed APAs. Though sharing many similarities in how the wires were laid and tested, the two sites used different apparatus and procedures to achieve the same results. The details of these techniques are laid out in the following section.

\subsection{Manual Wiring Apparatus}\label{subsec:US_Apparatus}
\begin{figure}[h!]
    \centering
    \includegraphics[width=0.88\textwidth]{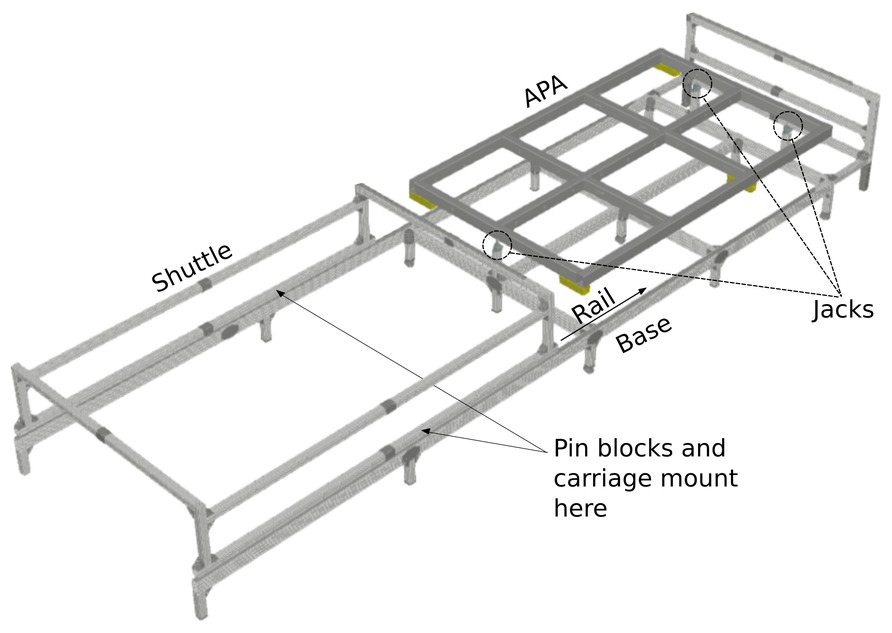}
    \caption{A model of the manual winding apparatus. The pin blocks and carriage are not shown in this model; they can mount at varying positions around the shuttle.}
    \label{fig:TRICERATops_Model}
\end{figure}

The manual wiring apparatus was used at Yale to lay wires by means of pin blocks mounted on a sliding shuttle. This setup allowed batches of wires to be wound and brought to tension in an open working space away from the APA, before being slid over the APA and soldered in place. An annotated model of the apparatus is shown in figure \ref{fig:TRICERATops_Model}.

\begin{figure}[h!]
    \centering
    \includegraphics[height=0.25\textheight]{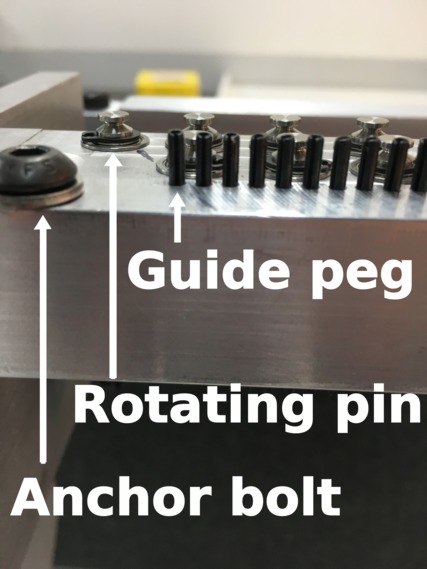}
    \includegraphics[height=0.25\textheight]{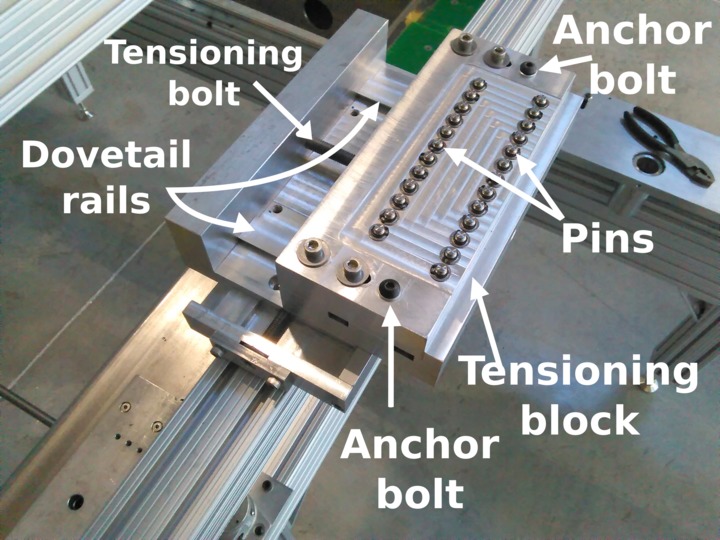}
    \includegraphics[width=0.49\textwidth]{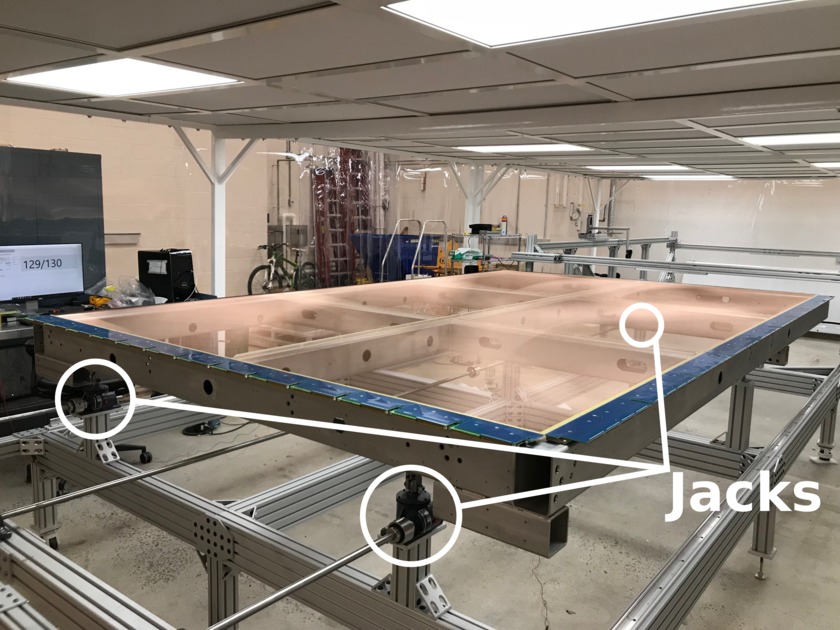}
    \includegraphics[width=0.49\textwidth]{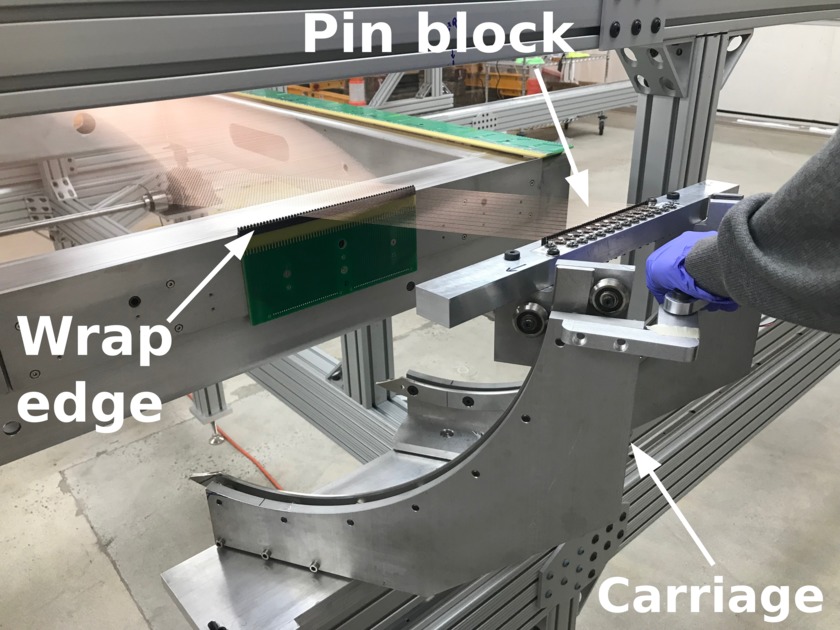}
    \caption{Photographs of the manual winding apparatus. Top left: a close-up of the non-tensioning pin block, showing the rotating pins around which the wires are wound. Top right: the tensioning block in its forward position, showing the dovetail rail on which its thread pulls it back to apply tension to the wires. Bottom left: a wired APA resting on the three jacks used to level the APA to the plane of the wires being transferred. Bottom right: the carriage being used to fold wires over the wrap boards.}
    \label{fig:TRICERATopsPhotos}
\end{figure}

The aluminium pin blocks between which the wires were wound are shown in figure \ref{fig:TRICERATopsPhotos} (top). These blocks carried arrays of rotating pins, around which the wires could be wound by hand by an operator walking back and forth between the blocks. One block of the pair was fitted with a thread on which it could be drawn back to apply tension to the wires; the rotation of the pins allows the added tension to be distributed evenly between all wires in the batch. The same pair of blocks could be mounted in varying positions on the shuttle to set the correct angle for each of the three wire planes (achieved by laying test wires and observing when they aligned on the appropriate solder pads).

A set of three jacks (shown in figure \ref{fig:TRICERATopsPhotos}, bottom left) supported the APA while it was mounted on the apparatus, allowing precise control of its height and tilt. These jacks allowed the APA to be lowered so that the shuttle could pass over it when being drawn back to receive a new batch of wires in the working area, then raised and levelled to the plane of the wires on the blocks before transfer. The gear ratio of the jacks used gave control of the jack height down to 0.25~mm per turn, translating to control of the APA pitch down to the level of 0.0001 radians.

In order to wire the U and V planes, the apparatus was also fitted with a ``carriage'': a manually-operated mechanical mounting allowing one pin block to be folded down over the wrap edge when winding the induction planes (shown in figure \ref{fig:TRICERATopsPhotos}, bottom right). This was the most complex part of the apparatus; its operation is detailed below.

\section*{The Carriage}\label{subsubsec:Carriage}
\begin{figure}[h!]
\centering
\includegraphics[height=0.16\textheight]{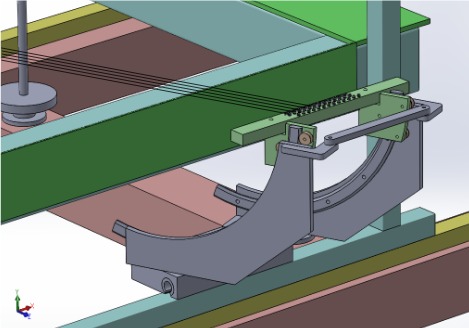}
\includegraphics[height=0.16\textheight]{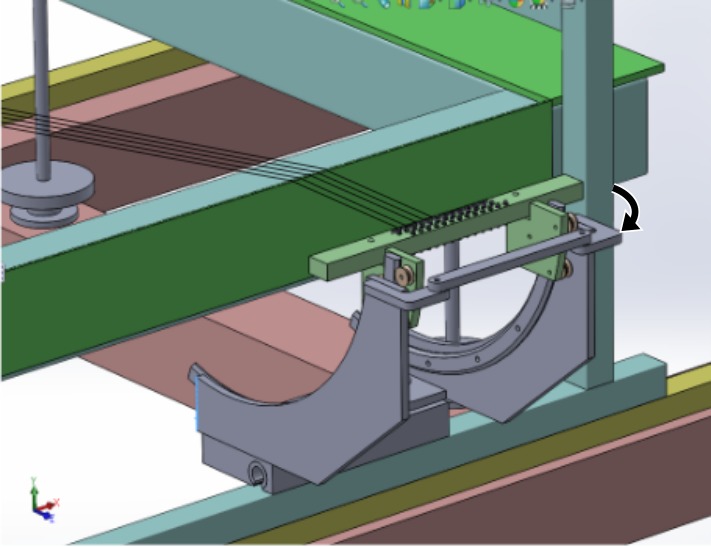}
\includegraphics[height=0.16\textheight]{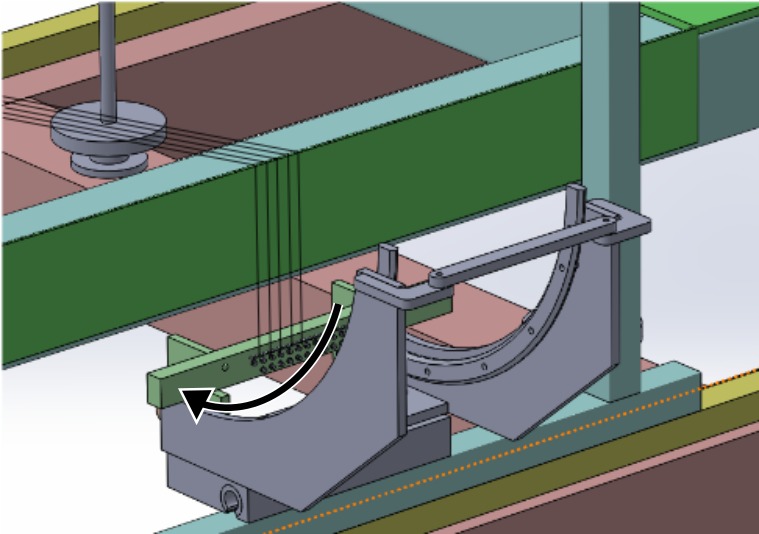}
\caption{The three steps by which the carriage folds wires down parallel to the wrap boards. Left: a batch of wires in its initial position before folding, stretched parallel to the angle of the wire plane between the carriage pin block and its static counterpart. Centre: the carriage is turned on bearings in its base to bring the wire lengths between its pin block and the plastic teeth of the board to subtend a right angle to the wrap edge. Right: the pin block is rolled down on a circular rail to bring the wires flush to the surface of the wrap boards.}
\label{fig:Carriage_Operation}
\end{figure}

The carriage is a device designed to allow wires in the U and V planes to be folded down parallel to the surface of the wrap boards without changing their tension. It achieves this by mounting a pin block on a mechanism that performs two rotations around the point where the wire touches the wrap edge itself. These rotations are shown in figure \ref{fig:Carriage_Operation}; as long as the rotations stay centred on the edge over which the wires are folded, the length of the wire between the wrap edge and the pins remains constant, and thus so does the tension.

The performance of the carriage in preserving the tension of the wires before the fold is extremely sensitive to how well the arc of the curved carriage rail is centred on the point of folding -- the spring constant of the wire used in SBND is 1.36~N/mm, meaning that even sub-millimetre misalignments could induce significant changes in tension. 

\begin{figure}[h!]
\centering
\includegraphics[height=0.235\textheight]{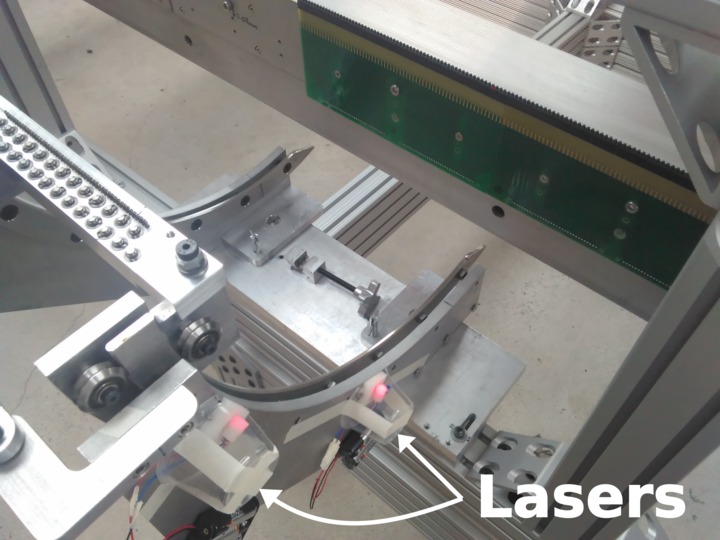}
\includegraphics[height=0.235\textheight]{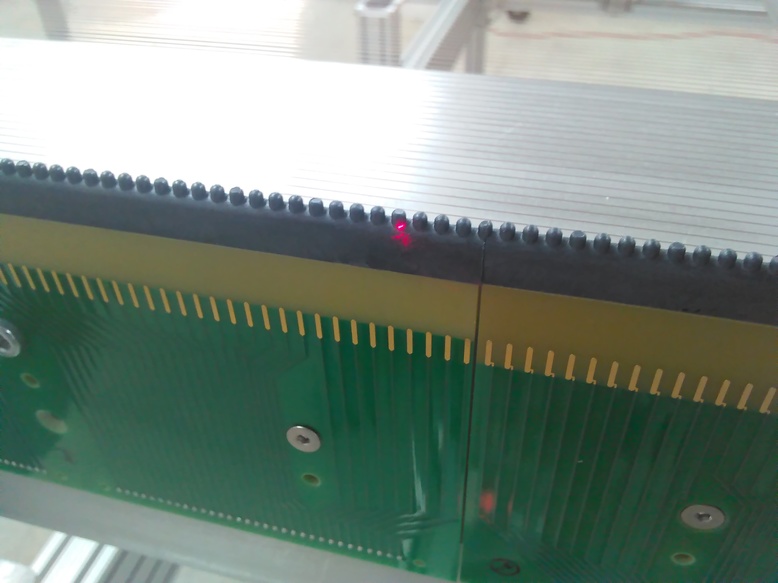}
\caption{Photographs of the carriage alignment lasers in use. On the left are shown the lasers in their housings on the side of the carriage rail. On the right, the laser dots appear superimposed on the folding edge, indicating the carriage is well-aligned.}
\label{fig:CarriageLasers}
\end{figure}

In order to better control the alignment, the carriage was fitted with two lasers (shown in figure \ref{fig:CarriageLasers}), aligned to the centre of the arc defined by the rail. When well-aligned, both laser dots should appear superimposed on the very edge of the board on which the wires are being folded. A dial indicator was used to ensure this alignment was accurate to within 0.01~mm (at which level other alignment effects from e.g. the flatness of the lab floor become dominant).

\subsection{Manual Winding Procedure}
The apparatus is operated according to the following procedure.
\begin{enumerate}
    \item With the shuttle slid away from the APA, a batch of wires is wound by hand between the two pin blocks.
    \item The shuttle is slid over the APA, and the APA is raised until the wires are in contact with the boards to which they will be soldered.
    \item The wires are brought approximately to tension using the thread on the tensioning pin block, using a previously-measured benchmark of how much travel is required on the thread to bring slack wires up to values near 7~N.
    \item For the U and V layers, the wires are folded over the wrap edge using the carriage.
    \item The tension of approximately $20\%$ of the wires in the batch is sampled with the laser apparatus described in section \ref{subsec:TensionTests}. If any wires are observed with tension outside the $7 \pm 1$~N range, the tension of the batch is adjusted (using the thread) and re-measured until the wires sampled are 100\% compliant with the specifications.
    \item The wires are soldered to the board, and the lengths behind each solder joint are cut.
    \item The tension of every wire in the batch is measured again. Any wires observed to be outside the specifications are cut and re-laid.
    \item The shuttle is moved back to its original position and the process is repeated.
\end{enumerate}

\subsection{Semi-Automated Wiring Apparatus}\label{subsec:UK-Apparatus}

\begin{figure}[!hbt]
  \centering
  \includegraphics[width=0.6\textwidth]{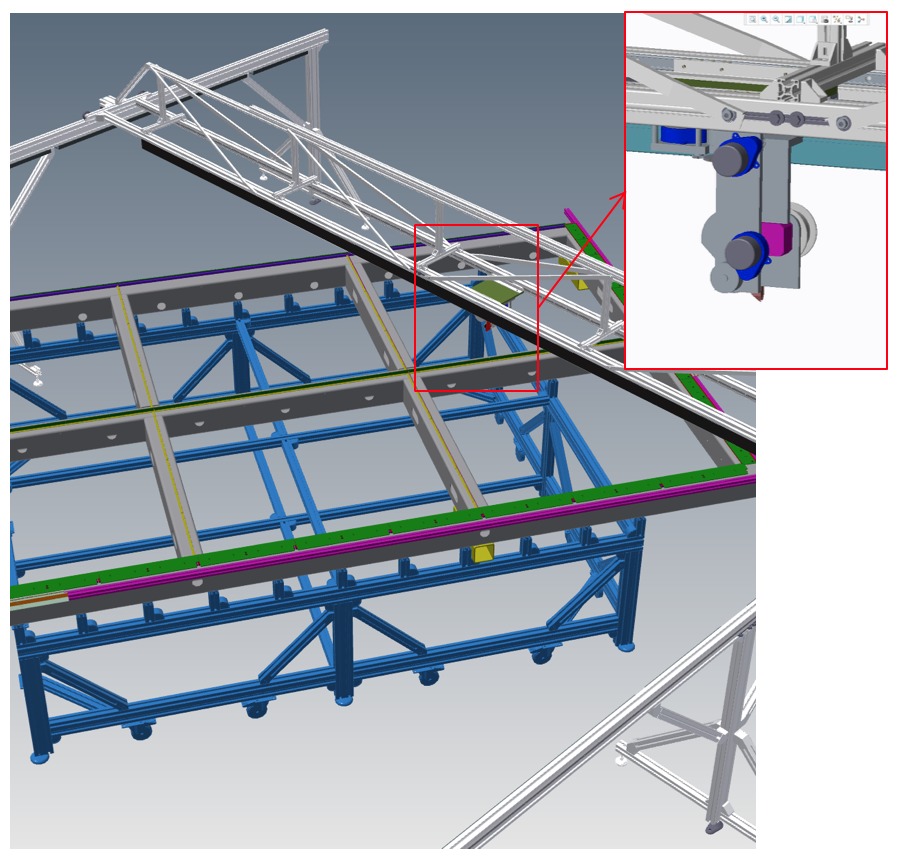}
  \caption{A drawing of the semi-automated wire-winding setup.}
  \label{fig:UK-apparatus-drawing}
\end{figure}

The semi-automated wiring apparatus lays wires one at a time using a semi-automatic wiring head which travels under remote control across a custom-built framework. A drawing of the apparatus can be seen in figure~\ref{fig:UK-apparatus-drawing}. 

During wiring, the APA frame rests on a turntable that is manually rotated to the required working angle and then fixed in position by lowering a number of supporting feet and retracting the rotating casters. The APA is supported on the turntable by a number of ``upside-down feet'' which are adjustable to allow the frame to be levelled (as measured with a sensor mounted on the wiring head), and to ensure that it is adequately and uniformly supported.

\begin{figure}[!h]
  \centering
  \includegraphics[height=0.18\textheight]{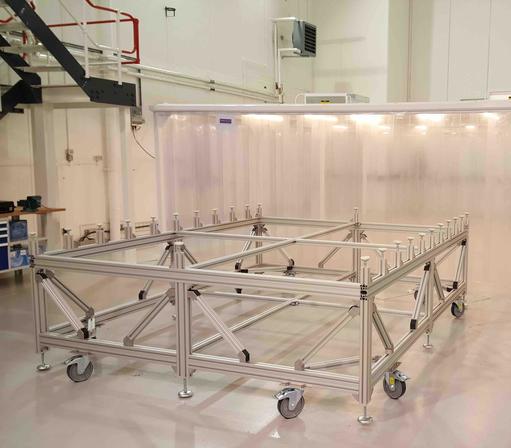}
  \includegraphics[height=0.18\textheight]{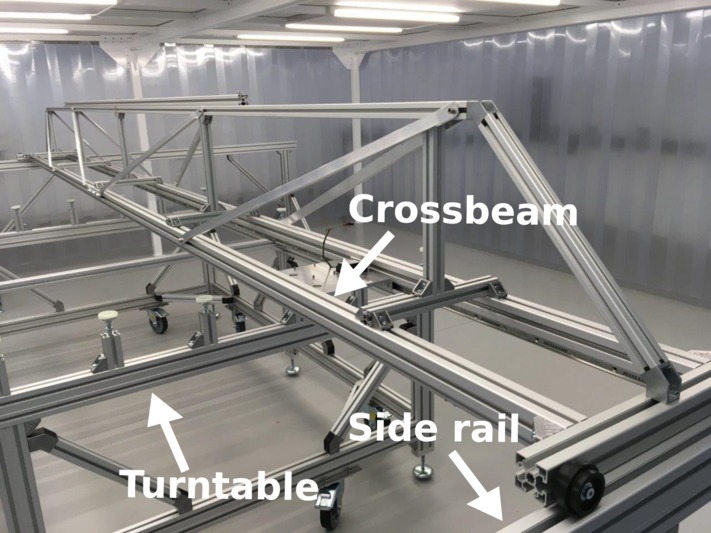}
   \includegraphics[height=0.18\textheight]{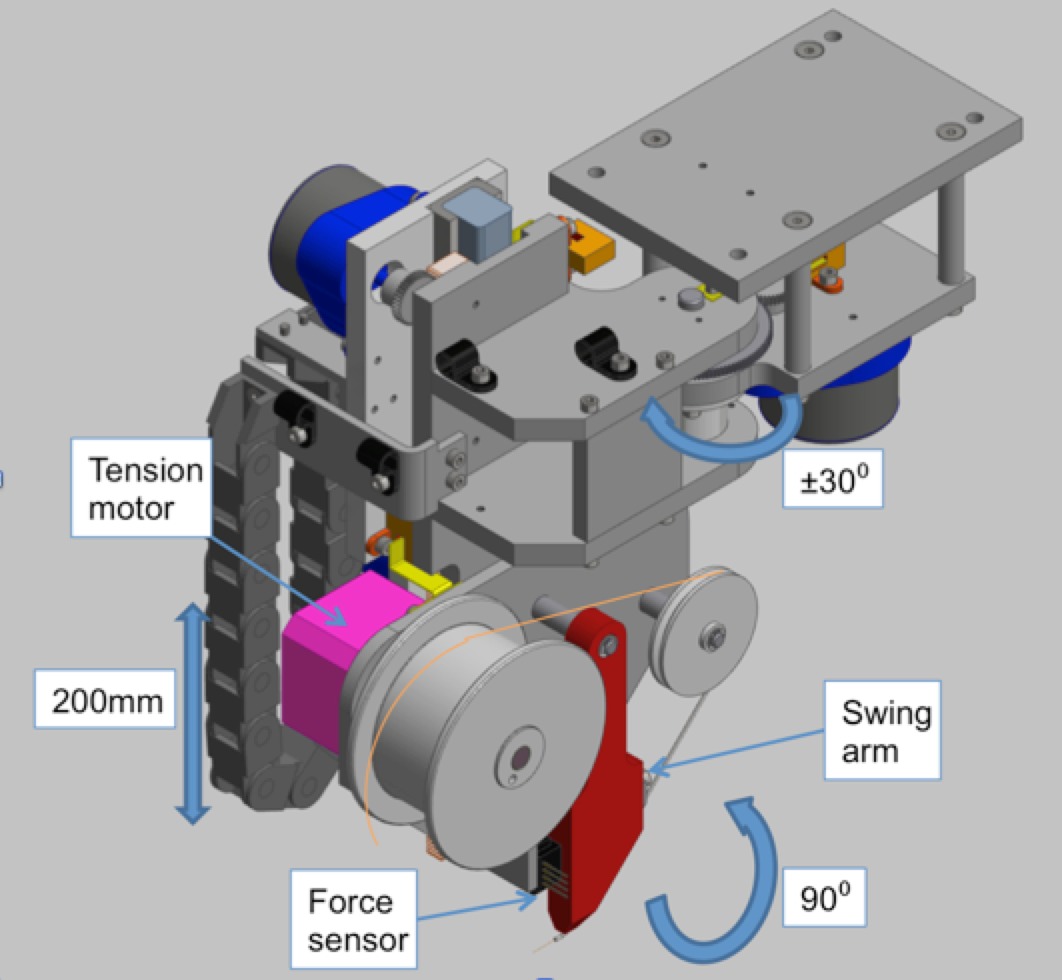}
  \caption{Left: the wiring turntable outside the clean
    tent at Daresbury. Centre: the turntable, rails and crossbeam inside the
    clean tent. Right: wiring head schematic.}
  \label{fig:UK-apparatus-photos}
\end{figure}

A cross beam carrying a semi-automatic wiring head is supported by two side rails, one on either side of the turntable, as is shown in figure \ref{fig:UK-apparatus-photos}. The wiring head is mounted on a linear stage which travels the length of the cross beam, which itself travels on the side rails, so that any part of the APA frame can be accessed by the wiring head. The head has a vertical motion range of 200~mm to allow it to reach the wrap boards, and it can swivel in two planes as required by the various wire layer geometries. The head is remotely controlled with LabView-based software, using keyboard commands. Wires are tensioned with a stepper motor which turns the wire spool backwards until the force sensor registers the desired tension.

\begin{figure}[!h]
  \centering
  \includegraphics[height=0.6\textwidth]{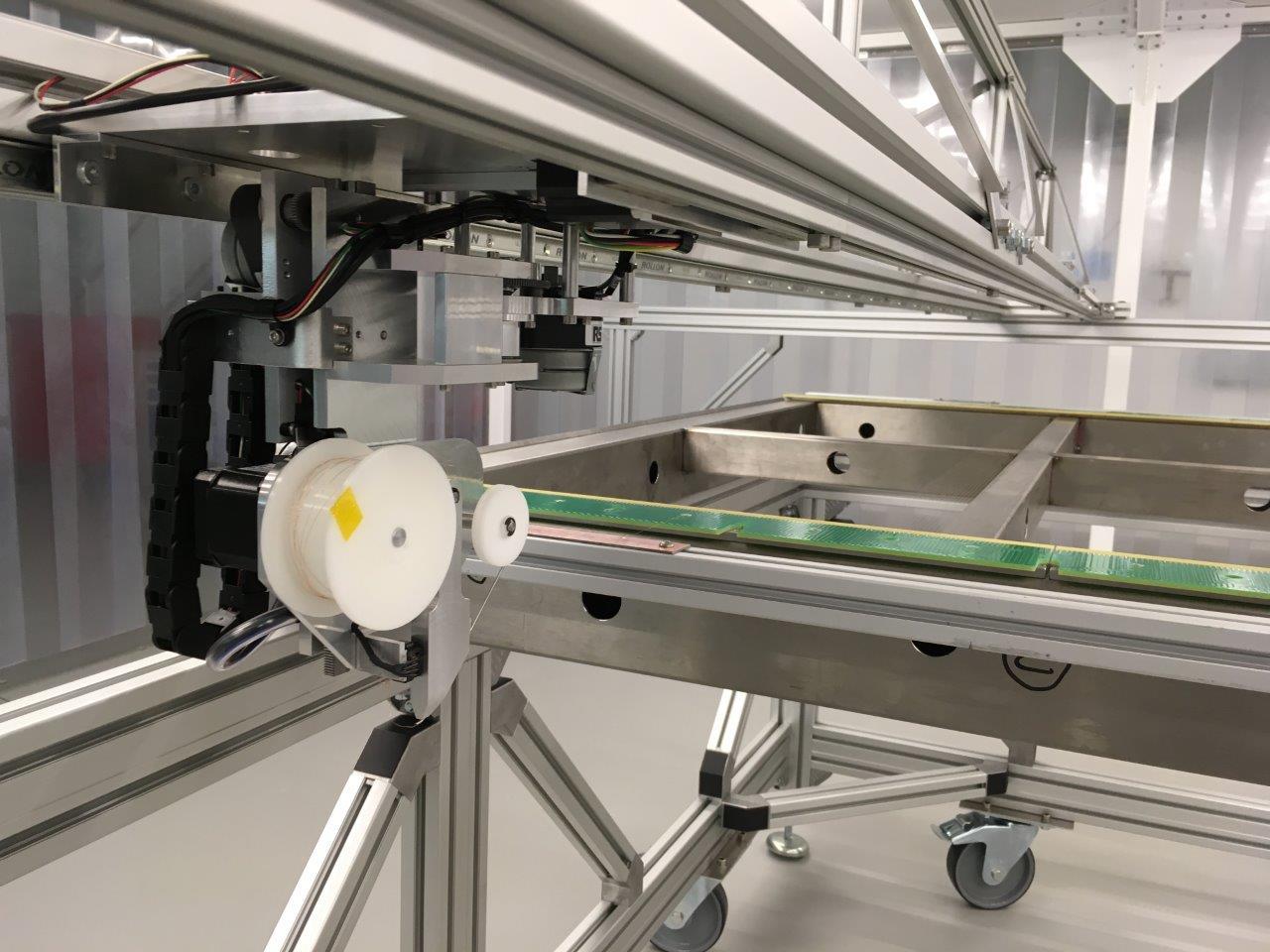}
  \caption{The wiring head, ready to start wiring trials.}
  \label{fig:UK-wiring-system}
\end{figure}

Figure~\ref{fig:UK-wiring-system} shows a photograph of the wiring system, including a small prototype APA frame, at the beginning of operational trials.

\subsection{Semi-Automated Winding Procedure}\label{subsec:UK-Winding-Procedure}

Before loading the APA onto the turntable, the turntable is levelled
(in place) to the winding machine, using the vertical axis of the wiring head as a reference.

Once a layer's electronics boards are in place, the wiring head (``robot'') control software finds the four corners of the frame; hereafter the software in principle knows the rough positions of the pads on the boards and how to pair them. Periodic checks are made to make sure the pairing is correct. 

Fine alignment to the solder pads is achieved using guide wires on a pair of temporary alignment boards, directly behind the permanent electronics boards at either side of the frame (shown in figure \ref{fig:Steps1-3}). The temporary boards are first roughly aligned to the electronics boards, then a guide wire is laid at each edge of the pair of temporary boards. The position relative to the solder pad on the electronics boards is checked by eye; the oval shape of the solder pad means that this is extremely reliable, even a small fraction of a millimetre offset is obvious. The positions of the temporary boards lateral to the side of the frame are then adjusted until the guide wires lie centrally on the corresponding solder pads. 

The guide wires are then removed, and the real wires are laid for the current pair of temporary boards. For the first wire of a layer a final pad count is done to ensure that the right pad pairing has been made.
    
The procedure to lay a single wire is as follows:
\begin{enumerate}
\item\label{item:UKstep1} The robot is advanced to the starting point for the desired wire. If the wire starts on a wrap board then the wiring head is rotated 90$^\circ$. The wiring head moves down to a point very close to the temporary board. A `tail' of wire protrudes from the wiring head.
\item\label{item:UKstep2} The wire is soldered to its solder pad on the temporary board.
\item\label{item:UKstep3} If the starting point is a horizontal board, the wiring head moves slightly up, away from the temporary board. For wrap boards, the wiring head moves slightly away from the temporary board, moves up to slightly above the guide pins, then rotates the wiring tip around the guide pins so that the wire falls into the correct slot.
  

\item\label{item:UKstep4} The robot moves across the frame to an end point which is in line with the solder pad on the other side of the frame, and slightly past the temporary board, leaving enough room for the operator to solder the wire (as shown in figure \ref{fig:Step4}).
    

\item\label{item:UKstep5} The robot moves down (as shown in figure \ref{fig:Steps5-6}) so that the wire just touches the surface of the copper on the temporary board, and the robot moves slowly to apply a ‘pre-tension’ to the wire of around 1~N. This slight tension removes any sag from the wire, and reveals if the wire has fallen into the correct slot in the alignment pins. If the wire is not properly seated between the pins then the operator can gently push the wire into place. 

\item\label{item:UKstep6} When the operator is satisfied that the wire is correctly positioned, the robot performs the full tensioning to the desired target value. 
 

 \item\label{item:UKstep7} The operator can now solder the wire to the solder pad on the temporary board (as shown in figure \ref{fig:Steps7-8}, left). When soldering is complete, the tension applied by the wiring head is released, allowing the operator to cut the slack wire behind the solder joint while the length of wire between the two boards remains soldered and taut.
 
\item\label{item:UKstep8} The operator can now cut the wire (as shown in figure \ref{fig:Steps7-8}, right). After cutting, the robot moves slowly away from the frame in the wire-laying direction for around 200~mm. This gives the operator time to react if the wire is not properly cut.


\item The wiring head then moves up to its highest position. If the ‘single wire’ option is selected in the software interface, then the robot remains in this position. If ‘continuous’ is selected then the robot moves perpendicular to the wires, along the cross beam, to the start position for the next wire and the process is repeated.
 
 \begin{figure}[!h]
  \centering
  \includegraphics[height=0.16\textheight]{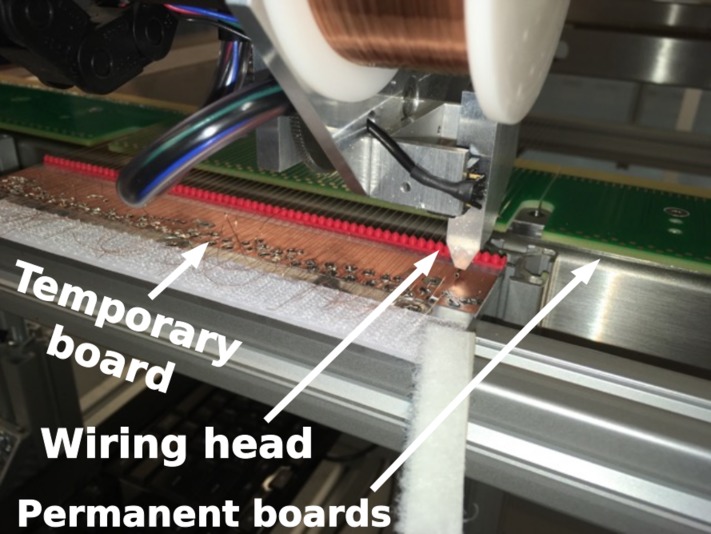}
  \includegraphics[height=0.16\textheight]{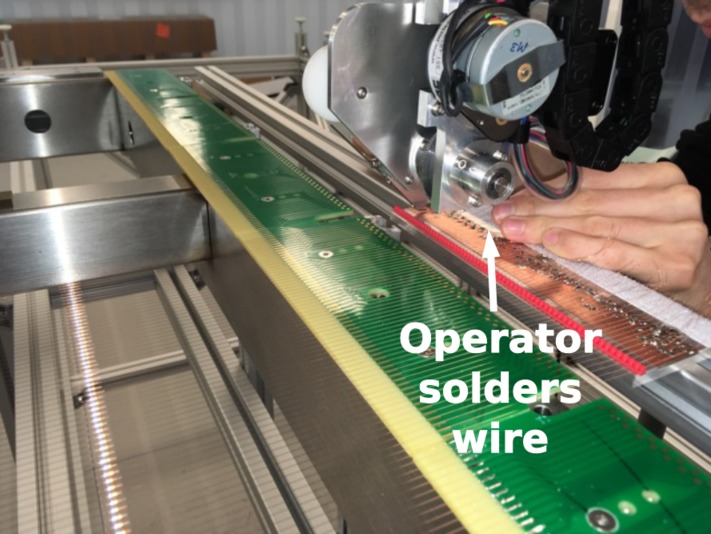}
  \includegraphics[height=0.16\textheight]{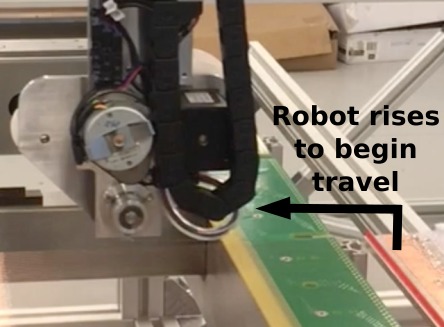}
 \caption{Steps~\ref{item:UKstep1}--\ref{item:UKstep3} of the wiring process. The wiring head is positioned and one end of the wire is soldered in place.}
  \label{fig:Steps1-3}
\end{figure}
\begin{figure}[!h]
  \centering
  \includegraphics[height=0.18\textheight]{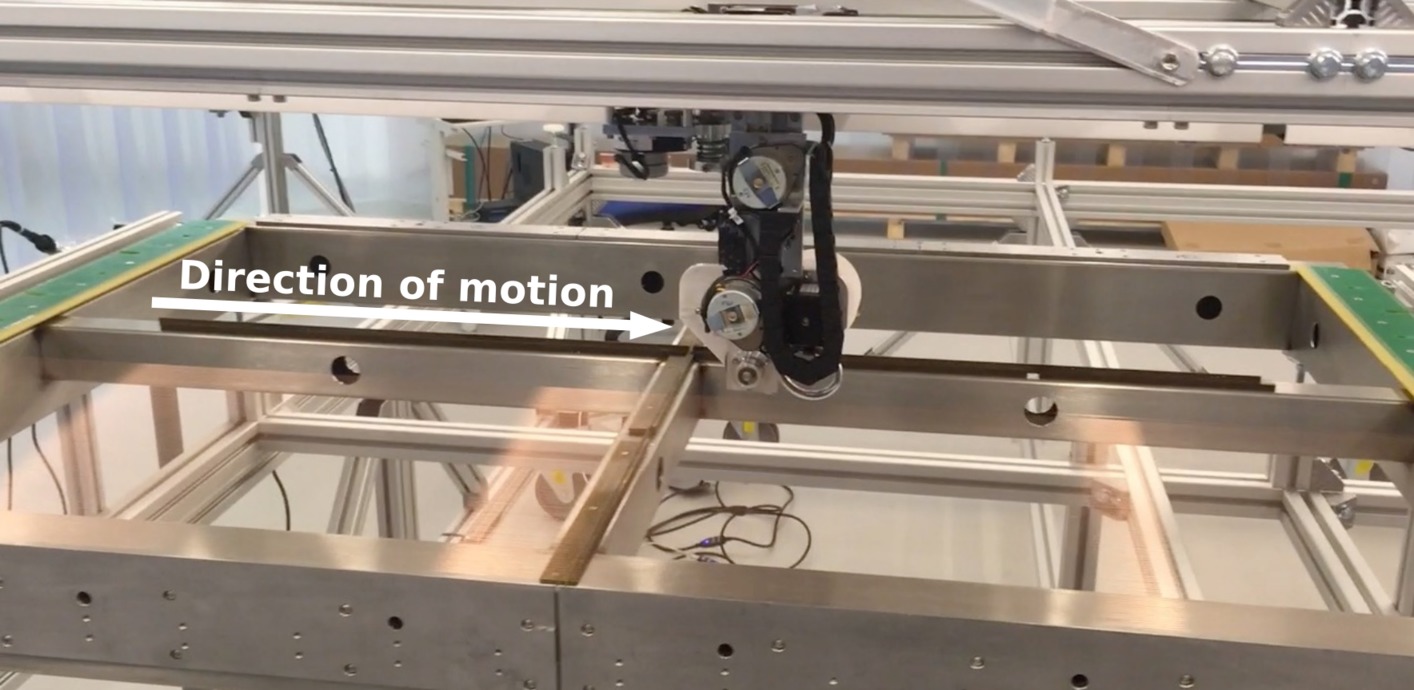}
  \includegraphics[height=0.18\textheight]{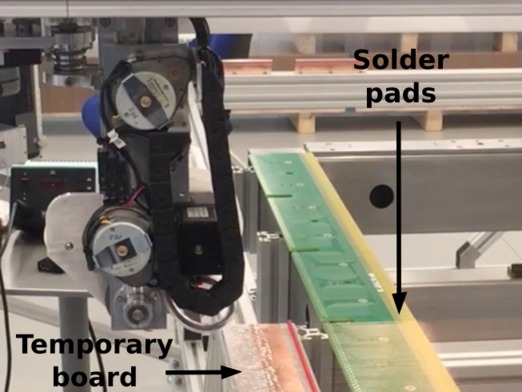}
 \caption{Step~\ref{item:UKstep4} of the wiring process. The wiring head draws the wire across to the opposing solder pad.}
  \label{fig:Step4}
\end{figure}
 \begin{figure}[!h]
  \centering
  \includegraphics[height=0.19\textheight]{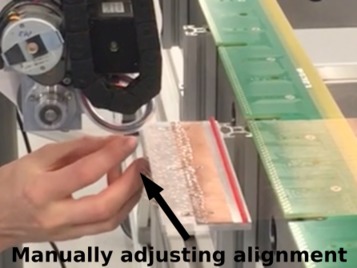}
  \includegraphics[height=0.19\textheight]{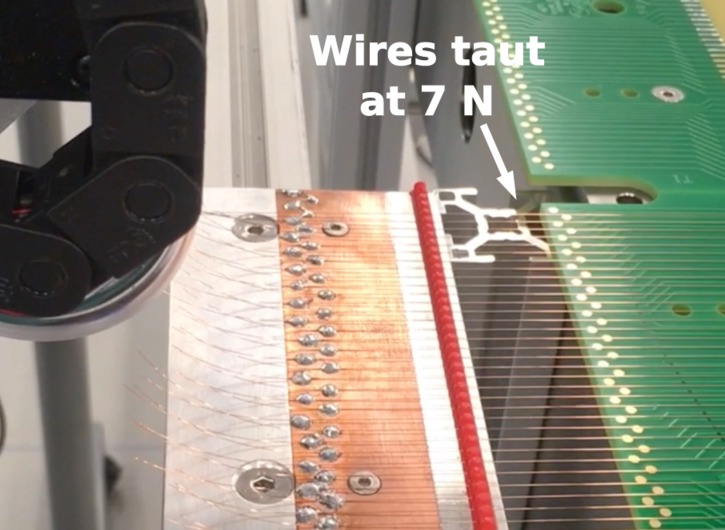}
 \caption{Steps~\ref{item:UKstep5}--\ref{item:UKstep6} of the wiring process. The wire is aligned to the opposing solder pad and tensioned.}
  \label{fig:Steps5-6}
\end{figure}
 \begin{figure}[!h]
  \centering
  \includegraphics[height=0.19\textheight]{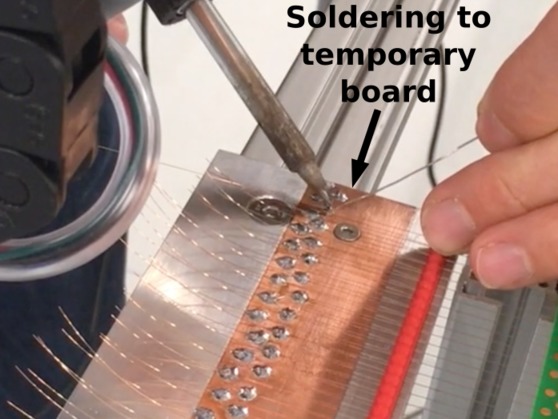}
  \includegraphics[height=0.19\textheight]{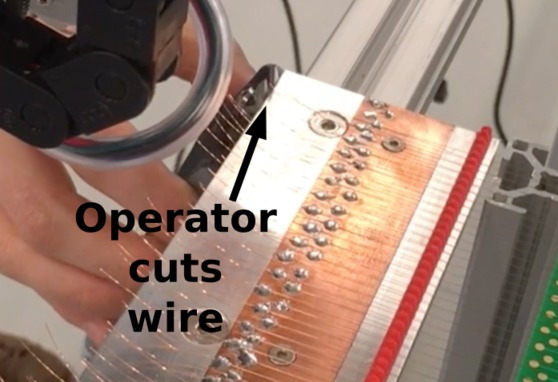}
 \caption{Steps~\ref{item:UKstep7}--\ref{item:UKstep8} of the wiring process. The wire is soldered to the opposing solder pad and cut.}
  \label{fig:Steps7-8}
\end{figure}
 \end{enumerate}
 
 \newpage
 
 If at any time an operator wishes to cancel the wiring operation, then a ‘cancel’ button is available on the interface. A warning is presented informing the operator that the wire must be cut, and when confirmed the robot will move to a safe location close to the starting point for the wire.

When the end of an aligned temporary board is reached, the wires are soldered by hand to the electronics board. The temporary boards are then lifted and the wires are carefully trimmed close to the solder joint on the electronics board. The temporary board is moved to the next position, and the above process is repeated.

When all the wires in a layer have been soldered to the electronics boards, the electrical quality control tests are performed, and the tension of each wire is checked with the laser. Any wires which are outside of specifications are replaced. This is done when a layer is complete rather than when a temporary board is finished, as experience has shown that it is extremely rare for a wire to fail these tests and the ratio of setup time to performance time for the tests is such that it is most efficient to perform them on all the wires in one go.

\subsection{Advantages and Disadvantages of Each Apparatus}

The manual apparatus had one main advantage: the ability to tension wires and attach them to the APA in batches of approximately fifty at a time. This helped reduce the production time required for each wire layer.

The manual apparatus also had two notable disadvantages. The first was that due to the small radius of the rotating pins, the wires would develop kinks where they wrapped around the pins, causing local variations in the tension within each batch. The second disadvantage was the high precision of the alignment between the carriage and the APA that was required to fold the wires without major changes in tension. These effects and their impact on the final tension distributions are discussed further in section~\ref{subsec:TensionResults}.

By contrast, the semi-automated apparatus was less sensitive to alignment. It also had the advantage of automated tensioning of each wire, eliminating the need to manually measure the tension before soldering. This saved time and produced more consistent tension distributions than the manual apparatus.

The main disadvantage of the semi-automated method was that it could only lay one wire at a time. This limited the speed at which wire layers could be produced, and also created an effect where as more wires were affixed to a given board, the progressive distortion of the board under strain would alter the tension of the wires that had been laid earlier. This effect and its impact on the final tension distributions is discussed further in section~\ref{subsec:TensionResults}.

A wiring apparatus can be envisioned that combines the advantages of both setups, using multiple semi-automatic wiring heads to lay batches of wires in parallel, or using a single wiring head to carry a pin block like those used in the manual apparatus, with pins of sufficient radius to avoid kinking. Such improvements would allow the production of APAs of similar scale in significantly less time, with no loss of precision.
\section{Quality Control Testing}\label{sec:Testing}

In order to verify that the wiring is consistently within the specifications outlined in table~\ref{tab:WireSpecs}, a series of quality control measurements were made at various stages throughout the construction process.  The techniques used for these measurements were identical at the two wiring sites, and an overview is presented in this section. 

\subsection{Tension Testing Method}\label{subsec:TensionTests}
\begin{figure}[h!]
    \centering
    \includegraphics[width=0.6\textwidth]{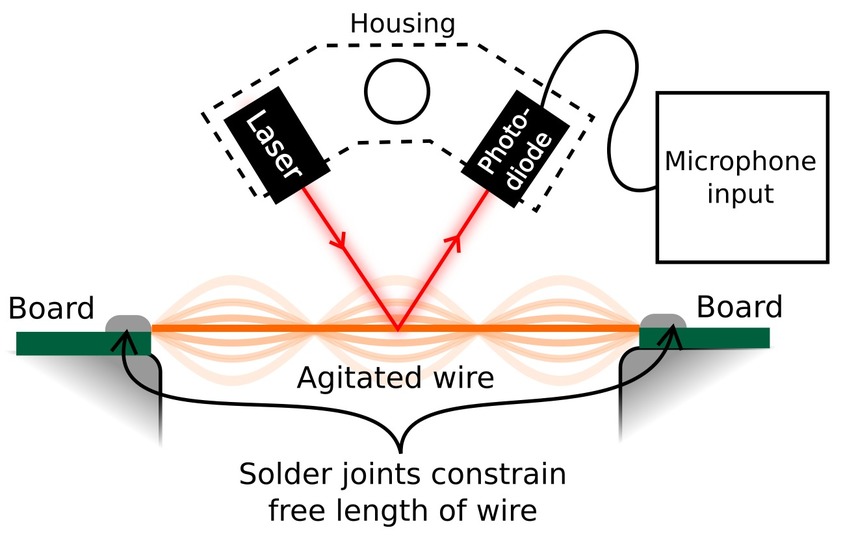}
    \caption{A cartoon showing the operation of the laser photodiode tension measurement apparatus.}
    \label{fig:LaserCartoon}
\end{figure}

\begin{figure}[h!]
    \centering
    \includegraphics[angle=270,width=0.3\textwidth]{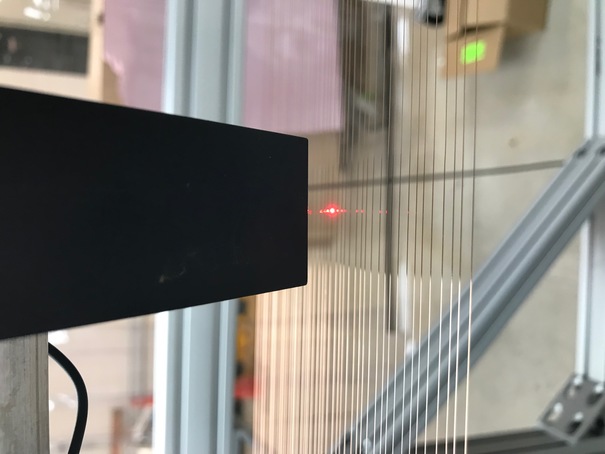}
    \includegraphics[angle=270,width=0.3\textwidth]{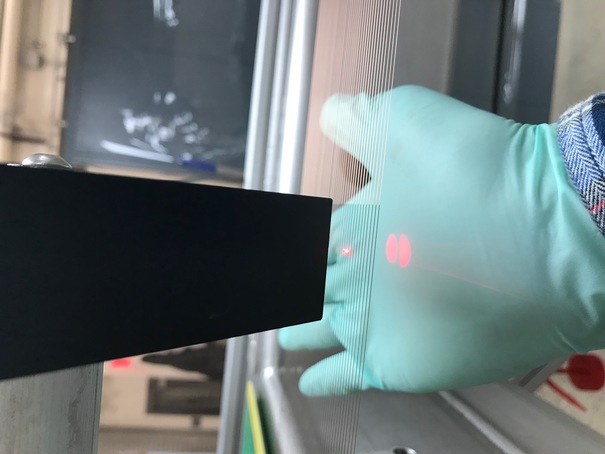}
    \caption{Photographs of the tension measurement laser in use at Wright Lab, focused on a tensioned wire in a batch ready for transfer. On the right, the shadow on the operator's hand shows the characteristic silhouette of the wire when the laser is focused. Since the wires lie in a plane, once the laser is focused on one wire it will be focused for the others.}
    \label{fig:LaserPhotodiode}
\end{figure}

The tension of each wire is measured using a laser photodiode setup provided by the University of Wisconsin-Madison Physical Sciences Laboratory. This process is illustrated in figure \ref{fig:LaserCartoon} and shown in action in figure \ref{fig:LaserPhotodiode}.

A laser is focused on the wire under inspection (using an adjustable stage to control the height above the wire), and the reflected light is measured by a photodiode. The wire is agitated to produce oscillations (manually, using a plectrum), and a fast Fourier transform is performed on the signal from the photodiode in order to extract the fundamental frequency of the oscillation. From this we extract the tension T according to equation \ref{eq:TensionEquation},

\begin{equation}
T = 4\rho(fl)^{2},
\label{eq:TensionEquation}
\end{equation}

\noindent where $f$ is the observed fundamental frequency of the oscillating wire, $l$ the length free to oscillate (whether between solder pads or between combs), and $\rho$ its linear density.

The accuracy of this measurement is dependent on the uncertainties on both the oscillation frequency and the length of the wire that is free to oscillate. The uncertainty on the wire length derives simply from the scale of the measuring tape used and user error in interpreting it, with an estimated value of 0.15~cm. The uncertainty on the oscillation frequency derives from the focus of the laser, the binning of the Fourier transform software, and the accuracy with which the peaks are identified (done manually for the manual apparatus and via a LabView algorithm for the semi-automated apparatus), and varies according to the length of the wire (with shorter wires having a larger uncertainty), with an estimated value of 0.5 Hz for the longest wires and 25 Hz for the shortest. On full-length wires, these length and frequency uncertainties produce a typical tension uncertainty of approximately 0.15~N for the U and V planes, and 0.25~N for the Y plane. For the very shortest wires in the corners of the U and V planes, these uncertainties grow as the proportional uncertainty on the length becomes more significant, producing uncertainties up to 15 times larger than the nominal values for longer wires. 

Such large uncertainties occur on 10--20 wires per layer. For these wires, visual inspection after laying provided a cross-check to ensure that the tension was not wildly different from the specification. In the manual wire-laying technique, the tension of these wires could also be measured before laying (when they were still of a length to span the sliding carriage), and was observed to be consistent with the tension of the other wires. Since the 5~N tension threshold was derived for the longest wires (approximately 4~m long) and shorter wires are less susceptible to oscillation (as discussed in section \ref{subsec:TensionOverTime}), this degree of uncertainty on the tension for wires on the order of 10~mm (i.e. approximately 400 times shorter than the wire length for which the specification was set) was judged acceptable.

\subsection{Tension Testing Results}
\label{subsec:TensionResults}

\begin{figure}[h!]
    \centering
    \includegraphics[height=0.225\textheight]{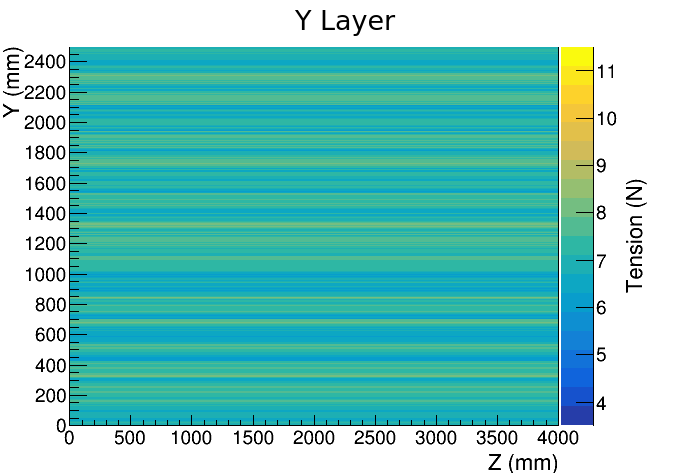}
    \includegraphics[height=0.225\textheight]{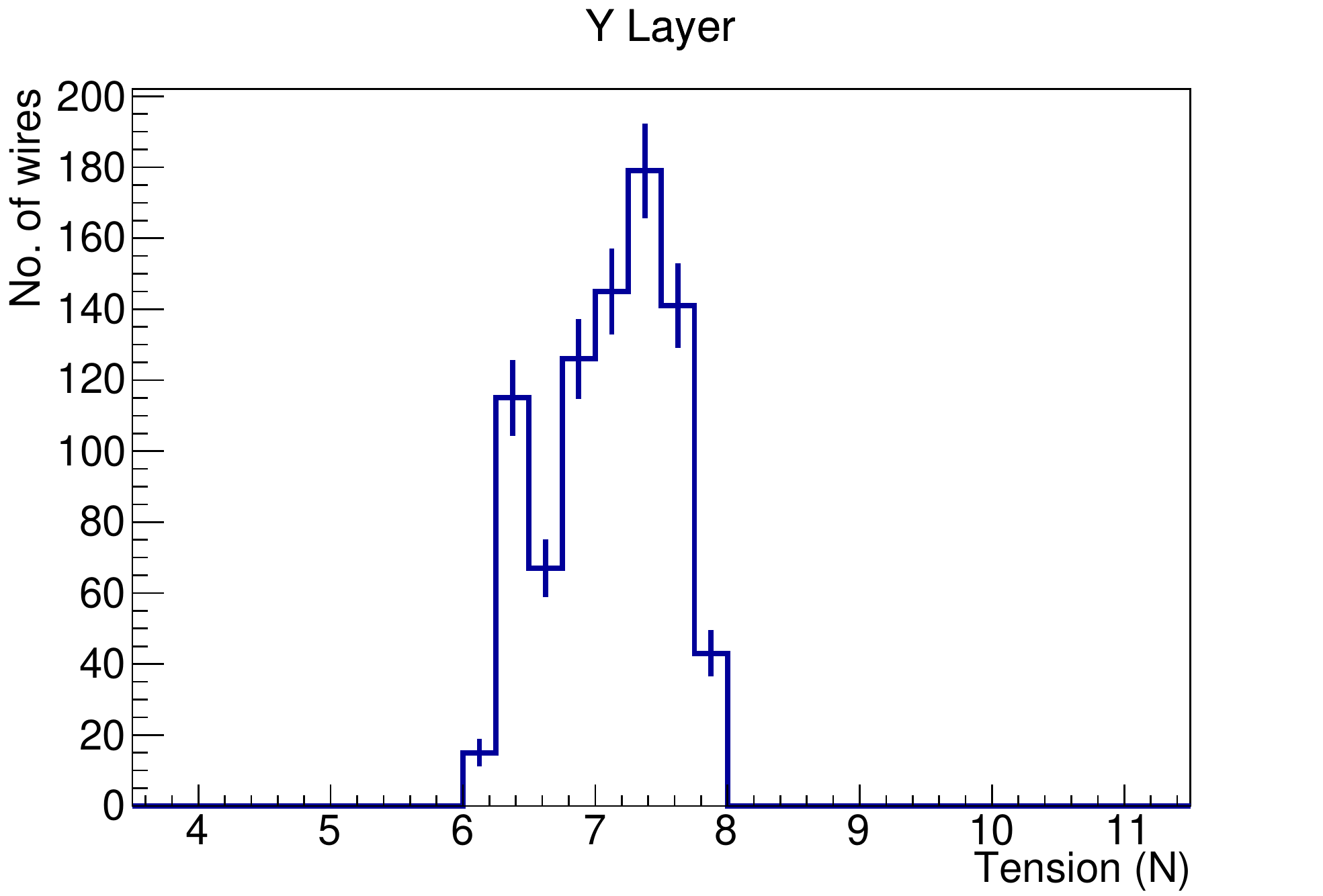}
    \includegraphics[height=0.225\textheight]{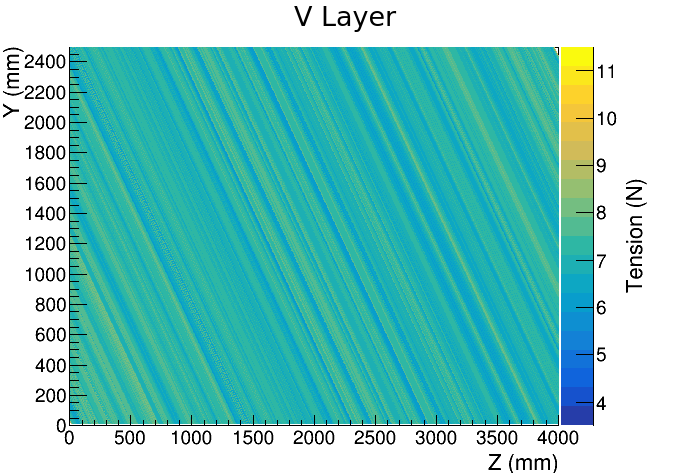}
    \includegraphics[height=0.225\textheight]{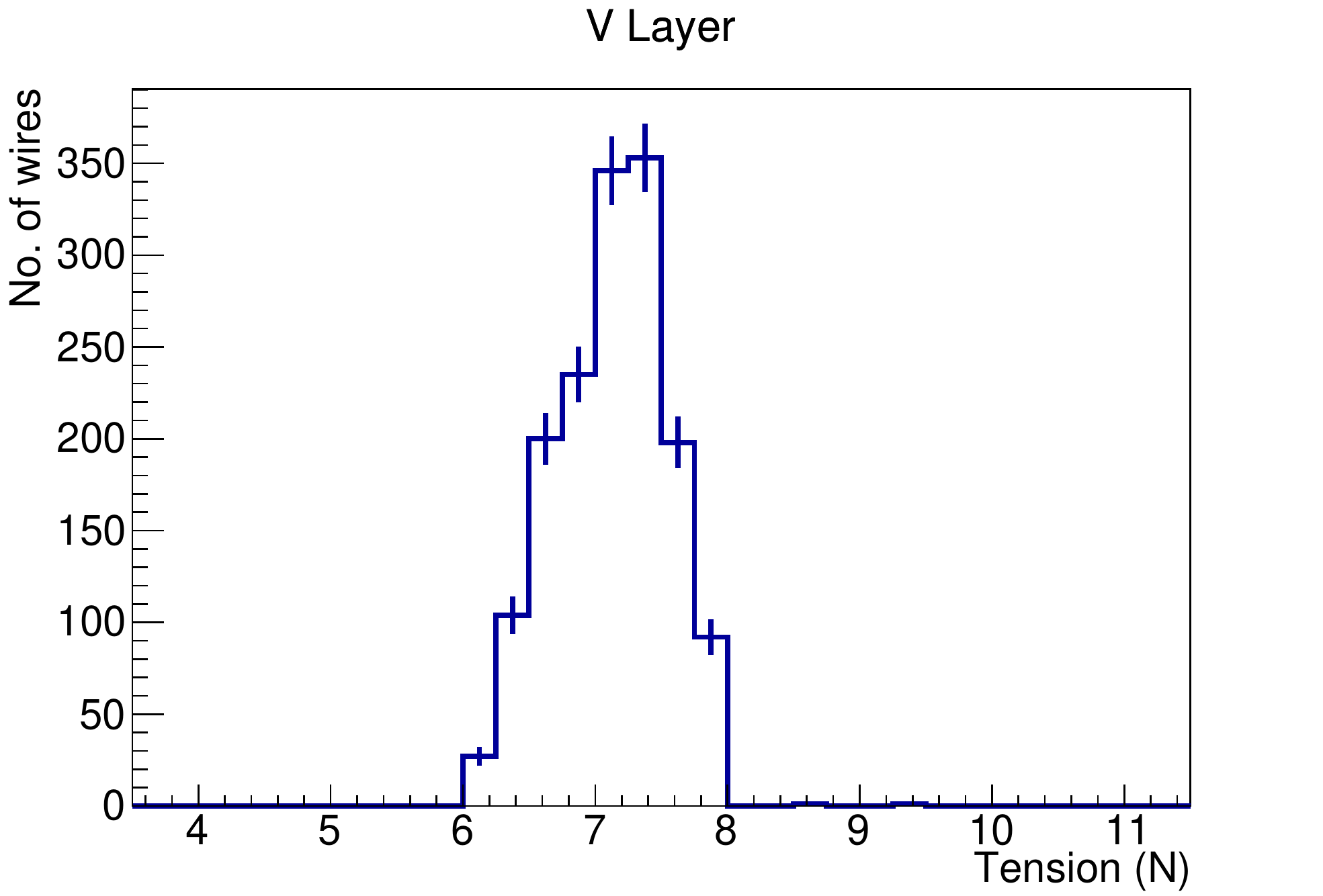}
    \includegraphics[height=0.225\textheight]{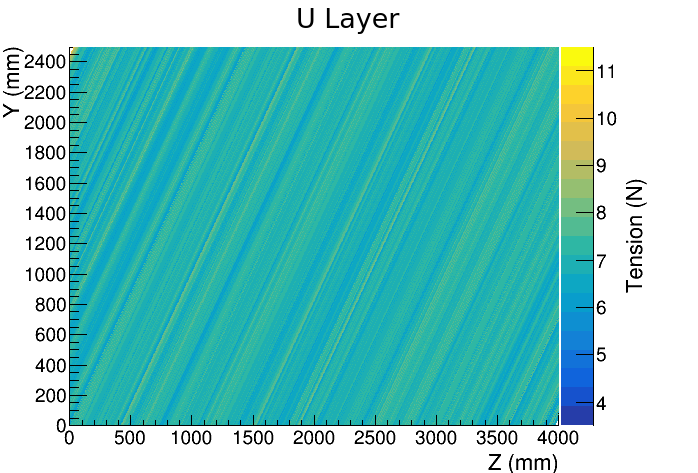}
    \includegraphics[height=0.225\textheight]{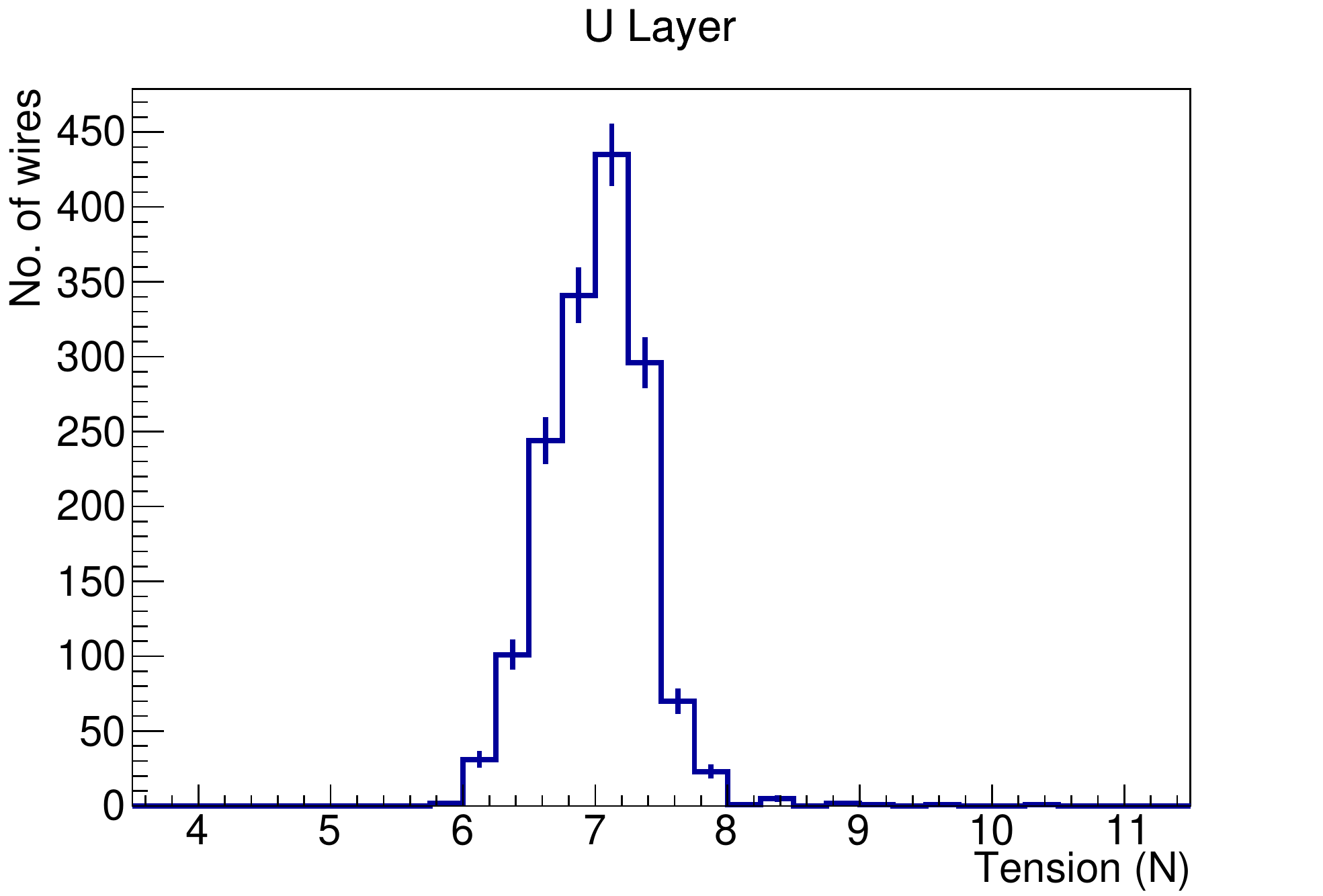}
    \caption{Tension distributions for the left-hand US APA, wired using the manual wiring method. In the left-hand column are overhead views of the wire layers, with tension represented by the colour scale. The `Y' axis corresponds to the short axis of each APA and the `Z' axis to the long axis (i.e. the axis of the wrap edge). In the right-hand column are the corresponding histograms of tensions.}
    \label{fig:US_LH_TensionDists}
\end{figure}

\begin{figure}[h!]
    \centering
    \includegraphics[height=0.225\textheight]{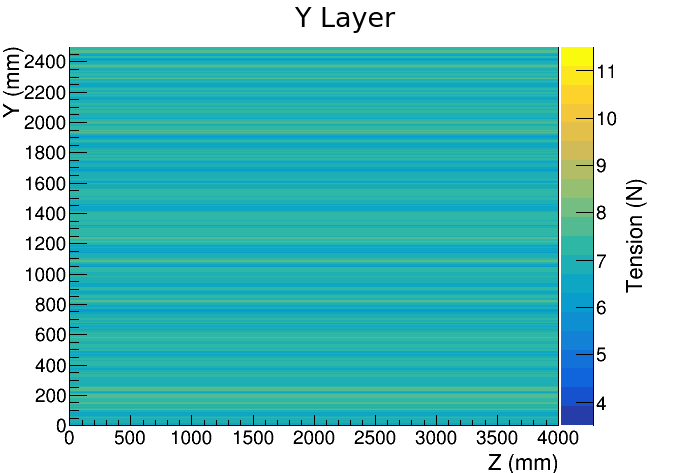}
    \includegraphics[height=0.225\textheight]{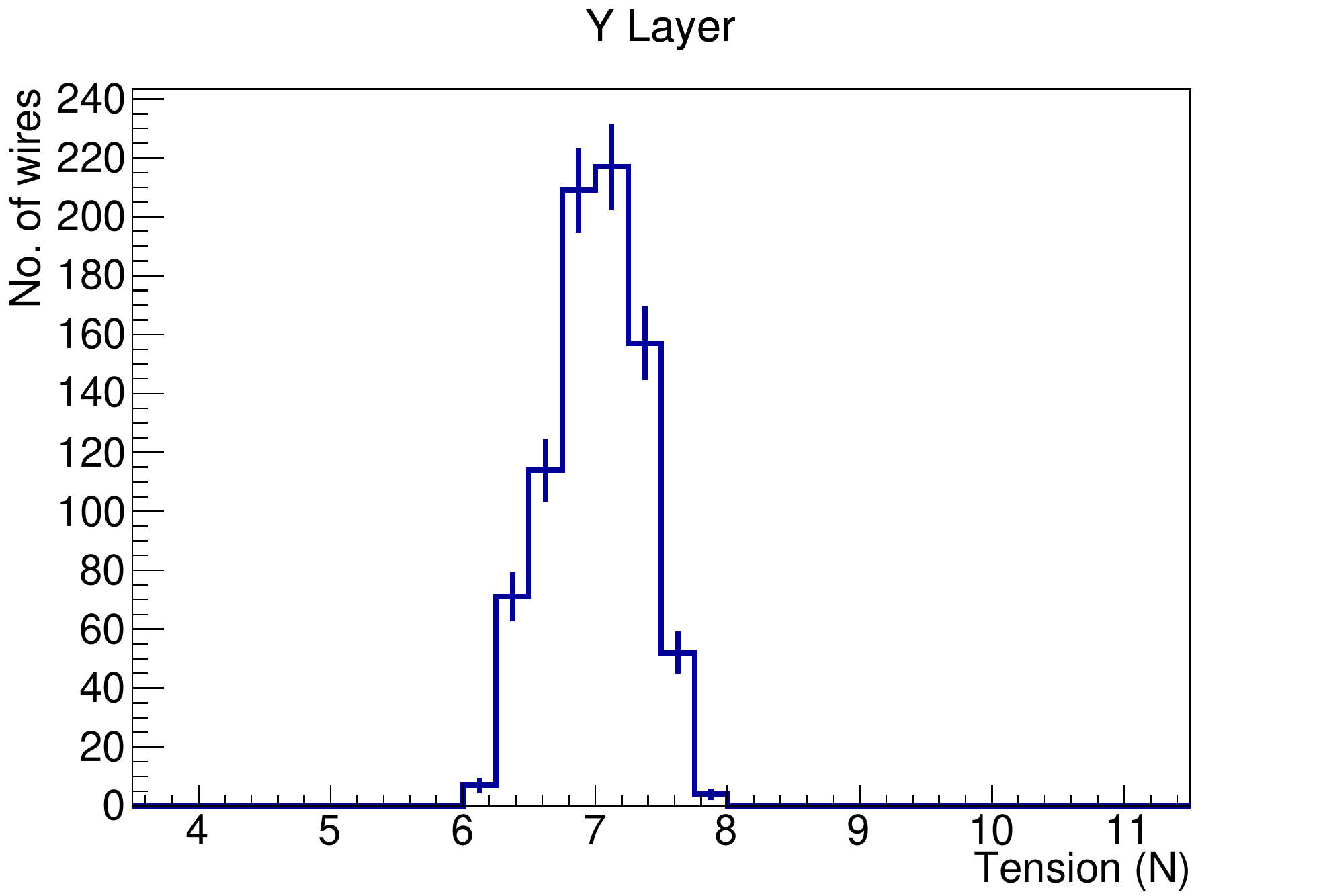}
    \includegraphics[height=0.225\textheight]{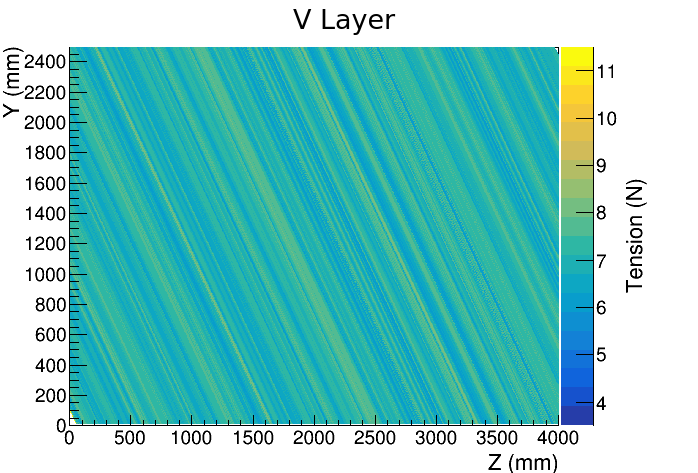}
    \includegraphics[height=0.225\textheight]{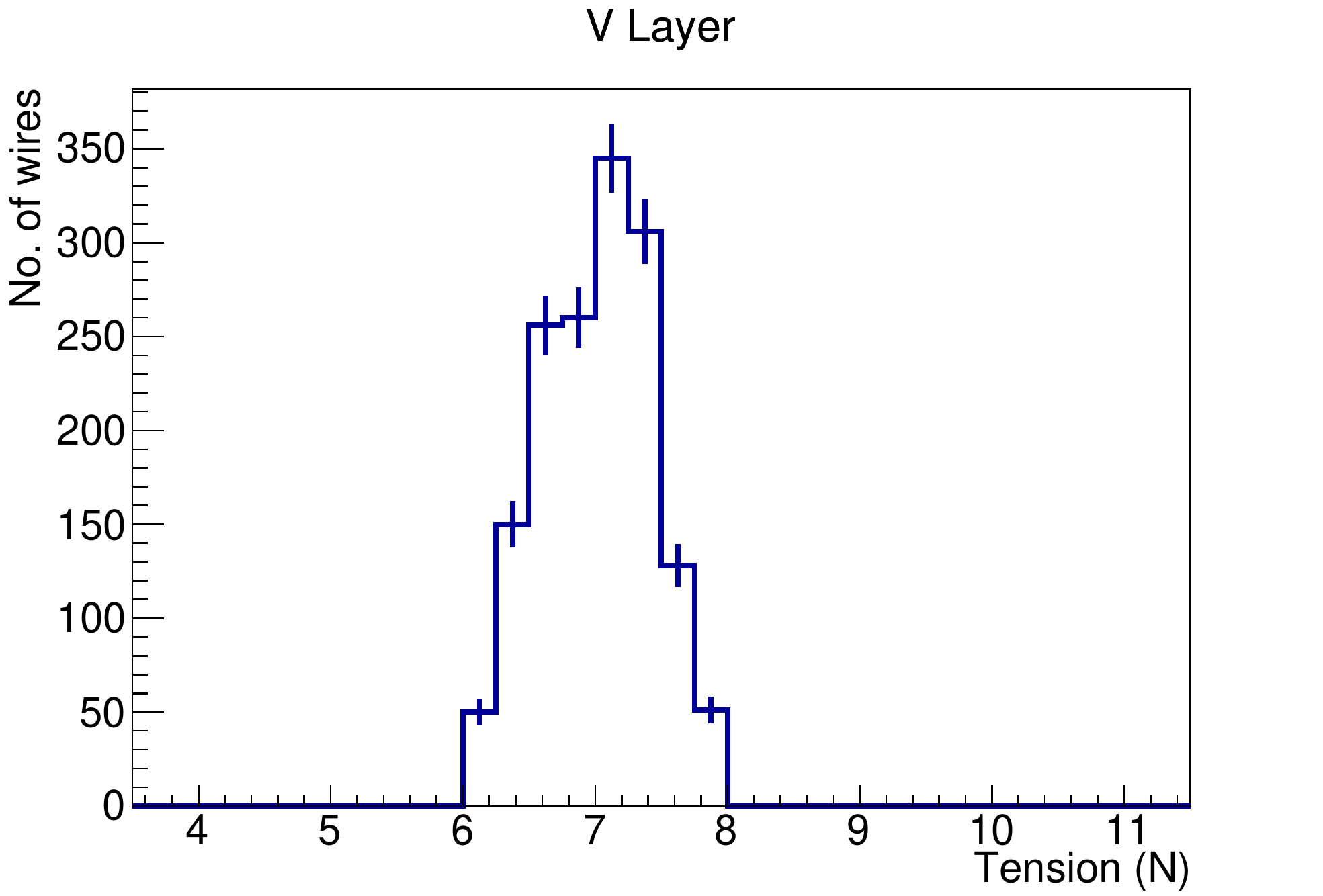}
    \includegraphics[height=0.225\textheight]{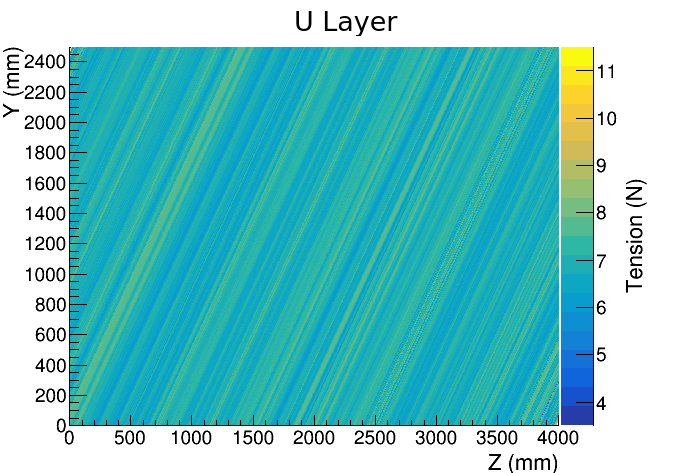}
    \includegraphics[height=0.225\textheight]{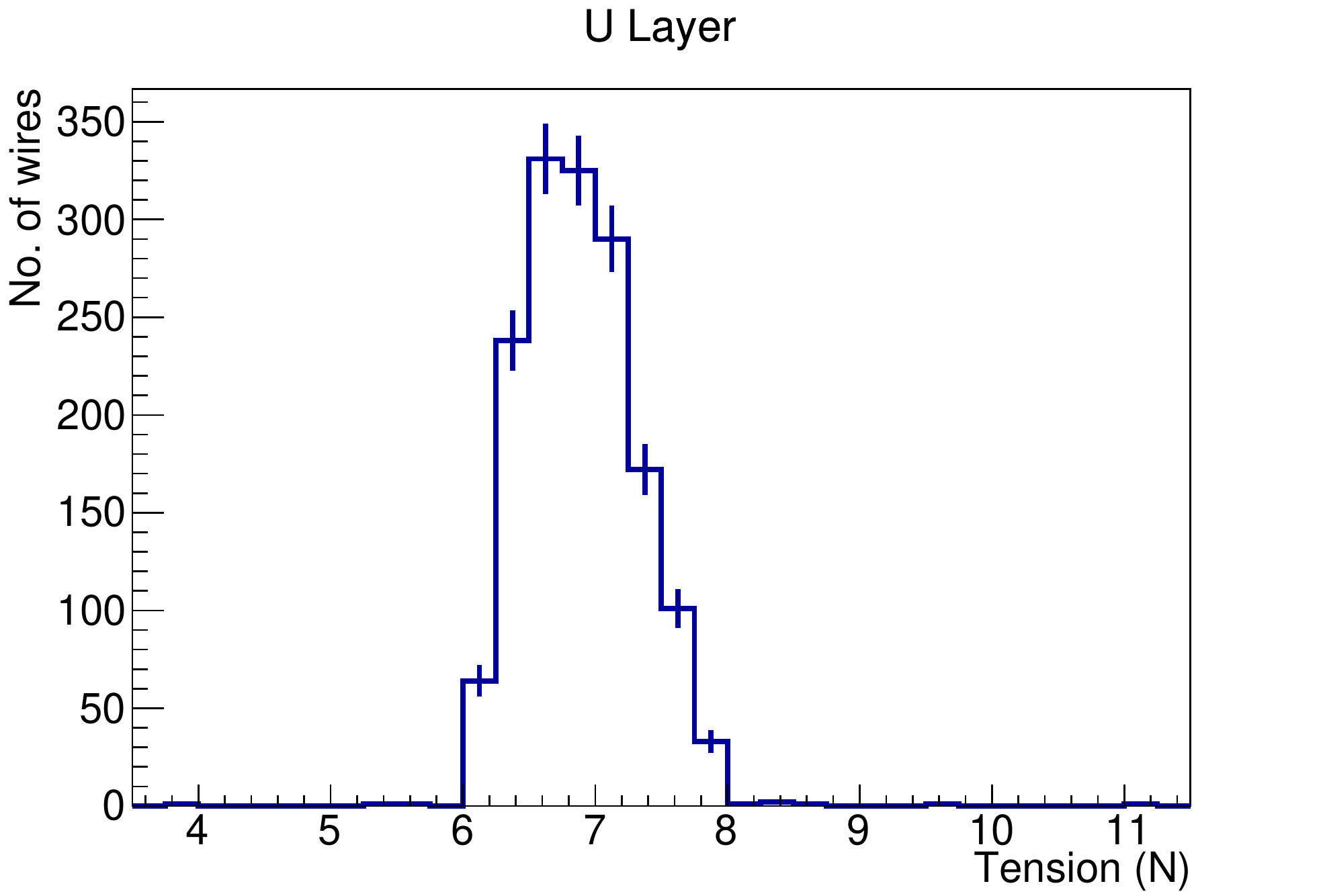}
    \caption{Tension distributions for the right-hand US APA, wired using the manual wiring method, following the same convention as figure \ref{fig:US_LH_TensionDists}.}
    \label{fig:US_RH_TensionDists}
\end{figure}

\begin{figure}[h!]
    \centering
    \includegraphics[height=0.225\textheight]{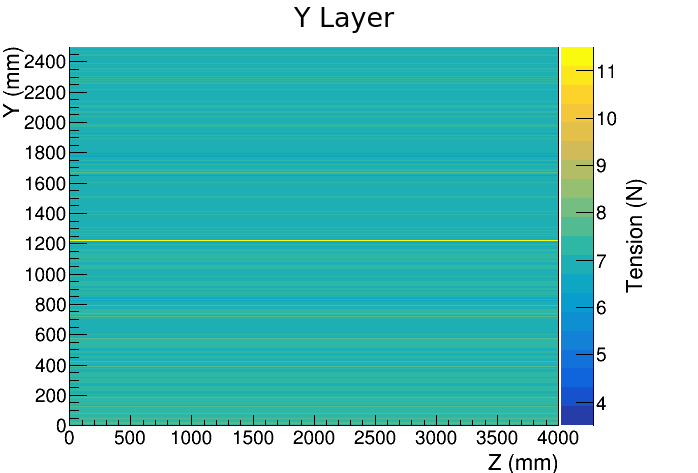}
    \includegraphics[height=0.225\textheight]{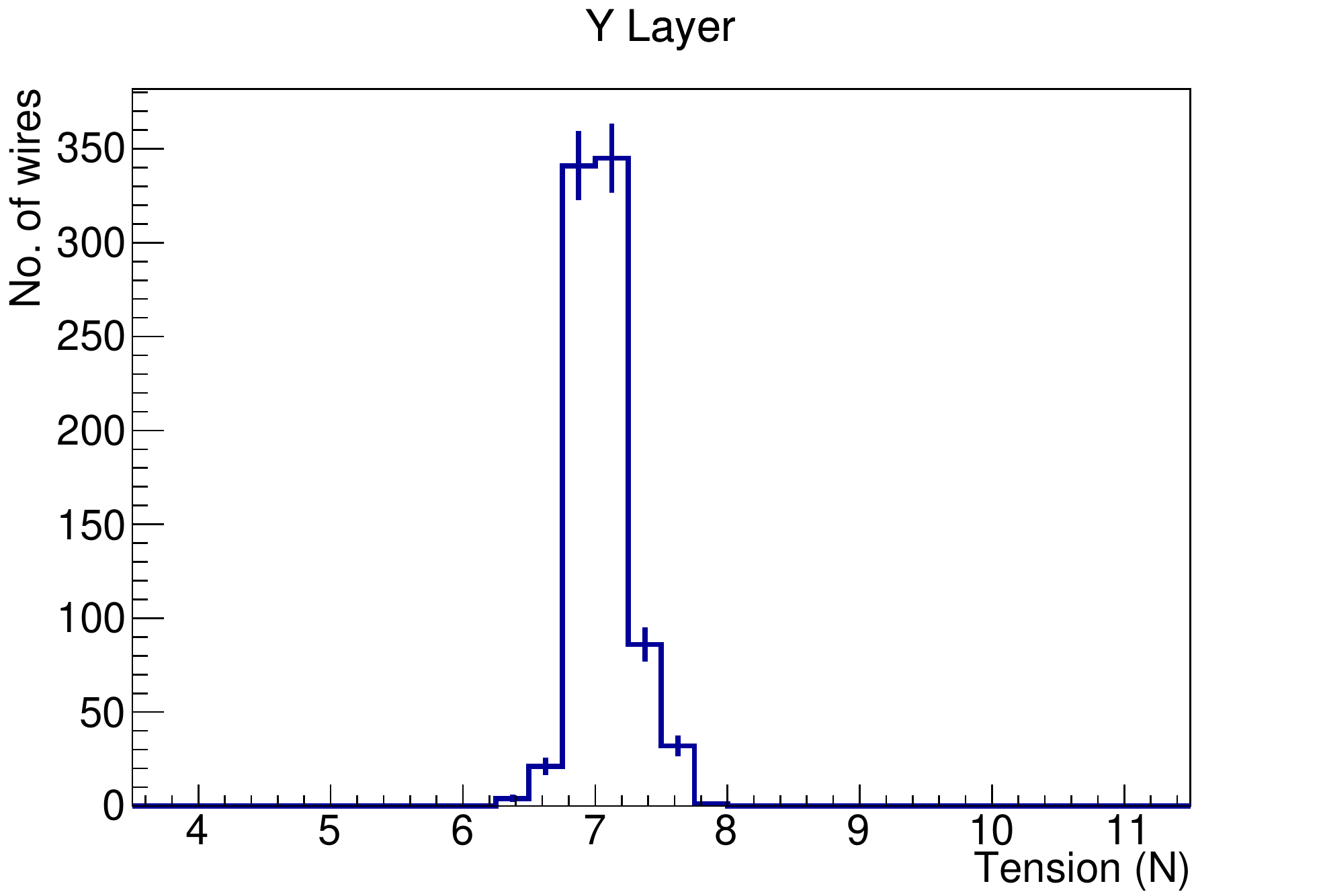}
    \includegraphics[height=0.225\textheight]{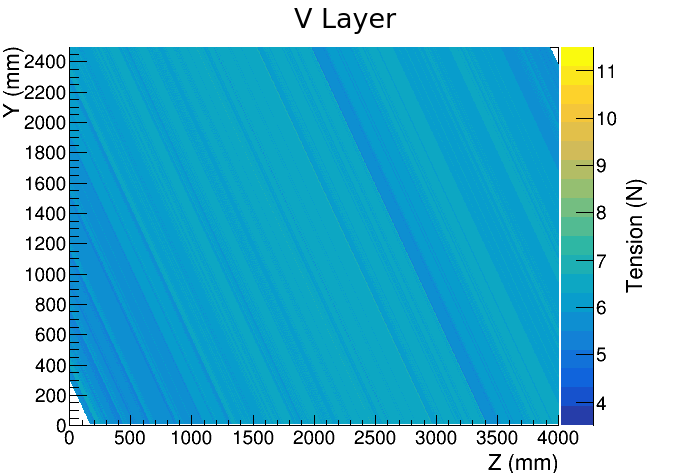}
    \includegraphics[height=0.225\textheight]{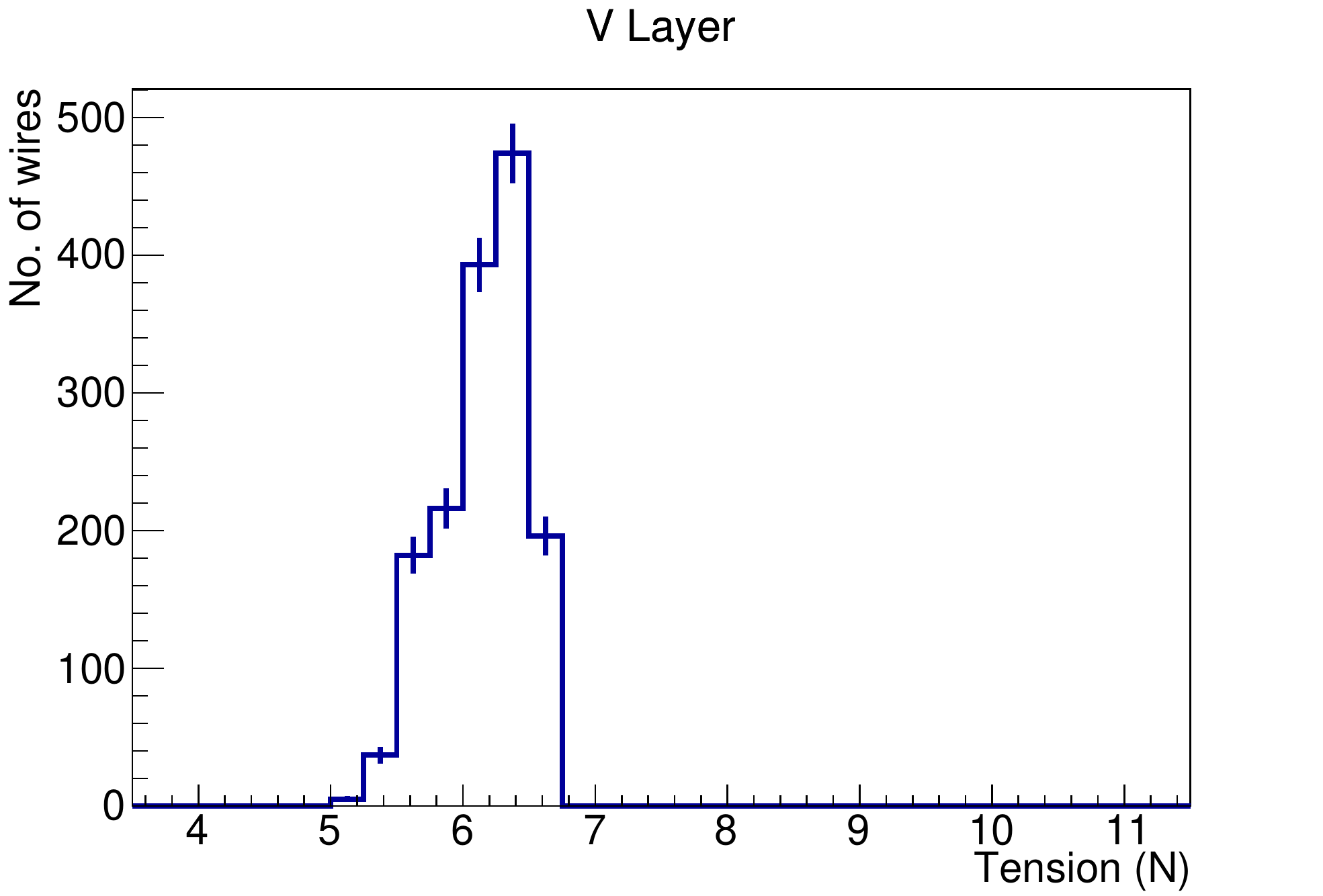}
    \includegraphics[height=0.225\textheight]{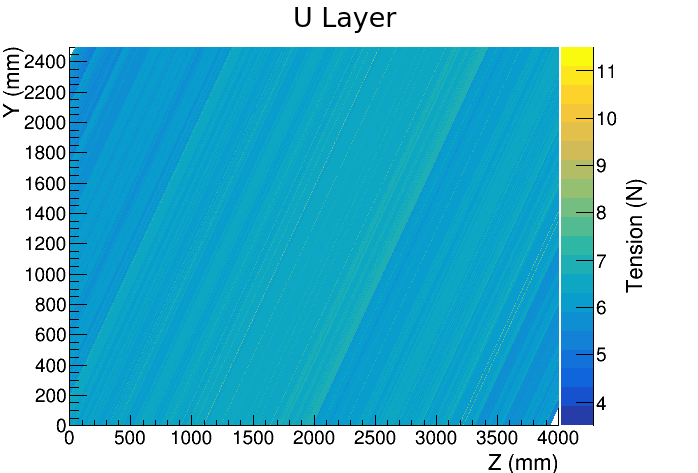}
    \includegraphics[height=0.225\textheight]{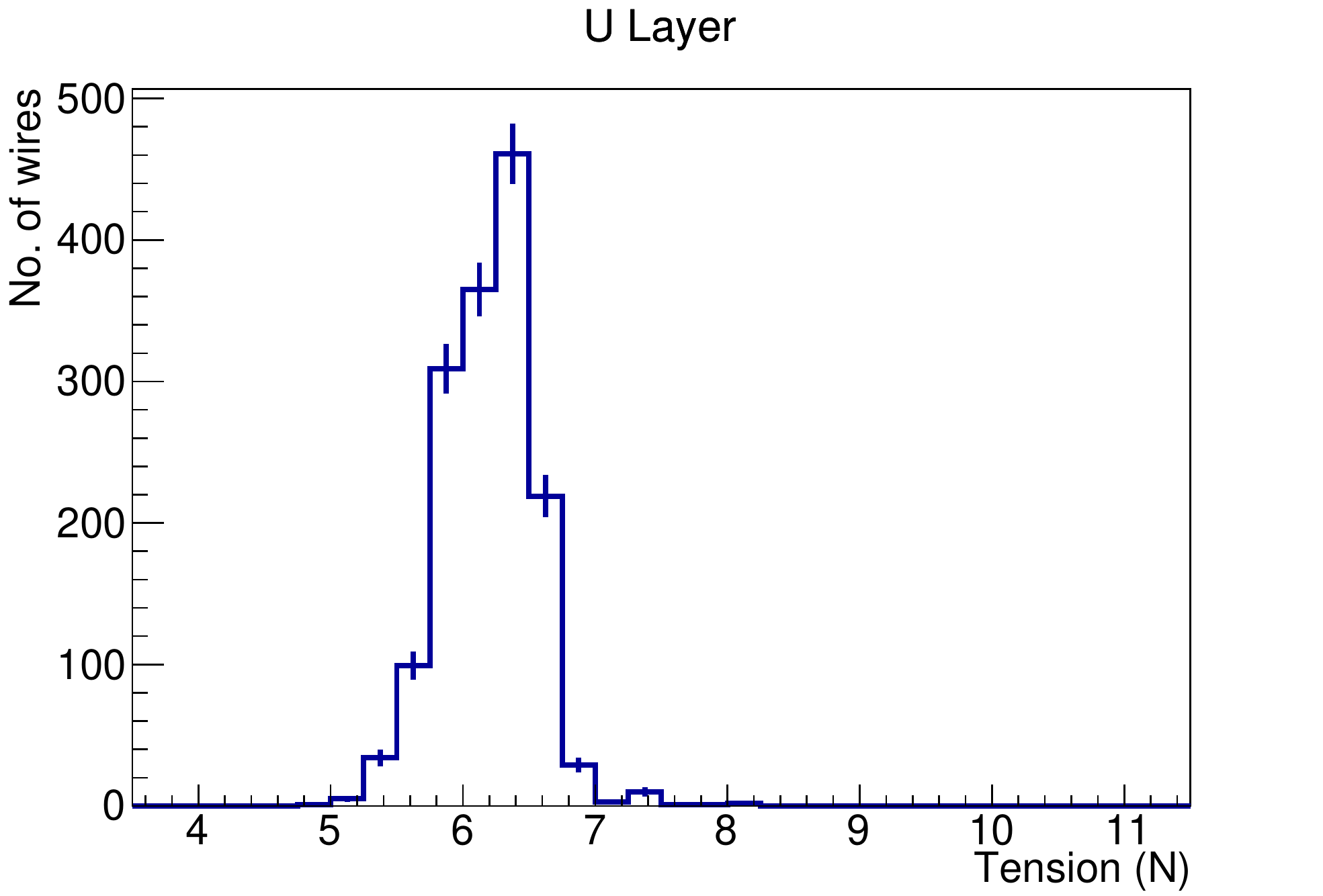}
    \caption{Tension distributions for the left-hand UK APA, wired using the semi-automated wiring method, following the same convention as figure \ref{fig:US_LH_TensionDists}.}
    \label{fig:UK_LH_TensionDists}
\end{figure}

\begin{figure}[h!]
    \centering
    \includegraphics[height=0.225\textheight]{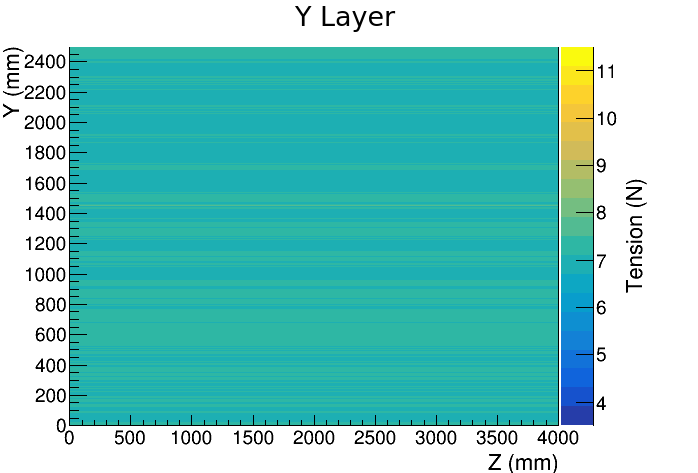}
    \includegraphics[height=0.225\textheight]{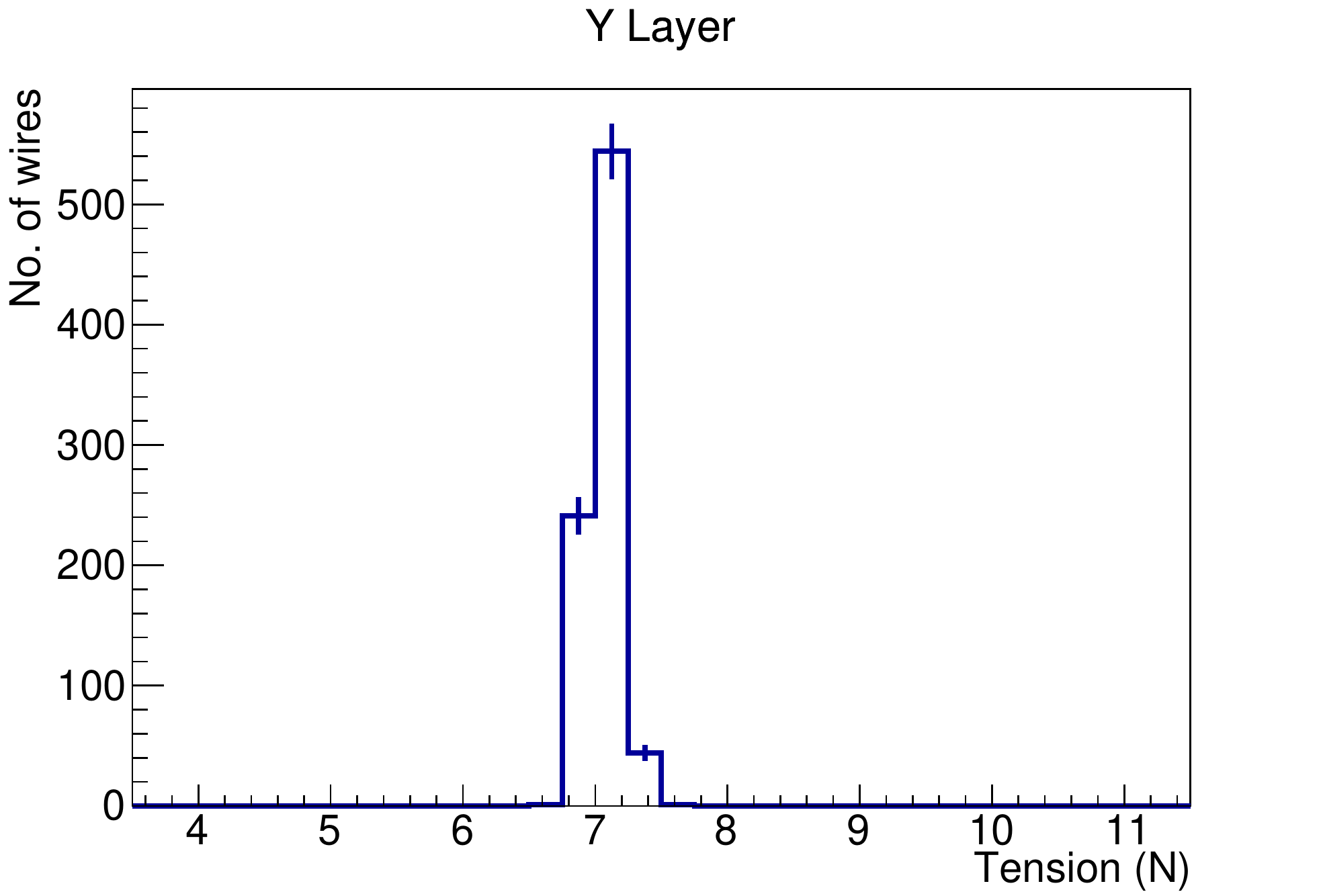}
    \includegraphics[height=0.225\textheight]{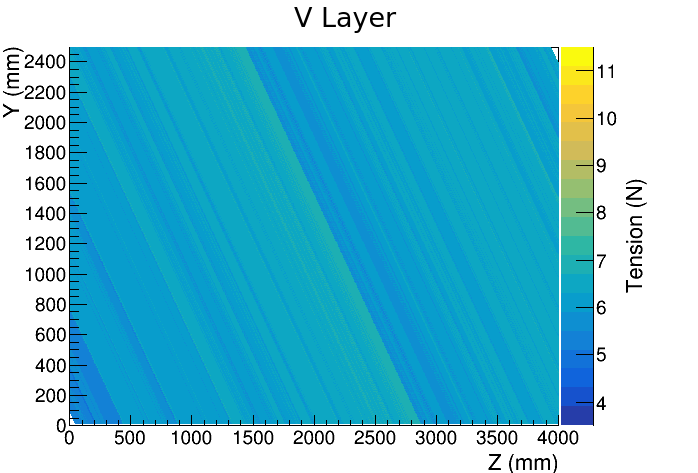}
    \includegraphics[height=0.225\textheight]{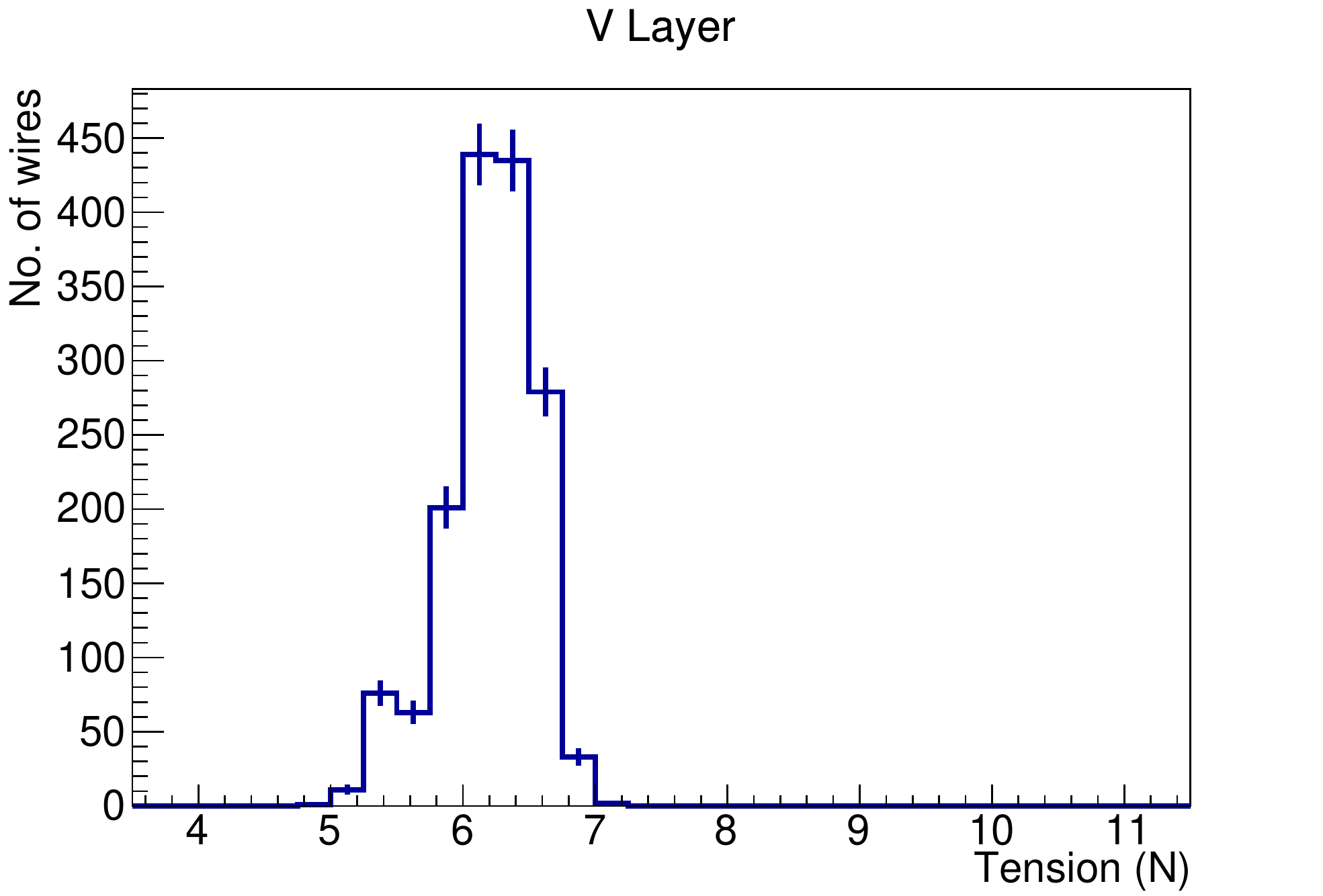}
    \includegraphics[height=0.225\textheight]{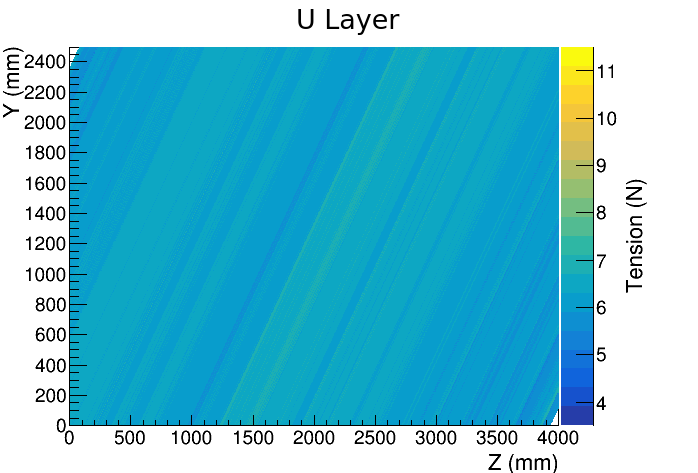}
    \includegraphics[height=0.225\textheight]{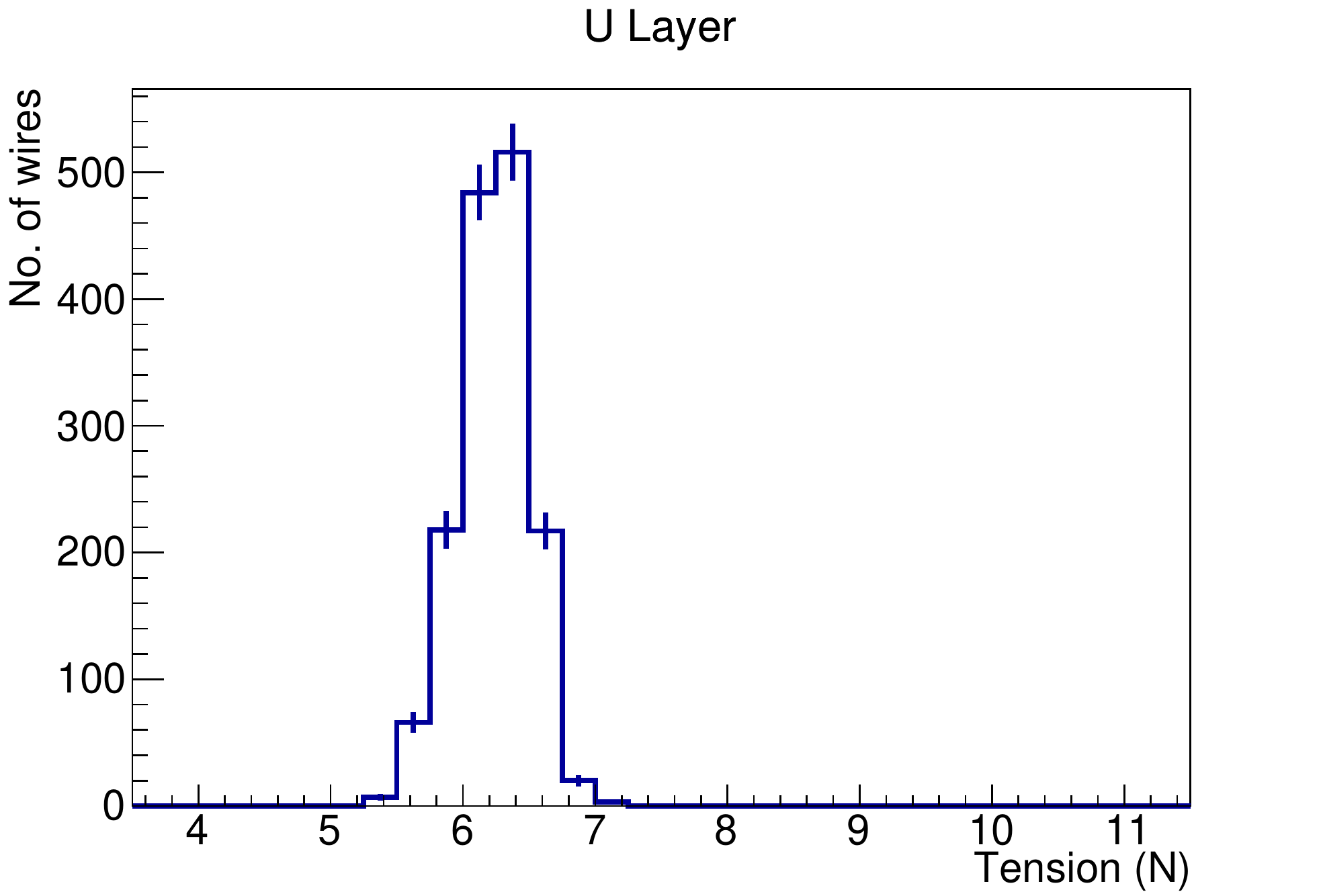}
    \caption{Tension distributions for the right-hand UK APA, wired using the semi-automated wiring method, following the same convention as figure \ref{fig:US_LH_TensionDists}.}
    \label{fig:UK_RH_TensionDists}
\end{figure}

Both wiring techniques produced tension distributions confined within a range of 2~N, as per the specifications. These distributions are shown in figures \ref{fig:US_LH_TensionDists} through \ref{fig:UK_RH_TensionDists}, and their statistical moments are summarised in tables \ref{tab:USTensionSummary} and \ref{tab:UKTensionSummary}. For every plane on every APA, the mean of the tension distribution lies within the nominal range of $7 \pm 1$~N, and only three wires across all four APAs fell below the post-slackening threshold of 5~N, which the 7~N specification was chosen to guarantee (as discussed in section \ref{sec:SBNDSpecs}). In all cases these were short corner wires where the measurement uncertainty was large and the 5~N specification was more than was needed (due to the length-dependence of the tautness requirement); as such these wires were allowed to remain. Similarly, a very small number of wires with tensions exceeding 8~N were measured in the corners, but were allowed to remain due to their short length.

\begin{table}[hbtp]
    \centering    
    \caption{\label{tab:USTensionSummary} Manual wiring tensioning results.}
    \begin{tabular}{|c|c|c|c|} 
        \hline
        Frame & Layer & Mean Tension (N) & RMS (N) \\
        \hline
        \hline
        US Left APA & Y     & 7.07             & 0.51    \\
              & V     & 7.13             & 0.42    \\
              & U     & 7.01             & 0.44    \\
        \hline
        US Right APA & Y     & 6.98             & 0.45    \\
              & V     & 7.01             & 0.48    \\
              & U     & 6.89             & 0.52    \\
        \hline
    \end{tabular}
\end{table}

\begin{table}[hbtp]
    \centering
    \caption{\label{tab:UKTensionSummary} Semi-automated wiring tensioning results.}
    \begin{tabular}{|c|c|c|c|} 
        \hline
        Frame & Layer & Mean Tension (N) & RMS (N) \\
        \hline
        \hline
        UK Left APA  & Y     & 7.05        & 0.21 \\
              & V     & 6.15        & 0.32 \\
              & U     & 6.20        & 0.35 \\
        \hline
        UK Right APA & Y     & 7.08        & 0.13 \\
              & V     & 6.20        & 0.34 \\
              & U     & 6.23        & 0.26 \\
        \hline
    \end{tabular}
\end{table}

\newpage

\newpage

\newpage

\newpage

The calibration procedures for each technique differ due to the processes involved, and the resultant variations in tension are characteristic to each technique.  Despite these differences, the resultant wire tension distributions fall within the precision specification as detailed in section \ref{sec:SBNDSpecs}.

The manual wiring method produced wider variation around the mean of its tension distributions than the semi-automated method. This enhanced variation was attributed to two main causes:
\begin{enumerate}
    \item Friction losses caused by wire kinking around the rotating pins of the pin blocks produced observable wire-to-wire tension differences in each batch of wires prior to soldering. This was unavoidable given the design of the pin blocks.
    \item For the U and V layers, the folding action of the carriage was highly sensitive to how well it was aligned to the plane of the wires, as described in section \ref{subsubsec:Carriage}. This alignment presented challenges since the precision was limited by the planarity of floor in the lab, which was not levelled. As such, the folding procedure did produce small, varying changes in tension from the initial measurement. Once the wires have been folded over the wrap edge the tension can no longer be adjusted, and so these deviations had to be accepted.
\end{enumerate}

Through anticipating these effects, and making manual adjustments to compensate, 99.98\% of the wires laid using the manual wiring technique were kept within $7 \pm 1$~N.

The APAs wired using the semi-automated setup have small gradients in tension across each board in the U and V layers, which is due to the wires being laid individually rather than in batches. As progressively more wires are laid, the lateral force applied to the wrap boards increases as a function of the number of wires. This causes a small inward distortion of the wrap edge board, thereby decreasing the tension on the initial wires laid. This created a sawtooth pattern in the tension distribution for these layers. During tests of the wiring setup, this variation was determined to be non-problematic, since the wires remained within the $7 \pm 1$~N wire tension tolerance at the time of wiring.

There is a noticeable baseline variation in tension between the APAs wired using the manual and semi-automated techniques. This is caused partially by the board distortion effect, and partially by the time at which the tension measurements were taken.  The wire tension measurements in the semi-automated wiring method were made upon completion of each full layer, whereas for the manual winding method the tension measurements were made after each small batch of wires were laid. The wire layers produced using the semi-automated method therefore had more time to relax before measurement (as discussed in section \ref{subsec:TensionOverTime}), particularly in the case of the U and V layers, which took the longest to lay. It should also be noted that the board layers on the UK frames were epoxied together in order to mitigate the board distortion effect, while the US boards were not. 

Repeat measurements of approximately 100 wires per plane were made on two APAs upon arrival at Fermilab, to assess the degree of slackening that had occurred. Over the 512 wires re-measured at Fermilab, not one wire was found to have fallen below 5N. This implies that at least $99.8\%$ of wires were delivered to Fermilab within specifications. With the youngest wire plane of those measured being over six months old at the time of measurement, there is no expectation that these tensions will experience any further decrease.


\subsection{Electrical Testing Methods}
\label{subsubsec:ElectricalQC}

Electrical continuity across each wire and electrical isolation between each pair of neighbouring wires are tested using a sourcemeter connected to a desktop LabView program. 

To measure continuity, a probe is placed manually at either end of the test channel, and a potential of 0.01~V is applied between the probes. Five consecutive measurements of the current under this voltage are averaged to produce a final measurement of the current (and thus the resistance across the channel).

To measure wire-to-wire isolation, a potential of 400V is sourced across probes placed on two adjacent wires. Once again, five consecutive measurements of the current are averaged, and the resistance between the two channels deduced from the mean current.

Due to the extremely low magnitude of the currents that flow between properly isolated channels, the absolute value of the resistance measured in the isolation tests was observed to fluctuate quite strongly with environmental conditions (e.g. local humidity). When the boards were tested submerged in liquid nitrogen, these surface currents were observed to subside to a uniform level of approximately 5~nA (roughly the sensitivity of the sourcemeter). Since even the largest fluctuations at room temperature left the measured resistance values well above the specified minimum, these fluctuations were not judged to be a cause for concern.

\subsection{Electrical Testing Results}
The measurements of electrical continuity and isolation gave similar results for all APAs. All channels were found to be electrically continuous and isolated from their neighbours, showing a resistance across each wire 10--100 times lower than the specified maximum for the continuity tests, and a resistance between each pair of neighbouring wires over 1,000 times higher than the specified minimum for isolation.

\begin{figure}[h!]
    \centering
    \includegraphics[width=\textwidth]{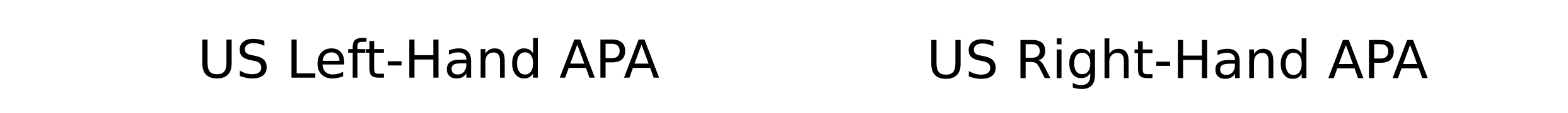}
    \includegraphics[height=0.21\textheight]{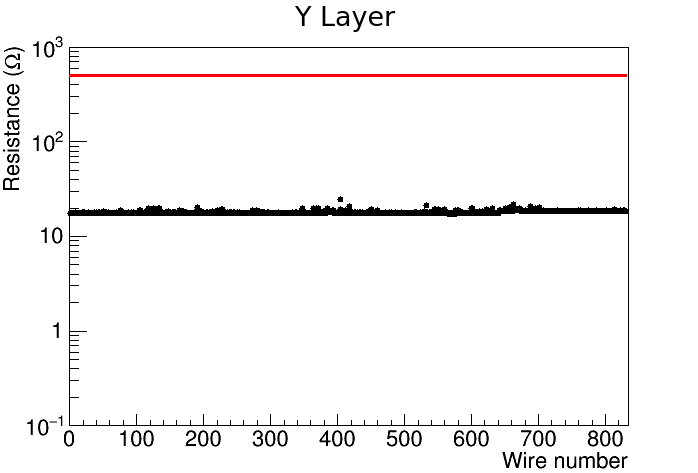}
    \includegraphics[height=0.21\textheight]{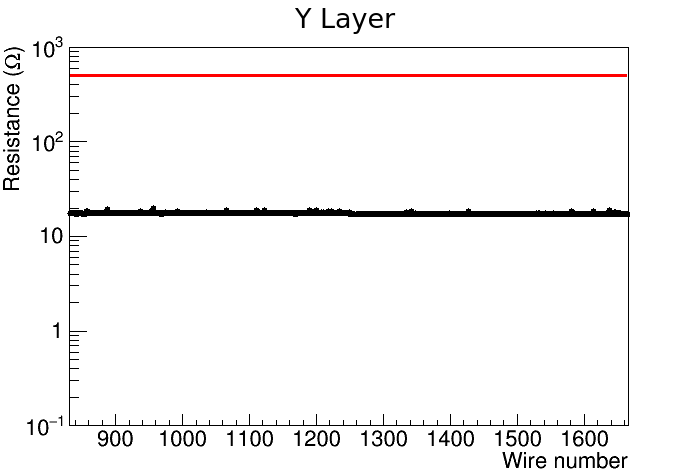}
    \includegraphics[height=0.21\textheight]{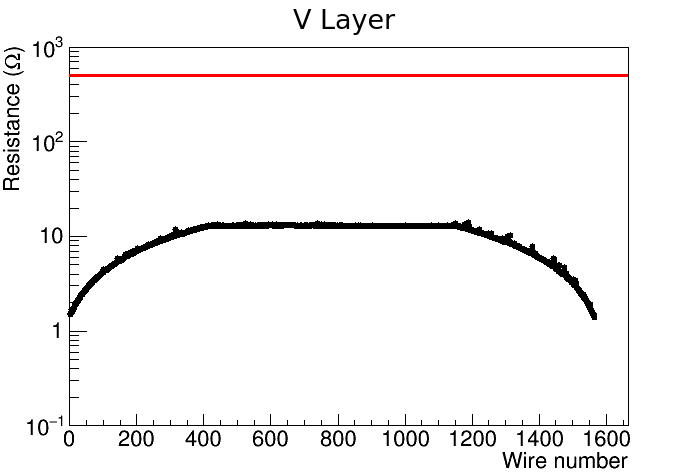}
    \includegraphics[height=0.21\textheight]{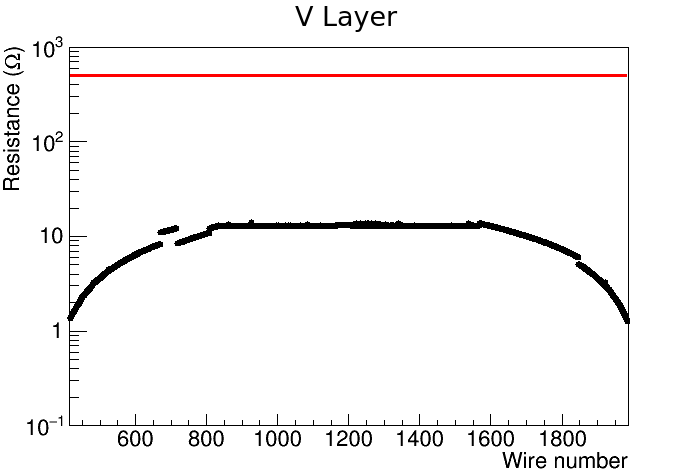}
    \includegraphics[height=0.21\textheight]{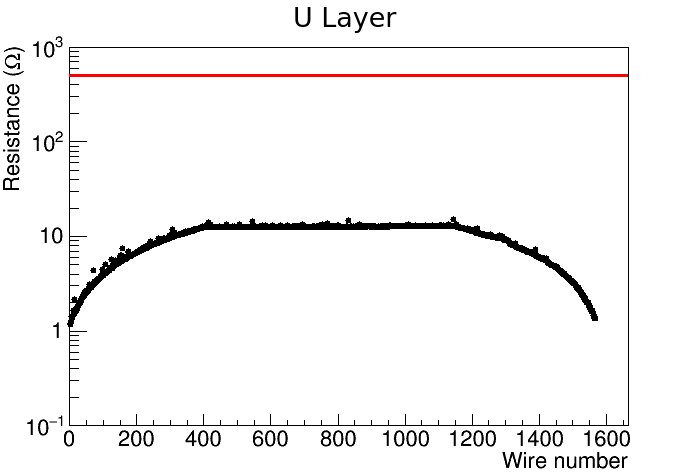}
    \includegraphics[height=0.21\textheight]{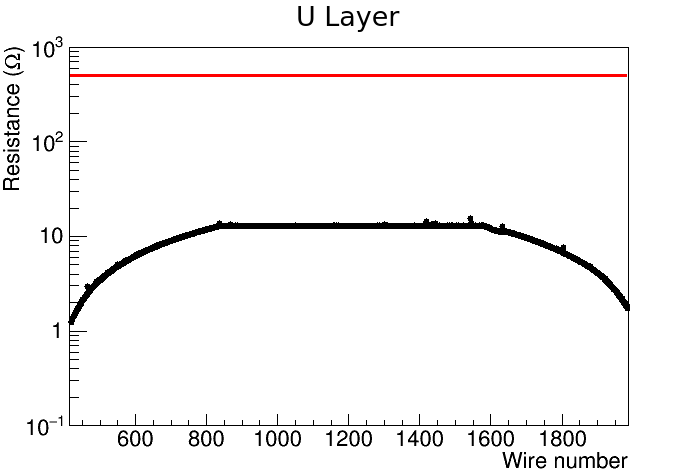}
    \caption{Continuity measurements for the wire planes wound using the manual wiring technique. Distributions for the left-hand US APA are shown in the left-hand column, the right-hand US APA on the right. The specification of 500~$\Omega$ is drawn in red: all measured resistances must be below this line.}
    \label{fig:US_ContDists}
\end{figure}

\begin{figure}[h!]
    \centering
    \includegraphics[width=\textwidth]{Plots/USHeading.png}
    \includegraphics[height=0.21\textheight]{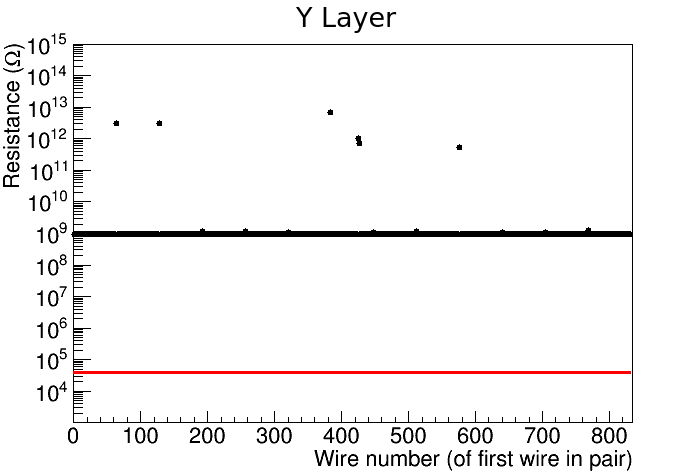}
    \includegraphics[height=0.21\textheight]{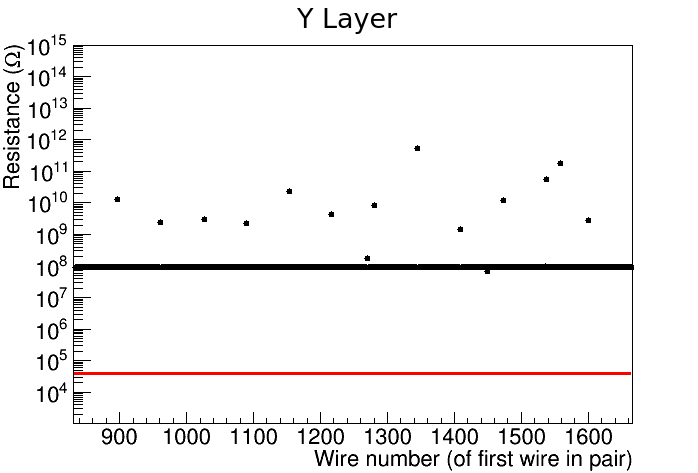}
    \includegraphics[height=0.21\textheight]{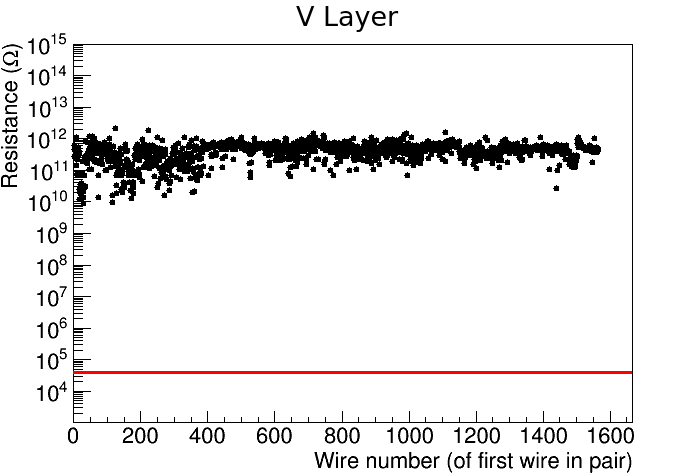}
    \includegraphics[height=0.21\textheight]{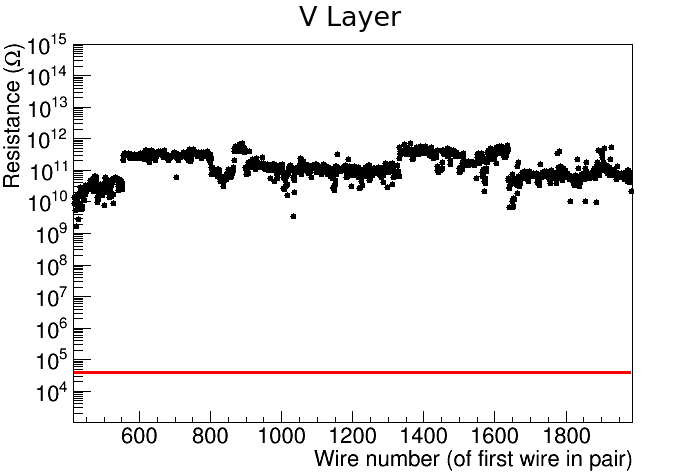}
    \includegraphics[height=0.21\textheight]{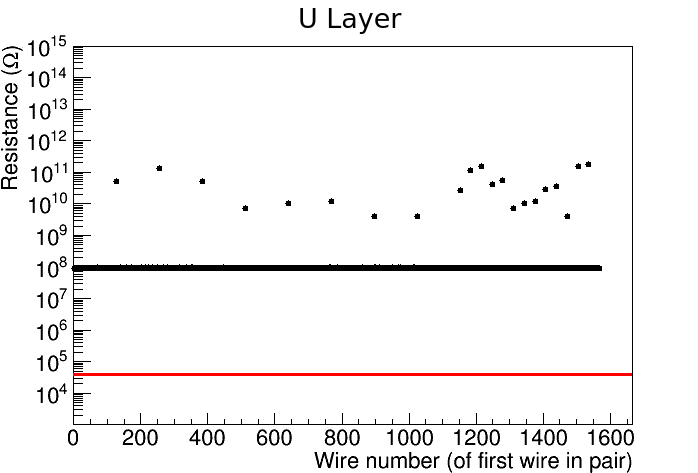}
    \includegraphics[height=0.21\textheight]{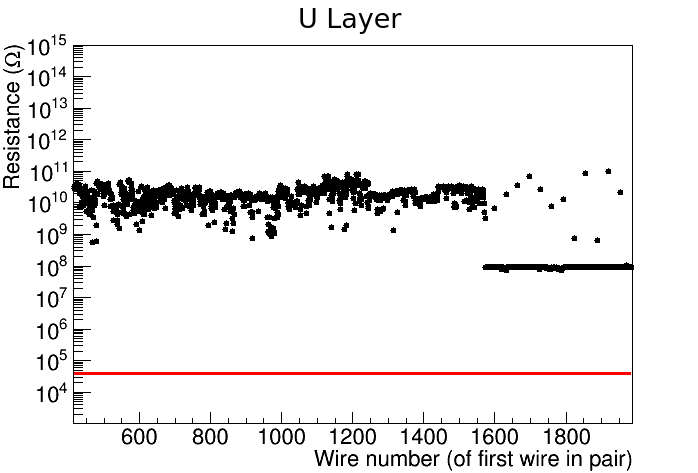}
    \caption{Isolation measurements for the wire planes wound using the manual wiring technique. Distributions for the left-hand US APA are shown in the left-hand column, the right-hand US APA on the right. The specification of 10~M$\Omega$ is drawn in red: all measured resistances must be above this line.}
    \label{fig:US_IsolDists}
\end{figure}

\begin{figure}[h!]
    \centering
    \includegraphics[width=\textwidth]{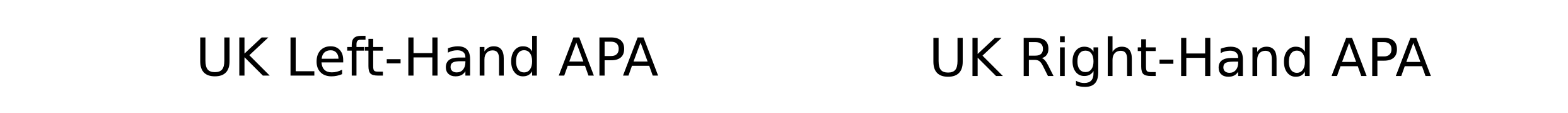}
    \includegraphics[height=0.21\textheight]{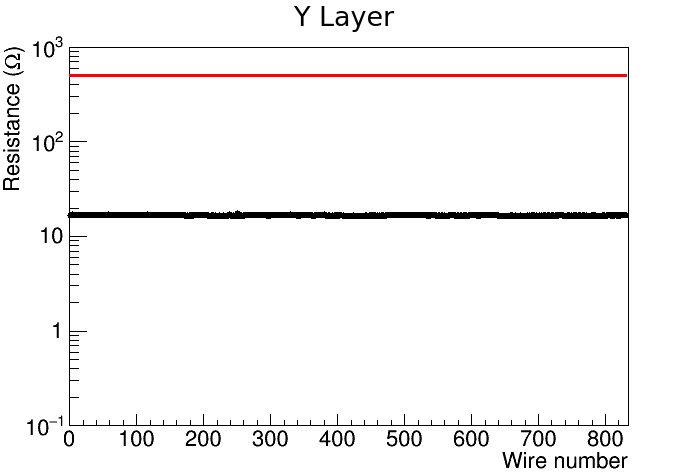}
    \includegraphics[height=0.21\textheight]{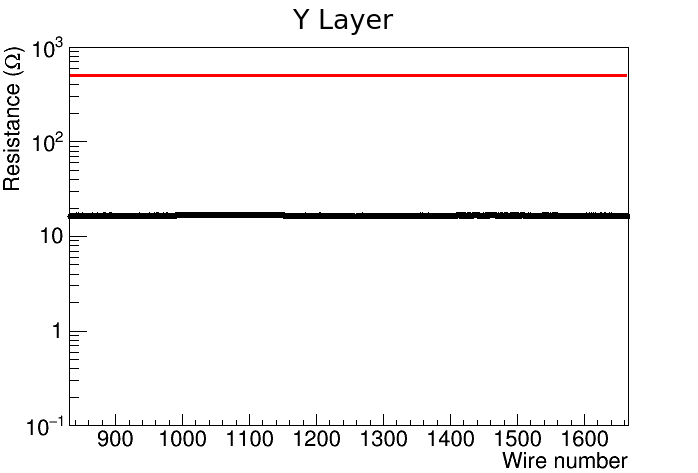}
    \includegraphics[height=0.21\textheight]{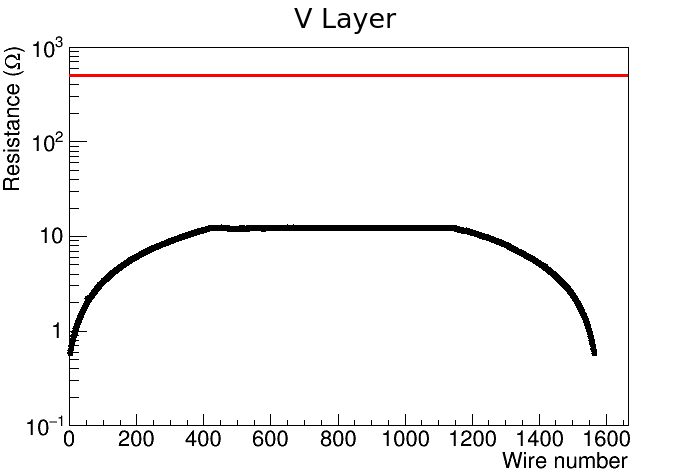}
    \includegraphics[height=0.21\textheight]{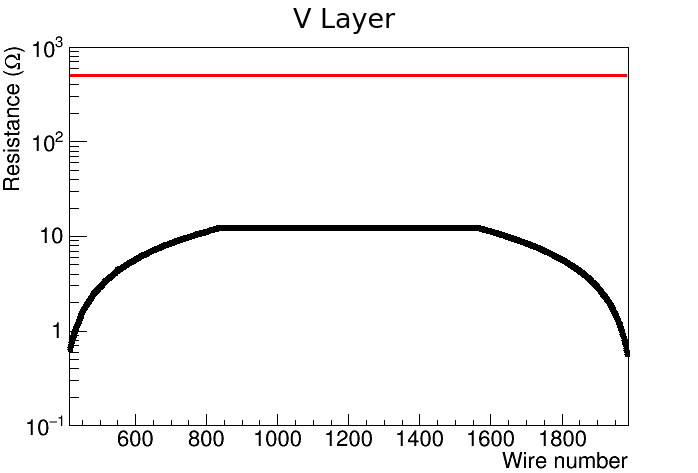}
    \includegraphics[height=0.21\textheight]{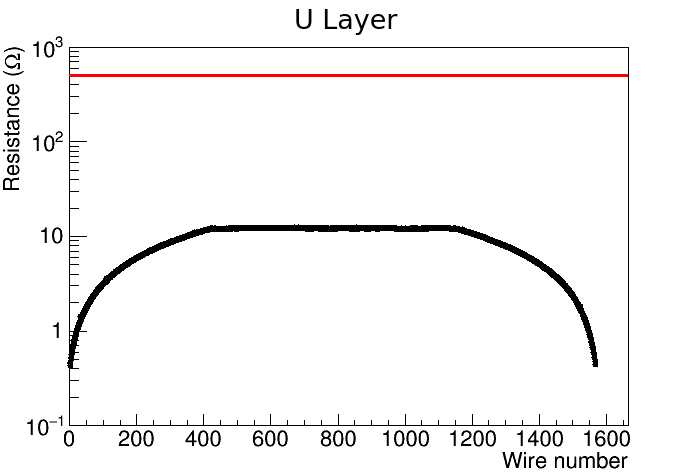}
    \includegraphics[height=0.21\textheight]{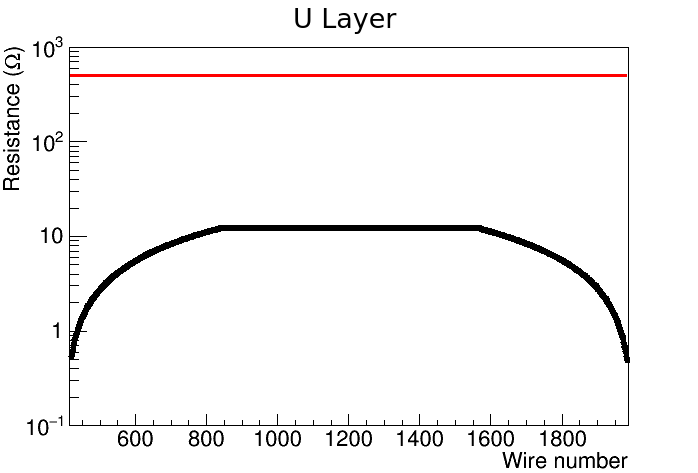}
    \caption{Continuity measurements for the wire planes wound using the semi-automated wiring technique. Distributions for the left-hand UK APA are shown in the left-hand column, the right-hand UK APA on the right. The specification of 500~$\Omega$ is drawn in red: all measured resistances must be below this line.}
    \label{fig:UK_ContDists}
\end{figure}

\begin{figure}[h!]
    \centering
    \includegraphics[width=\textwidth]{Plots/UKHeading.png}
    \includegraphics[height=0.21\textheight]{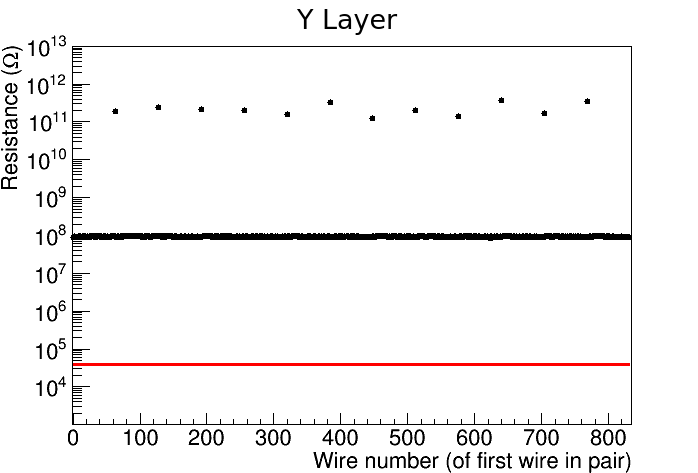}
    \includegraphics[height=0.21\textheight]{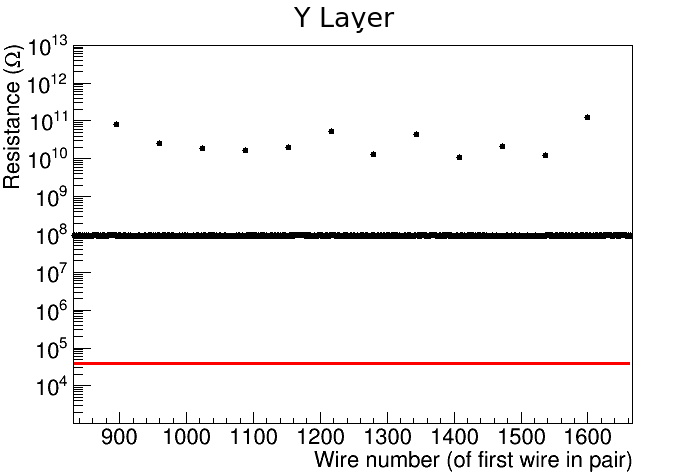}
    \includegraphics[height=0.21\textheight]{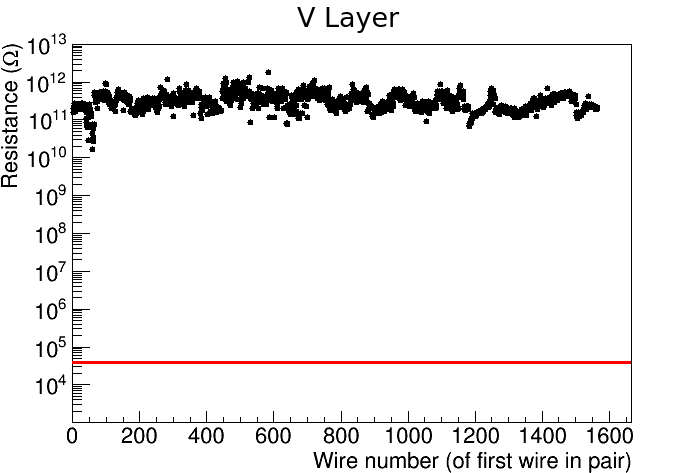}
    \includegraphics[height=0.21\textheight]{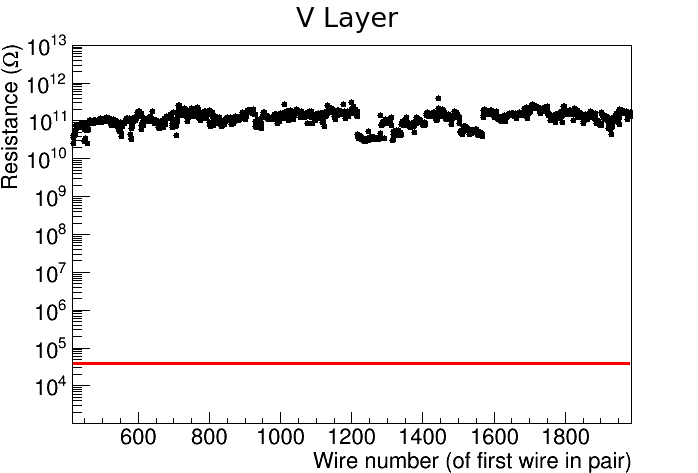}
    \includegraphics[height=0.21\textheight]{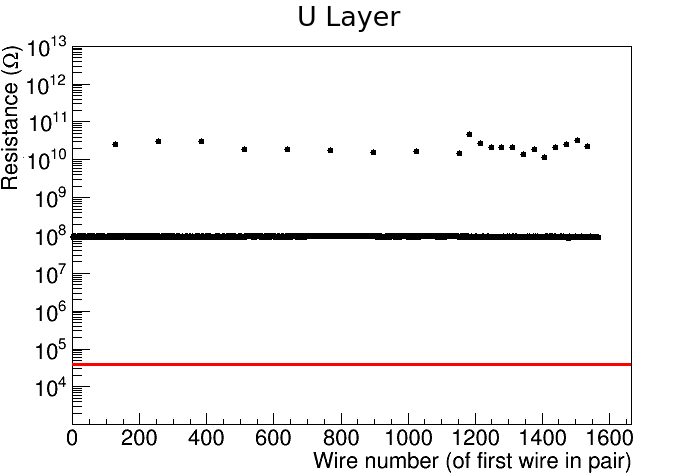}
    \includegraphics[height=0.21\textheight]{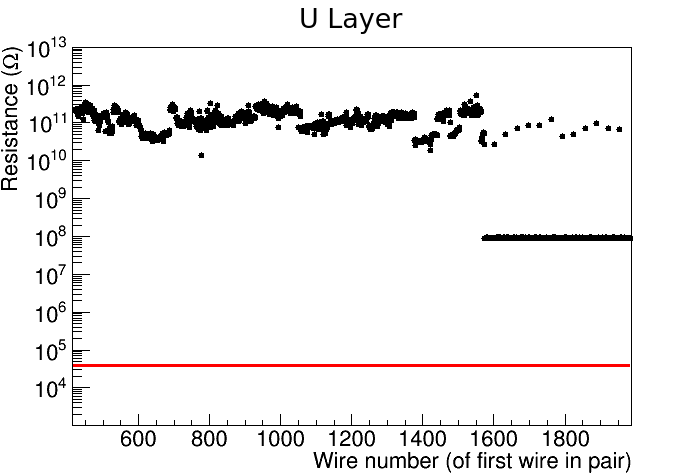}
    \caption{Isolation measurements for the wire planes wound using the semi-automated wiring technique. Distributions for the left-hand UK APA are shown in the left-hand column, the right-hand UK APA on the right. The specification of 10~M$\Omega$ is drawn in red: all measured resistances must be above this line.}
    \label{fig:UK_IsolDists}
\end{figure}

The full distributions of these resistances are shown in figures \ref{fig:US_ContDists} and \ref{fig:US_IsolDists} for the APAs wound using the manual wiring method, and in figures \ref{fig:UK_ContDists} and \ref{fig:UK_IsolDists} for the APAs wound using the semi-automated method. 

The continuity graphs show the expected distributions of resistance scaling linearly with wire length, giving flat distributions in the Y planes and distributions that rise linearly from the corners of the U and V planes to a plateau in the centre (where the wire length is uniform).

The isolation graphs also show the expected behaviour, with a uniform high resistance between channels on boards without resistor and capacitor components (as in the lower board shown in figure \ref{fig:GeomBoardLayouts}), and a lower, still uniform resistance on boards with those components (as in the upper board shown in the same figure). For the latter boards, gaps between boards appear as single data points with a much higher resistance than the mode.

Some strong variation is visible in the isolation resistance between channels on boards without components, largely due to the varying environmental conditions discussed in section \ref{subsubsec:ElectricalQC} and the fact that the current flowing approached the sensitivity of the sourcemeter ($\sim$5~nA). The values of resistance measured were within the specifications by better than an order of magnitude, therefore the inconsistencies wire-to-wire were deemed acceptable and were not further investigated.

Similarly to the tension measurements, the continuity and isolation measurements were re-sampled on two APAs after arrival at Fermilab, over approximately 100 wires per layer. The results were observed to be consistent with those measured at the wiring sites, with every re-measured wire falling well within the specifications.

\newpage

\newpage

\newpage

\newpage

\section{Cold Test}
\label{sec:ColdTest}

In order to verify the stability of an APA during cooldown to cryogenic temperatures, the first fully wired SBND APA frame was subjected to a quality control test in which the APA frame was gradually cooled to $\sim$150~K.  The procedure provides an integration test of a fully wired APA frame in an environment similar to a LArTPC, while also providing the opportunity to test key sub-component requirements (such as tension) after exposure to cryogenic temperatures.  The initial period of cooldown places the APA under the most stress, due to the non-uniform cooling rate of the components; the wires cool rapidly, whereas the more massive frame cools slowly.  This motivates the design of equipment and procedure which ensure a slow cool down rate of $\sim$50~K/hour.  After reaching equilibrium, the frame cools at a uniform rate, therefore reaching the absolute temperature of liquid argon is not necessary in this test. Particular care was taken to maintain a dry environment throughout, to avoid moisture buildup on the electrical circuitry.

A custom designed cryogenic vessel was constructed for this purpose, which consisted of an open-top stainless steel vessel of dimensions 6.5~m $\times$ 3.0~m $\times$ 0.5~m, surrounded by 15~cm thick insulating foam on all sides, structurally supported by an outer aluminium frame.  The vessel has a removable roof composed of polyisocyanurate foam sheets, which are supported on rails across the vessel top.


The cryogenic cooling system consists of an extensive closed copper coil which is punctured with 2~mm holes at various points along its length and rests on the floor of the cryogenic vessel. The liquid nitrogen is injected into the vessel interior via the coil's puncture holes where it flash boils; the boil-off nitrogen gas drives the cool down of the vessel.  The cryogenic vessel is purged of moisture using compressed air and nitrogen gas, prior to and after injection of the cryogenic liquid to avoid moisture condensing on the APA structure.  

\begin{figure}[h!]
  \centering
  \includegraphics[height=0.25\textheight]{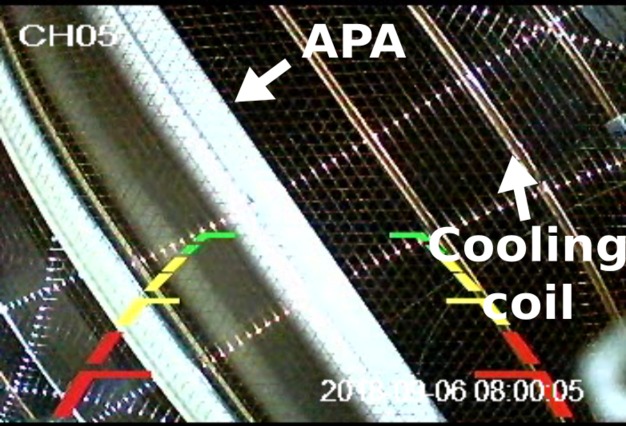}
  \caption{Camera footage taken during the wired APA cold test.  The stainless steel APA frame, the APA wires and a small section of the cryogenic vessel's copper coil cooling system are visible. The overlaid reversing lines should be ignored.}
  \label{fig:ColdTestCam5PreCooldown}
\end{figure}

The vessel is equipped with temperature, pressure and relative humidity monitoring equipment, as well as cameras for monitoring the vessel environment. The fifteen cameras (commercial car reversing cameras of a kind previously demonstrated to work at liquid nitrogen temperatures \cite{McConkey:2016spe}) are attached to the vessel's roof support rails, and during operation the vessel is illuminated by LEDs mounted on the vessel walls.  Camera footage taken during the wired APA cold test is shown in figure \ref{fig:ColdTestCam5PreCooldown}. 

\begin{figure}[h!]
    \centering
    \includegraphics[width=0.99\textwidth]{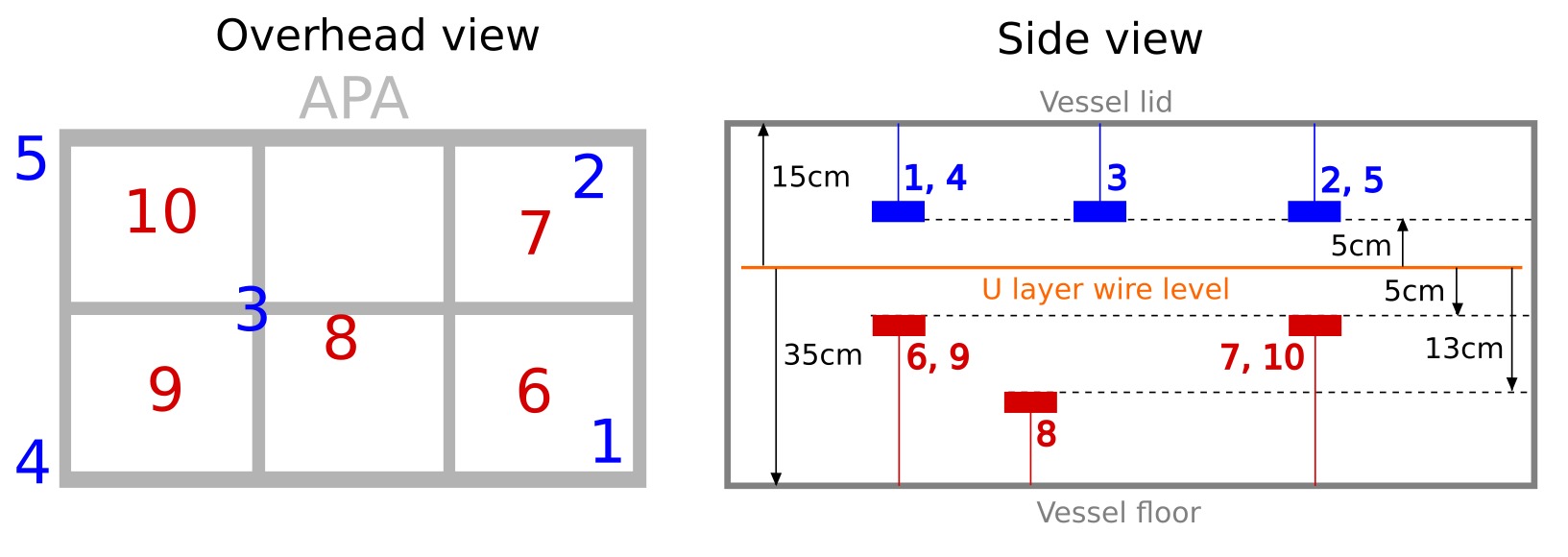}
    \caption{The positioning of the ten numbered RTDs in the plane of the APA (left), and their heights with respect to the wire plane (right). RTDs suspended from the top of the vessel are shown in blue, while those attached to the bottom are shown in red.}
    \label{fig:thatDomNeedsToMake}
\end{figure}

The temperature of the vessel is continually measured by ten Resistance Temperature Detectors (RTDs) located above and below the APA as shown in figure \ref{fig:thatDomNeedsToMake}. These systems are operated through a National Instruments USB-6002 data acquisition system running a LabVIEW script.

\begin{figure}[h!]
  \centering
  \includegraphics[height=0.25\textheight]{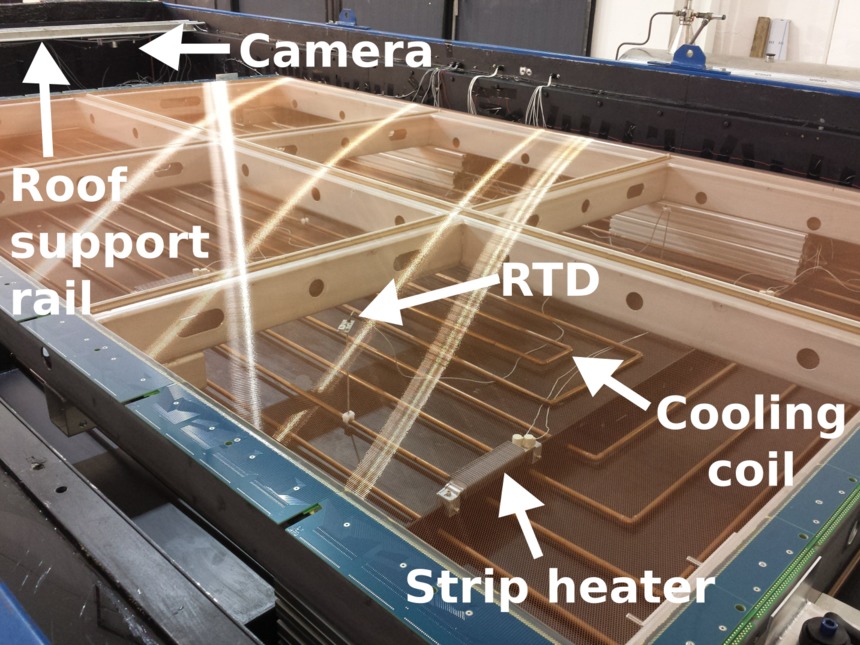}
  \caption{The first wired APA resting in the cryogenic vessel.  Several key features of the cryogenic vessel are visible: the cooling coil, an RTD mounted to the cooling coil and the finned strip heaters attached to the L-shaped cross-section floor support.  A roof support rail with a mounted camera are visible in the top left corner of the photo.}
  \label{fig:ColdTestAPAInVessel}
\end{figure}

The APA was supported inside the cryogenic vessel by levelling feet and a set of aluminium struts. A photograph of the wired APA in the cryogenic vessel is shown in figure \ref{fig:ColdTestAPAInVessel}.

\begin{figure}[t!]
    \centering
    \includegraphics[height=0.35\textheight]{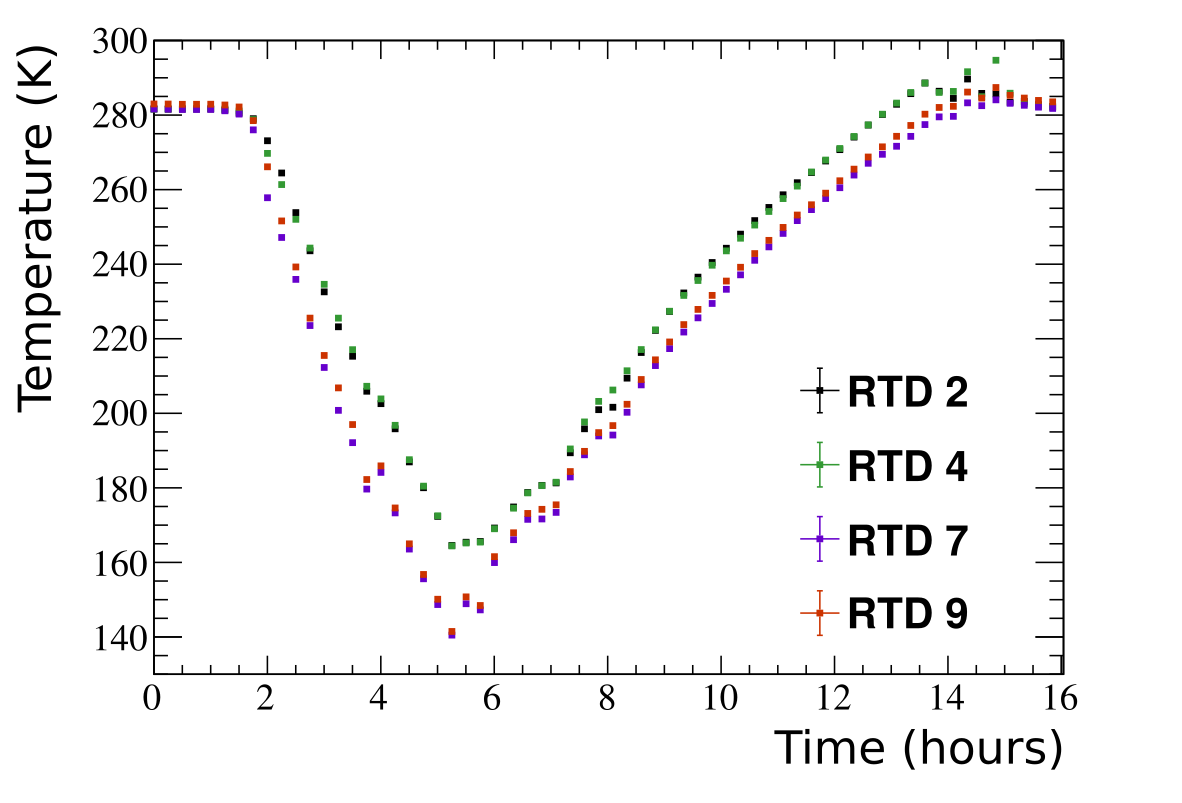}
    \caption{Temperature of the cryogenic vessel during the wired APA cold test as measured by selected RTDs, positions shown in figure \ref{fig:thatDomNeedsToMake}.  The RTD measurements have been re-sampled (4 samples/hour) to make the overlaid features of the distribution as clear as possible.}
    \label{fig:ColdTestTemperatures}
\end{figure}

The vessel environment was purged of moisture using compressed air until the relative humidity reached $\sim$10$\%$ where it levelled out.  Room-temperature nitrogen gas was then flushed through the piping to purge any final moisture from the copper cooling coil. Liquid nitrogen was then pushed into the copper coil to initiate the cool down.  The nitrogen flow was manually controlled on the dewar to ensure a slow cooling rate was maintained.  The nitrogen flow was stopped once the vessel environment temperature was $\sim$150~K.  The cooling phase took 3.5 hours and maintained a cooling rate of $\sim$40~K/hour.  To warm-up the setup, eight 500~W finned-strip heaters mounted on the vessel floor were powered up, and in order to prevent moisture entering the vessel and condensing on the cold wires, vaporised room temperature nitrogen gas continually purged the vessel while it was heated over a $\sim$8 hour period. As can be seen from figure \ref{fig:ColdTestTemperatures}, there was effectively no horizontal temperature gradient and only a small vertical temperature gradient in the vessel environment.  


\begin{figure}[h!]
    \centering
    \includegraphics[width=0.4\textwidth]{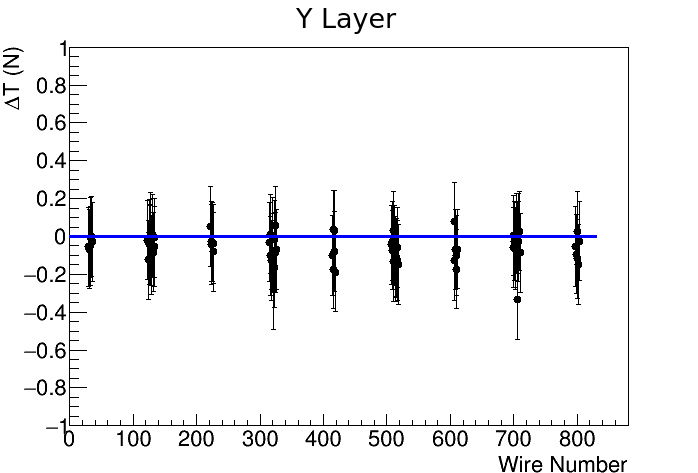}
    \includegraphics[width=0.4\textwidth]{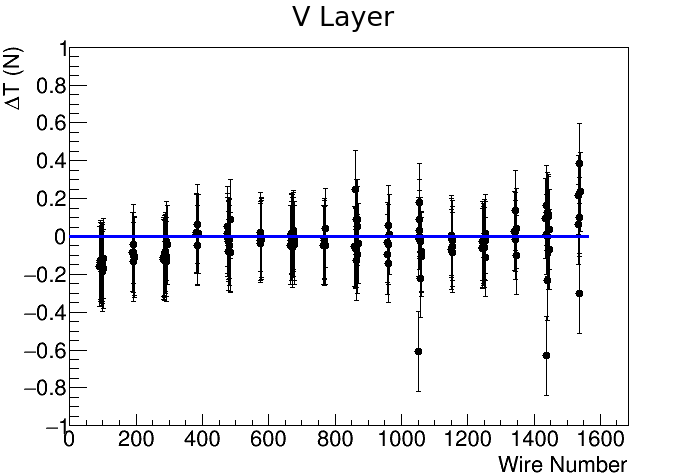}
    \includegraphics[width=0.4\textwidth]{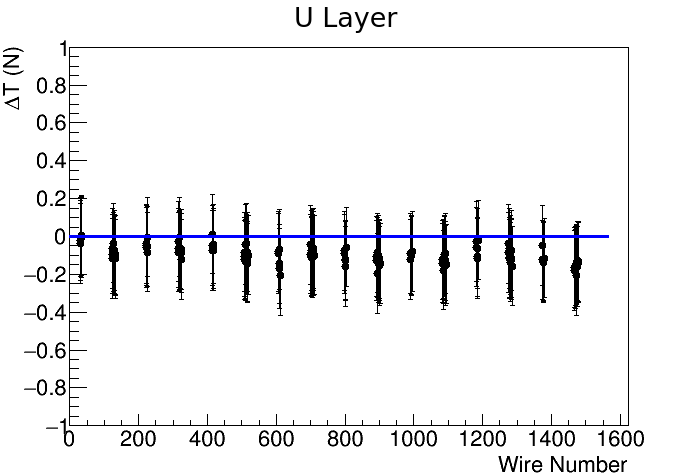}
    \caption{The change in the tension before and after the cold test for all three layers of the UK left-hand APA. The blue line shows the line of zero change.}
    \label{fig:ColdTestTension}
\end{figure}

\begin{figure}[h!]
    \centering
    \includegraphics[width=0.4\textwidth]{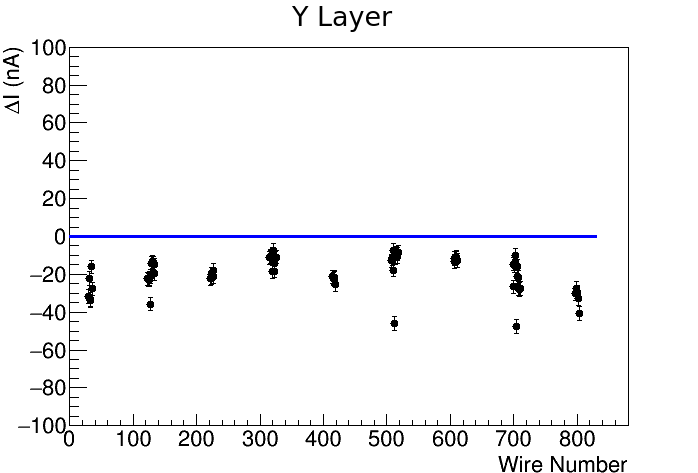}
    \includegraphics[width=0.4\textwidth]{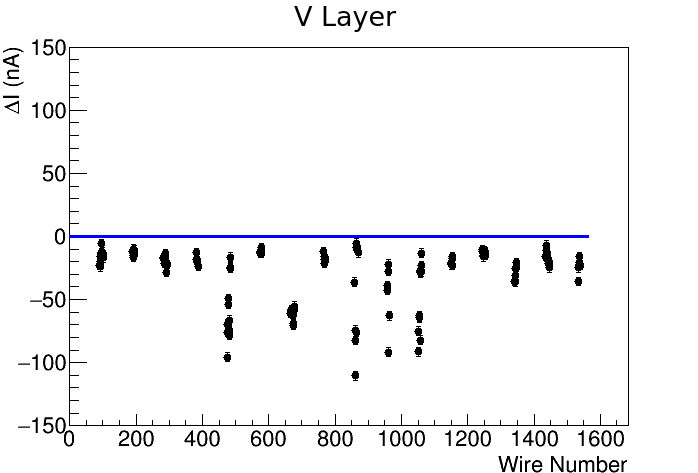}
    \includegraphics[width=0.4\textwidth]{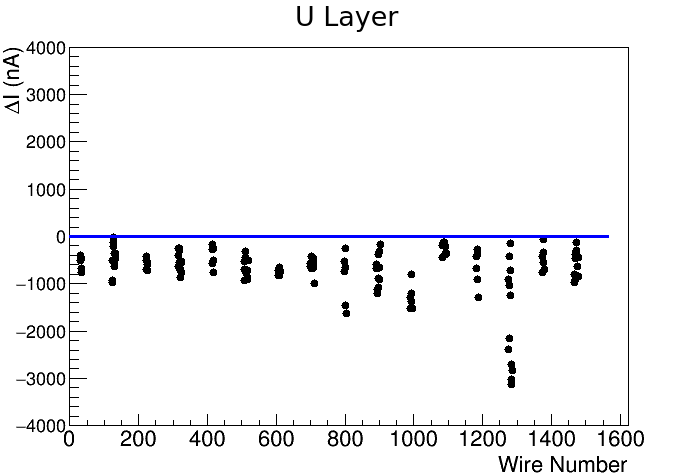}
    \caption{The change in the channel-to-channel isolation currents before and after the cold test for all three layers of the UK left-hand APA. The blue line shows the line of zero change.}
    \label{fig:ColdTestIsolation}
\end{figure}

\begin{figure}[h!]
    \centering
    \includegraphics[width=0.4\textwidth]{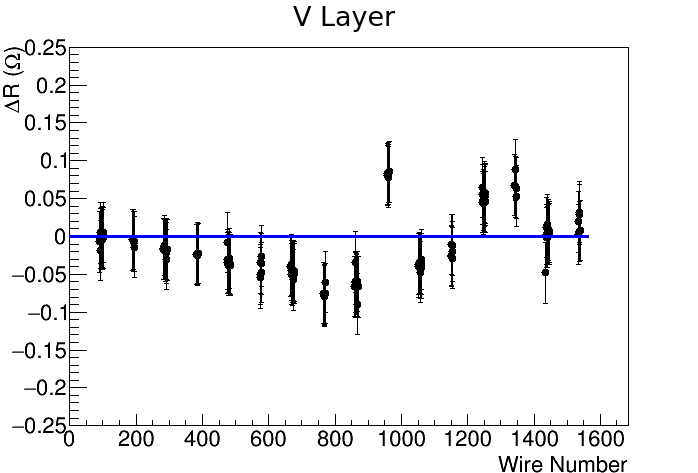}
    \includegraphics[width=0.4\textwidth]{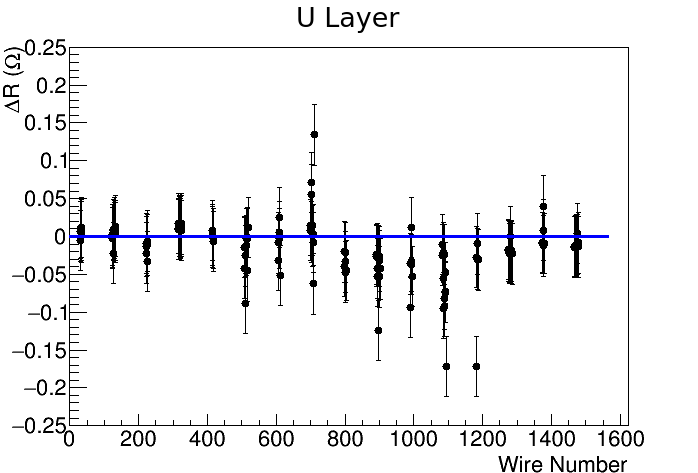}
    \caption{The change in the resistance across each wire before and after the cold test for the U and V layers of the UK left-hand APA. The Y plane was no longer accessible for this measurement at the time of measuring. The blue line shows the line of zero change.}
    \label{fig:ColdTestContinuity}
\end{figure}

Prior to removal of the APA from the vessel, it was visually assessed, and appeared unchanged. Tension and electrical isolation were measured on a subset of the wires of each plane both before the APA was cooled, and after it had returned to room temperature. Measurement of Y-layer continuity using the measurement techniques described in section \ref{sec:Testing} was not possible, due to certain test-points being  inaccessible after the U and V layer installation. Continuity was re-measured for the induction planes. The results of these tests are shown in figures \ref{fig:ColdTestTension} through \ref{fig:ColdTestContinuity}, and summarised in table \ref{tab:ColdTestChanges}.

\begin{table}[hbtp]
\centering
\caption{\label{tab:ColdTestChanges} Changes observed after the cold test.}
\begin{tabular}{|c|c|c|c|}
\hline
    Plane & Variable & Mean change & RMS of change \\
\hline
\hline
    Y & Tension    & -0.06~N  & 0.07~N \\
      & Isolation  & -19~nA   & 9~nA   \\
\hline
    V & Tension    & -0.04~N       & 0.12~N       \\
      & Isolation  & -31~nA        & 26~nA        \\
      & Continuity & -12~m$\Omega$ & 42~m$\Omega$ \\
\hline
    U & Tension    & -0.10~N       & 0.04~N       \\
      & Isolation  & -692~nA       & 534~nA       \\
      & Continuity & -16~m$\Omega$ & 36~m$\Omega$ \\
\hline
\end{tabular}
\end{table}

The observed change in tension was at most 0.1~N, while changes in continuity were no larger than 16 m$\Omega$ (and statistically consistent with 0). Isolation currents were observed to decrease (taking them further within specifications) due to the freeze-drying of moisture on the electronics boards. These results were consistent with the expectation of minimal mechanical or electrical change, and clearly showed that the performance of the APAs would not be negatively affected by immersion in liquid argon.

In summary, the quality control testing of the APA planes has verified that the tension, electrical continuity and electrical isolation of all wires are within specification, and furthermore they remain so after cryogenic testing and transportation to Fermilab.  

\newpage

\section{Lessons Learned}
The construction of the SBND APAs provided several insights to inform and improve the construction of future TPC wire planes -- particularly for the DUNE experiment, which will require the manufacture of 150 APAs of similar scale (6~m $\times$ 2.3~m) per 10~kt module~\cite{Abi:2018alz}\cite{Brailsford:2018dzn}. In this section are summarised three of the lessons learned that are most relevant to other LArTPC experiments: the potential improvements to be made to the wire tension measurement method, the importance of rigorous alignment of the APA components, and the susceptibility of the wires to kinking when bent.

\subsection{Tension Measurement Method}

The laser photodiode apparatus detailed in section \ref{subsec:TensionTests} successfully provided tension measurements which were accurate to within 5\% for over 98\% of the wires sampled. However, the process of taking these measurements was time-consuming: reducing the time taken per measurement would have greatly reduced the total time required for APA construction.

The most time-consuming part of operating the laser photodiode was the fine adjustment of its position that was required to achieve clear focus on each measured wire. Secondary vibrational modes on the wires, sources of vibration from the environment, and obstruction by intervening wire layers above the focal plane also had the potential to complicate these measurements, making their results require careful interpretation (and sometimes repetition). The time taken for tension measurements could be greatly reduced by automating the process of focusing the laser on a given wire, and by parallelising the measurement procedure to measure several wires at a time.

It should also be noted that the laser photodiode method was not suitable for measuring the tension of extremely short wires, as discussed in section \ref{subsec:TensionTests}. These wires make up fewer than 2\% of the wires measured, are less susceptible to movement in the argon at low tension than longer wires, and will usually fall outside the fiducial volume of any analysis performed on the TPC data due to their positions in the extreme corners of the wire layers. Nevertheless, it would be desirable to develop a complementary means of tension measurement for these short wires.

\subsection{APA Component Alignment}

In order to maintain the positioning and tension of the APA wires and spacing of the wire planes, precision measurements of alignment were made at each stage of construction, as each component was added. The successful outcome of this process demonstrates how important it is for any TPC of this size to verify layer-to-layer spacing at every stage of construction, as millimetre-scale displacements over metre-scale objects are very easy to produce.

There are a few notable instances, for example the alignment of the geometry boards that bear the wires and the levelling bars that support those boards was critical to ensuring the uniformity of the wire-to-wire and layer-to-layer spacing, as discussed in section \ref{subsec:APAFrame}. As discussed in that section, verification measurements at each stage of production highlighted the need for minor adjustments.  Implementation of these corrections allowed us to maintain the specified layer-to-layer spacing.  

Another notable effect was the positions of the wrap edge boards when wires were laid using the semi-automated method.  As discussed in section \ref{subsec:TensionResults}, the tension on the wires created small displacements of the boards holding them, which resulted in small tension gradients across each board.  For any apparatus which is laying wires singly, this gradient should be anticipated, and could be countered by making the boards thicker, and therefore more robust against the nominal load tension of 224~N for a 32 wire geometry board.  Another solution would be to lay all wires attached to a given board simultaneously, as demonstrated by the absence of this effect in the results from the manual apparatus.

\subsection{Wire Kinking}

Care should be taken at all times when handling wire for TPC construction to ensure that wire is never bent around a short enough radius of curvature to produce a kink. The ease with which the wire used on SBND can be kinked was demonstrated by the strong local variation in the tension on the APAs wired with the manual apparatus, visible in figures \ref{fig:US_LH_TensionDists} and \ref{fig:US_RH_TensionDists}. 

This variation is due to the wire kinking around the rotating pins of the tensioning blocks, preventing the pins from evenly transmitting tension as discussed in section \ref{subsec:US_Apparatus}. Operators of the manual apparatus could observe permanent kinks being created where wire was wound around these pins even before any tension had been applied using the tensioning bolt. Any kink on a length of wire subjected to cryogenic conditions is a weak point that carries a strong risk of breakage, and wires with kinks in the straight length between the tensioning blocks were observed to break extremely easily when tensioned using the manual apparatus.

The kinks around the rotating pins on the manual apparatus were on lengths of wire that were cut away during the process of wire-winding, and operators at both sites were vigilant to inspect the wire layers for kinks as they were being laid. Future TPCs might consider procedures for wire handling that minimise the opportunities for the wire to be bent.

\section{Conclusions}
The SBND collaboration has produced four APAs to rigorous specifications, succesfully demonstrating two different wiring techniques. To summarise how well the specifications have been met:
\begin{itemize}
    \item The tension distributions of the wires are consistent and single-peaked for each plane at values close to the target of 7~N, with a standard deviation not exceeding 0.6~N, and above the specification of 5~N at the level of $99.8\%$.
    \item The resistance between neighbouring channels exceeds the specified lower limit of 10~M$\Omega$ by at least a factor of $10^3$ on all 15,860 wires.
    \item The resistance across a single wire falls below the specified maximum of 500~$\Omega$ by at least a factor 25 on $100\%$ of channels.
    \item Through detailed laser surveys, the APA frame flatness has been successfully controlled to within 0.5~mm.
    \item A fully-wired APA was cooled to 150~K, and after returning to room temperature showed no changes in its characteristics that would adversely impact its performance.
\end{itemize}


The consistency and precision with which these wire planes have been manufactured provides confidence for the large-scale manufacture of wire planes for DUNE, which will require 150 APAs per 10~kt module~\cite{Abi:2018alz}. In particular, the measurement techniques used for tension, isolation and continuity have direct applications to the DUNE project.


\acknowledgments

This material is based upon work supported by the National Science Foundation under Grant No. PHY-1428753. The US team would also like to thank Lee Greenler for his work designing and prototyping the manual winding apparatus, Tom Hurteau and Frank Lopez for their invaluable support, Aditi Shetty, Annie Polish, Anna Gumberg, Brendon Bullard and London Cooper-Troendle for their contributions to the wire-winding effort at Yale, and all the staff at Wright Lab for their tremendous hospitality to our wire-winding team. 

This work was supported in the UK by the Science and Technology Facilities Council (STFC), part of United Kingdom Research and Innovation, and the Royal Society of the United Kingdom. The UK team would like to thank Jonathan Mercer for his work in designing and fabricating the wire support combs; Simon Dixon for the APA frames; and Mike Perry for his support. We also thank the staff of the Daresbury Laboratory (UK), in particular Alan Grant, for the provision of Laboratory space and significant technical support.

The collaboration would also like to thank Reidar Hahn at Fermilab for the photography in figures \ref{fig:DoesntExist}, \ref{fig:APAphoto} and \ref{fig:Slats}, and Erik Voirin for his studies on wire forces and vortex induced motion from argon flow.

\medskip
\bibliographystyle{unsrtnat}
\bibliography{sbnd_wiring.bib}

\end{document}